\documentclass[10pt]{article}
\usepackage{graphicx} 
\usepackage{authblk} 
\usepackage[left=3cm, right=3cm, top=3cm, bottom=3cm]{geometry}
\usepackage{mathtools} 
\usepackage{amssymb}
\usepackage{amsthm}
\usepackage{multirow}
\usepackage{array}
\usepackage{hyperref}
\usepackage{enumitem}
\usepackage{setspace}

\usepackage[linesnumbered, ruled]{algorithm2e}
\SetKwInOut{Input}{input} 
\SetKwInOut{Output}{output}

\usepackage{pgf,tikz} 
\usetikzlibrary{arrows, fit, shapes, arrows.meta, snakes,
positioning, matrix}
\usepackage{tkz-graph}
\pgfdeclarelayer{bg}
\pgfsetlayers{bg,main}

\newcounter{example}
\newenvironment{example}[1][]{\refstepcounter{example}\par\medskip
   \noindent \textbf{Example~\theexample. #1} \rmfamily}{$\circ$}

\newcommand{\G}{\mathcal{G}}
\newcommand{\D}{\mathcal{D}}
\newcommand{\C}{\mathcal{C}}
\newcommand{\I}{\mathcal{I}}
\newcommand{\K}{\mathcal{K}}
\newcommand{\X}{\mathbf{X}}
\newcommand{\V}{\mathbf{V}}
\newcommand{\E}{\mathbf{E}}
\newcommand{\R}{\mathcal{R}}
\newcommand{\F}{\mathcal{F}}

\newcommand{\adj}[2]{\mathrm{adj}_{{#1}}({#2})}

\newcommand{\pa}[2]{\mathrm{pa}_{{#1}}({#2})}
\newcommand{\nd}[2]{\mathrm{nd}_{{#1}}({#2})}
\newcommand{\an}[2]{\mathrm{an}_{{#1}}({#2})}
\newcommand{\de}[2]{\mathrm{de}_{{#1}}({#2})}
\newcommand{\possan}[2]{\mathrm{possan}_{{#1}}({#2})}
\newcommand{\indep}{\perp\mkern-10mu\perp}
\newcommand{\any}{*\mkern-6mu-\mkern-7mu*}

\newtheorem{definition}{Definition}
\newtheorem{proposition}{Proposition}

\definecolor{silvergray}{RGB}{240, 240, 240}
\definecolor{lightblue}{RGB}{164,221,255}

\definecolor{myorange}{RGB}{79, 81, 254}
\newcommand{\myorange}[1]{\textcolor{myorange}{#1}}
\definecolor{myblue}{RGB}{140, 30, 146}
\newcommand{\myblue}[1]{\textcolor{myblue}{#1}}
\definecolor{mypurple}{RGB}{255, 78, 11}
\newcommand{\mypurple}[1]{\textcolor{mypurple}{#1}}

\title{Improving Finite Sample Performance of Causal Discovery\\ by Exploiting Temporal Structure}

\author[1,2]{Christine W. Bang}

\author[1]{Janine Witte}

\author[1]{Ronja Foraita}

\author[1,2]{Vanessa Didelez\footnote{Corresponding author: didelez@leibniz-bips.de}}

\affil[1]{Leibniz Institute for Prevention Research and Epidemiology -- BIPS}
\affil[2]{Faculty of Mathematics and Computer Sciences, University of Bremen}

\date{June 27, 2024}

\begin{document}

\maketitle

\begin{abstract}
    Methods of causal discovery aim to identify causal structures in a data driven way. Existing algorithms are known to be unstable and sensitive to statistical errors, and are therefore rarely used with biomedical or epidemiological data. We present an algorithm that efficiently exploits temporal structure, so-called \emph{tiered background knowledge}, 
    for estimating causal structures. Tiered background knowledge is readily available from, e.g., cohort or registry data. When used efficiently it renders the algorithm more robust to statistical errors and ultimately  increases accuracy in finite samples. We describe the algorithm and illustrate how it proceeds. Moreover, we offer formal proofs as well as examples of desirable properties of the algorithm, which we demonstrate empirically in an extensive simulation study. To illustrate its usefulness in practice, we apply the algorithm to data from a children's cohort study investigating the interplay of  diet, physical activity and other lifestyle factors for health outcomes. 
\end{abstract}


\section{Introduction}

Discovering causal structures in data is a challenging task. Ideally, we would like to input the data into an algorithm that outputs one or more plausible causal \emph{directed acyclic graphs} (DAGs) linking the variables in the data. Such data driven approaches for estimating a causal DAG are known as \emph{causal discovery} (or causal search, structure learning etc.).  While   algorithms for  causal discovery were first developed in the field of computer science more than twenty years ago \cite{spirtes2000} and have, since then, continually been generalised and refined \cite{vowels:21}, their use with biomedical or epidemiological data is still rare (other than in genetics \cite{maathuis:nature}). Exceptions are, for example, two applications of causal discovery to data from cohort studies finding that modifiable risk factors in early childhood or early life have mostly indirect, if any, causal relations with later health outcomes\cite{petersen2021,foraita2022}.  Similarly, in an analysis of  healthcare data considering cardiac surgery it was found that many of the known predictors were in fact only indirect causes of  postoperative length of stay \cite{lee2022causal}. Also, it has been suggested to use  causal discovery to improve the quality of care for hip replacement patients by investigating the complex clinical performance of implants with data from large patient registries \cite{cheek2018application}.  

Typical causal analyses often aim at  causal effect estimation. In contrast to the above, such analyses typically assume the causal structure, i.e.\ the DAG,  to be given, usually derived from domain expertise, which may entail `confirmation bias'.
Moreover, the question whether the expert knowledge is correct or compatible with the data is typically not addressed \cite{tennant2021}. Data driven methods for estimating causal DAGs provide an alternative, or supplement, to relying solely on expert knowledge \cite{didelez2024invited}.
For instance, in the context of building a life course model for depression in early old age, it was found that  different experts' causal DAGs  and  the output of a causal discovery algorithm  exhibited considerable disagreement \cite{petersen2023constructing}. But the comparison also suggested that the discovery algorithm detected several causal relations which were, post hoc, found to be plausible by the experts.

While promising and potentially useful, algorithms for causal discovery face a number of  issues in practical applications (see recent overview  \cite{didelez2024invited}). One major problem is that they can be unstable or 
ambiguous with finite data due to sensitivity to sampling variation. 
In this paper, we describe how temporal structure of the data, or more generally \emph{tiered background knowledge}, can efficiently be  exploited to improve causal discovery algorithms, rendering the methods more stable and robust  for practical applications. This is very useful as temporal information is often available in biomedical or epidemiological data, e.g., when using  patients' records or   cohort data   as in life course epidemiology \cite{kuh2004}, and it is usually unambiguous and correct.

We will focus on the PC algorithm \cite{spirtes2000}, which is a so-called constraint-based causal discovery algorithm. It has the advantages that it relies on  relatively few assumptions, and that it is  flexible, allowing for various extensions and adaptations to different data types. However, the PC algorithm also suffers from the  limitations mentioned above: One is a lack of informativeness and the other is sensitivity to statistical errors. The first problem is due to the limited assumptions, which  imply that some causal directions cannot be identified even from `infinite data'. In contrast, the second problem occurs in practice when we apply the algorithm to finite samples. 
While the PC algorithm has desirable asymptotic properties \cite{kalisch2007estimating, harris2013pc}, with finite samples it is prone to errors due to erroneous conclusions based on statistical tests propagating through the algorithm. It was previously shown that tiered background knowledge improves the informativeness of the estimated graphs from infinite data \cite{bang2023we}. In the present  paper we specifically address finite sample properties and  show how making efficient use of tiered background knowledge increases the robustness towards statistical errors yielding more accurate outputs. 

We consider an extension of the PC algorithm that exploits tiered background knowledge, and we refer to this as the tiered PC (tPC) algorithm.
Such an extension was previously suggested \cite{spirtes2000}, and versions of the tPC algorithm have been implemented in the TETRAD software \cite{scheines1998} and R packages \cite{witte2022b, petersen2023, andrews2023}, and have been applied  \cite{petersen2021,foraita2022}. However, it has not yet been addressed how tiered background knowledge is  used most efficiently, and to our knowledge, this paper is the first to thoroughly examine the finite sample properties of the tPC algorithm.

The outline of the paper is as follows. Section \ref{sec:theory} introduces  the basic concepts of the paper, which we illustrate with a toy example of a children's cohort study. Section \ref{sec:discovery} describes versions of the PC algorithm, and Section \ref{sec:backgroundknowledge} formally introduces the concept of (tiered) background knowledge. In Section \ref{sec:alg} we introduce the tPC algorithm,  and  show important asymptotic and finite sample properties.
In Section \ref{sec:simstudy} we conduct an extensive simulation study exploring the finite sample performance of the tPC algorithm. In Section \ref{sec:application} we show how the algorithm can be used in practice by applying it to data from an actual children's cohort study  investigating the interplay of  diet, physical activity and other lifestyle factors with regard to several health outcomes \cite{ahrens2017}. 
Finally, Section \ref{sec:discussion} provides a discussion of the results.

\section{Basic Concepts}\label{sec:theory}

In this section we  introduce the formal graphical framework and describe the algorithms  which ours extends. Detailed background can be found in Appendix \ref{app:theory}. As (constraint-based) causal discovery reconstructs the causal structure from the pattern of conditional independencies in the data, we focus on the relations between graphs, causal relations and conditional (in)dependencies. 

\subsection{Graphs}

A graph $\G=(\V, \E)$ consists of a set of nodes $\V$ and edges $\E$. In this paper we allow for edges that are either directed ($\rightarrow$) or undirected (--). In some special cases we will use bidirected ($\leftrightarrow$) edges, as explained in Section \ref{sec:alg}. We use $\any$ as a placeholder for an arbitrary edge type. We consider directed acyclic graps (DAGs), partially directed acyclic graphs (PDAGs), completed PDAGs (CPDAGs), and maximally oriented PDAGs (MPDAGs). The \emph{skeleton} of each of these is given as the undirected graph obtained by dropping any edge orientations.

Graphical models link graphs and probability distributions, and we let the nodes represent random variables: With $\G=(\V,\E)$, we consider a set of random variables $\mathbf{X}_\mathbf{V}$ corresponding to the node set $\V$, with individual  random variables  $X_V$ represented by $V\in\V$, and subsets $\mathbf{X}_{\V'}$  of random variables represented by $\V'\subseteq\V$. If  the distribution of $\textbf{X}_\textbf{V}$ satisfies the Markov properties with respect to $\G$ we can use \emph{d-separation}  to determine (conditional) independencies among the random variables  (this we refer to as an \emph{independence model}): Any d-separation of  two  nodes implies that the corresponding variables are conditionally independent given the {\em separating set}  (see Definition \ref{def:dsep} and Definitions \ref{def:globalmarkov} and \ref{def:localmarkov} in Appendix \ref{app:theory}).  \emph{Adjacent} nodes are those connected by an edge and they are not d-separated by  any subset. 

We are interested in causal relations underlying the data, and we therefore focus on DAGs, which only contain directed edges and have no directed cycles. 
A directed edge $V_i\rightarrow V_j$ represents a direct causal relation in the sense that intervening on $X_{V_i}$  leads  to a change in the distribution of $X_{V_j}$  while  fixing all other direct causes of $X_{V_j}$. A DAG with such a causal interpretation is referred to  as a \emph{causal} DAG and 
will be assumed to satisfy the \textit{causal Markov property}: A random variable is conditionally independent of its non-effects given its direct causes \cite{spirtes2000}.

In order to represent causal structures using DAGs, we need to make the assumption of \emph{causal sufficiency}: For each pair of nodes (variables), all their common causes are nodes in the DAG.

\begin{figure}[!htbp]
\centering
\begin{tikzpicture}

\node[align=center] (A) at (0,1.125) {\small \textsf{parental}\\ \small \textsf{education}};
\node (B) at (0,-.25) {\small \textsf{breastfeeding}};
\node (C) at (2.5,1.75) {\small \textsf{screen time}};
\node (D) at (2.5,0.5) {\small \textsf{sleep}};
\node (E) at (2.5,-0.75) {\small \textsf{well-being}};
\node[align=center] (F) at (5,1.125) {\small \textsf{physical}\\ \small \textsf{activity}};
\node (G) at (5,-.25) {\small \textsf{BMI}};

\node[align=center] (A') at (8,1.125) {\small \textsf{parental}\\ \small \textsf{education}};
\node (B') at (8,-.25) {\small \textsf{breastfeeding}};
\node (C') at (10.5,1.75) {\small \textsf{screen time}};
\node (D') at (10.5,0.5) {\small \textsf{sleep}};
\node (E') at (10.5,-0.75) {\small \textsf{well-being}};
\node[align=center] (F') at (13,1.125) {\small \textsf{physical}\\ \small \textsf{activity}};
\node (G') at (13,-.25) {\small \textsf{BMI}};

\node (a) at (-0.25,-1.25) {(a)};
\node (b) at (7.75,-1.25) {(b)};

\tikzset{dir/.style = {->, -{To[length=6, width=7]}, thick}}
\draw[dir]
(A) edge (B)
(A) edge [bend left] (C)
(B) edge [bend right] (E)
(C) edge (D)
(D) edge (E)
(C) edge [bend left] (F)
(F) edge (G)

;

\tikzset{undir/.style = {-, thick}}
\draw[undir]
(A') edge (B')
(A') edge [bend left] (C')
(B') edge [bend right] (E')
(C') edge (D')
(D') edge (E')
(C') edge [bend left] (F')
(F') edge (G')

;

\end{tikzpicture}
\caption{Toy example of a graph representing the variables in a cohort study. (a) A DAG $\D=(\V, \E)$, where the nodes $\V$ represent the variables. (b) The \emph{skeleton} of $\D$ representing its \emph{adjacencies}.}
\label{fig:dag}
\end{figure}
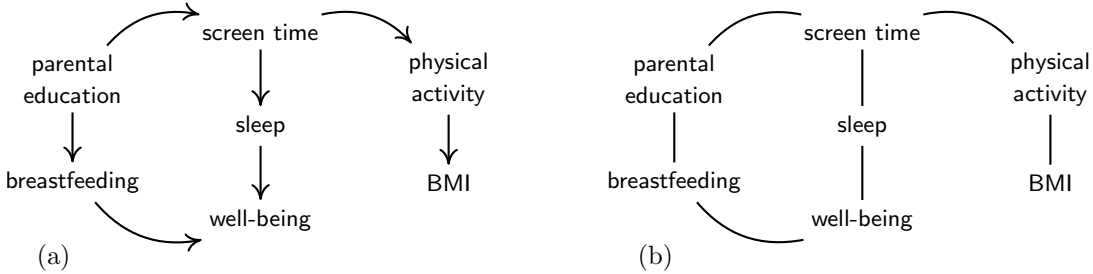

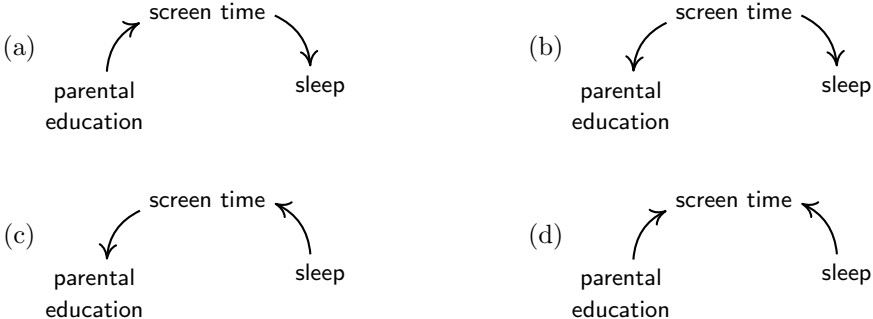
\begin{figure}[!htbp]
\centering
\begin{tikzpicture}

\node[align=center] (a1) at (0,0.75) {\small \textsf{parental}\\ \small \textsf{education}};
\node (a2) at (1.5,2) {\small \textsf{screen time}};
\node (a3) at (3,1) {\small \textsf{sleep}};

\node (a) at (-1,1.5) {(a)};

\node[align=center] (b1) at (7,0.75) {\small \textsf{parental}\\ \small \textsf{education}};
\node (b2) at (8.5,2) {\small \textsf{screen time}};
\node (b3) at (10,1) {\small \textsf{sleep}};

\node (b) at (6,1.5) {(b)};

\node[align=center] (c1) at (0,-1.75) {\small \textsf{parental}\\ \small \textsf{education}};
\node (c2) at (1.5,-.5) {\small \textsf{screen time}};
\node (c3) at (3,-1.5) {\small \textsf{sleep}};

\node (c) at (-1,-1) {(c)};

\node[align=center] (d1) at (7,-1.75) {\small \textsf{parental}\\ \small \textsf{education}};
\node (d2) at (8.5,-.5) {\small \textsf{screen time}};
\node (d3) at (10,-1.5) {\small \textsf{sleep}};

\node (d) at (6,-1) {(d)};

\tikzset{dir/.style = {->, -{To[length=6, width=7]}, thick}}
\draw[dir]
(a1) edge [bend left] (a2)
(a2) edge [bend left] (a3)

(b2) edge [bend right] (b1)
(b2) edge [bend left] (b3)

(c2) edge [bend right] (c1)
(c3) edge [bend right] (c2)

(d1) edge [bend left] (d2)
(d3) edge [bend right] (d2)
;

\end{tikzpicture}
\caption{Examples of DAGs: (a) Is a subgraph of $\D$ from Figure \ref{fig:dag}. (b) and (c) are Markov equivalent to (a). The graph in (d) constitutes a v-structure and is not Markov equivalent to (a), (b) and (c).}
\label{fig.eq}
\end{figure}

\begin{example}[(Graph terminology and interpretation).]\label{ex:dag}
    In the DAG $\D$ in Figure \ref{fig:dag}(a), the node \textsf{screen time} is an \emph{ancestor} of the node \textsf{well-being} (\textsf{well-being} is a \emph{descendant} of \textsf{screen time}). They are not adjacent, so the variable $X_\textsf{screen time}$ indirectly causes the variable $X_\textsf{well-being}$. The causal Markov property states that $X_\textsf{screen time}$ and $X_\textsf{well-being}$ are marginally dependent, but conditionally independent given the separating set  $\{X_\textsf{sleep},X_\textsf{par.edu}\}$ or $\{X_\textsf{sleep},X_\textsf{br.feed}\}$. In contrast, the node \textsf{screen time} is a \emph{parent} of the node \textsf{sleep} (\textsf{sleep} is a \emph{child} of \textsf{screen time}). The corresponding variables $X_\textsf{screen time}$ and $X_\textsf{sleep}$ are marginally dependent, and they are conditionally dependent regardless of which variables we  condition on.

    Since the node \textsf{well-being} has two non-adjacent parents, the three nodes form a \emph{v-structure}: The variables $X_\textsf{br.feed}$ and $X_\textsf{sleep}$ are conditionally independent given either $X_\textsf{par. edu}$, $X_\textsf{screen time}$, or both;  conditioning on $X_\textsf{well-being}$ may induce a dependence (also known as \emph{collider} effect). This illustrates how v-structures encode a particular conditional independence pattern, which is different from the alternative structures, as exemplified by Figure  \ref{fig.eq}. From $\D$, we can conclude that $X_\textsf{par. edu}$ and $X_\textsf{sleep}$ are conditionally independent given $X_\textsf{screen time}$, which is encoded by the structure in Figure \ref{fig.eq} (a). However, Figure \ref{fig.eq} (b) and (c) encode the same conditional independence relation.  Figure \ref{fig.eq} encodes that $X_\textsf{par. edu}$ and $X_\textsf{sleep}$ are marginally independent but become dependent conditionally on $X_\textsf{screen time}$. Note that while (a), (b) and (c) encode the same d-separations, the four structures all have different causal interpretations.
\end{example}

\bigskip

As we have seen in the example of Figure 2, different DAGs  can induce the same set of d-separations and we then refer to them as \emph{Markov equivalent}. We refer to a class of Markov equivalent graphs as a  \emph{(Markov) equivalence class}. 
An equivalence class of DAGs is represented by a CPDAG (a full characterisation of CPDAGs can be found in Andersson et al.\ \cite{andersson1997}). A CPDAG is partially directed in the sense that it contains both directed and undirected edges: The undirected edges represent edges that are directed in one direction in some DAGs in the equivalence class, and in the opposite direction in other DAGs, while the directed edges represent edges that are common to all DAGs in the class. If a CPDAG represents a class of causal DAGs, we will refer to it as a causal CPDAG, and in a causal CPDAG, we will interpret undirected edges $V_i-V_j$ as unidentified causal direction: $X_{V_i}$ could be a direct cause of $X_{V_j}$, but $X_{V_j}$ could also be a direct cause of $X_{V_i}$. Similarly, any PDAG represents a set of DAGs, but not necessarily an equivalence class.  A CPDAG is completed in the sense that it is not possible to orient more edges based on the independence model and the acyclicity alone \cite{meek1995} (see Appendix \ref{app:meekrules}). We refer to this property as the graph being \emph{maximally informative}.

\subsection{Constraint-based causal discovery}\label{sec:discovery}

Assuming the Markov properties, graphs encode conditional independencies among a given set of variables. constraint-based causal discovery aims to solve the inverse problem: To estimate a graph from the conditional independencies in a given dataset by performing a series of statistical tests for conditional independence. Other types of causal discovery methods operate under different assumptions and use different strategies such as, e.g., optimising a score function, but these will not be discussed here.

When estimating graphs based on conditional independence tests, a key assumption is the reverse of the global Markov property: A distribution is {\em faithful} to a graph if any conditional independence in the distribution  corresponds to d-separation  encoded by the graph (see Definition \ref{def:globalmarkov} and Definition \ref{def:faithfulness} in Appendix \ref{app:theory}).
Even under faithfulness, and given the correct independence model, we are still limited by the fact that multiple DAGs can be Markov equivalent. 

The basic PC algorithm \cite{spirtes2000} is a constraint-based algorithm that estimates a CPDAG. The main steps are as follows.

\subsubsection*{The basic PC algorithm}

\begin{itemize}
    \item[(I)] Start with a complete undirected graph over the given nodes.
    \item[(II)] Visiting every pair $(A,B)$ of nodes, remove an edge between them if and only if there exists a subset $\mathbf{S}$ of the nodes adjacent to $A$ such that $X_A$ and $X_B$ are conditionally independent given ${X}_\mathbf{S}$. Continue until no more edges can be removed. The algorithm considers the candidate separating sets in rounds, starting with the empty set and letting the size of the set increase by one each round.
    \item[(III)] If $A$ and $C$ are not adjacent, orient any $A-B-C$ as $A\rightarrow B\leftarrow C$ (v-structure) if and only if $X_A$ and $X_C$ are dependent when conditioning on $X_B$. 
    \item[(IV)] Orient further edges according to Meek's rules 1-3 \cite{meek1995} (see Figure \ref{fig:meeksrules} in Appendix \ref{app:meekrules}) until none of the rules apply anymore.
\end{itemize}

The basic PC algorithm makes use of faithfulness in phase (II) to construct the skeleton, and of the particular conditional independencies encoded by v-structures in phase (III). 

When analysing the properties of causal discovery methods,  the \emph{oracle} version  and the \emph{sample} version should be  distinguished. The oracle version refers to the algorithm when the input are the true  conditional independencies. Under faithfulness and causal sufficiency, 
the oracle version of the basic PC algorithm is sound \cite{spirtes2000} and complete \cite{meek1995}, i.e.  the correct and most informative CPDAG is obtained. The sample version refers to the case where the input consists of conditional independencies found by statistical tests on finite data. These tests make statistical errors so that the input is not necessarily correct;  early errors can  propagate through all iterations of the algorithm as updated adjacencies are used in each round of the skeleton phase (II). The sample version of the basic PC algorithm therefore does not typically output the correct CPDAG, it may  not even  output a valid CPDAG.

Moreover, the output of the  sample version of the basic PC algorithm  depends on the sequence (order) in which the variables are visited. To account for this,  a modification has been introduced, the so-called LMPC-stable algorithm, which is order independent through three phase specific modifications to (II-IV). We provide a short summary of the modifications and refer to the literature for details \cite{colombo2014}.

\subsubsection*{The LMPC-stable algorithm}

\begin{itemize}
    \item[(II)] At the beginning of each round, for each node $A$, save the set of nodes adjacent to $A$ and consider subsets of this set for separation.
    \item[(III)] Majority rule: Determine whether each $A-B-C$ is a v-structure by considering every set of nodes separating $A$ and $C$. If $B$ is not contained in the majority of these sets, it is oriented as a v-structure. Else, the edges are not oriented. If it is contained in exactly half of the sets, the triple is listed as ambiguous.
    \item[(IV)] Orient remaining undirected edges with Meek's rules only if they are not ambiguous. If conflicts occur, indicate these using bidirected edges.
\end{itemize}

Like the basic PC, the oracle version of LMPC-stable is sound and complete, but its sample version has better properties, e.g.\ it is order independent and it can be expected to propagate fewer errors. An alternative  modification is the conservative PC (CPC) algorithm \cite{ramsey2006conservative}, where phase (III) requires $B$ to be in none of the separating sets;  if $B$ is in some but not all separating sets, then the triple is listed as ambiguous.
The sample version of the LMPC-stable algorithm  might output two types of special edges or structures:

\begin{itemize}[align=left]
    \item[\textbf{Conflicts}] Consider the path $A-B-C-D$, where $A$ and $C$, and $B$ and $D$ are not pairwise adjacent. Suppose we found that $A\not\indep C\mid B$ and $B\not\indep D\mid C$. Then we would orient $A\rightarrow B\leftarrow C$ and $B\rightarrow C\leftarrow D$; however, they cannot both be v-structures and this creates a conflict. This is represented by a bidirected edge, $A\rightarrow B\leftrightarrow C\leftarrow D$. Similarly, conflicts (i.e. bidirected edges) can also arise if Meek's rules lead to conflicting orientations.
    
    \item[\textbf{Ambiguous triples}] Consider the triple $A-B-C$ where $A$ and $C$ are not adjacent. Suppose the triple is ambiguous, i.e.  it could not be decided whether it should be oriented as a v-structure or not. Moreover, suppose that in phase (IV) $A\rightarrow B$ were  oriented due to Meek's rules. It would then follow (by Meek's 1st rule) that $B\rightarrow C$. However, this rule presupposes that $\langle A, B, C\rangle$ is not a v-structure which was ambiguous. Hence, the LMPC-stable algorithm leaves $B-C$ undirected in such cases (without further visualization in,  e.g., the \texttt{pcalg} implementation \cite{kalisch2012causal}). 
\end{itemize}

Conflicting edges and ambiguous triples make the estimated graph difficult to interpret since it no longer necessarily represents an equivalence class. However, they  reflect the uncertainty entailed by finite data.

\subsection{Tiered background knowledge} \label{sec:backgroundknowledge}

Often external background knowledge on some aspects of the true causal structure is  available. 
Such background knowledge implies that graphs have more in common than just the independence model. In this section we  show how incorporating background knowledge can improve the informativeness, i.e. help us orient otherwise undecidable edges. In Section \ref{sec:alg} and \ref{sec:simstudy}, we show how background knowledge can reduce  errors and conflicts. 

We define \emph{background knowledge} as a pair $\K=(\R, \F)$, where $\R$ is a set of \emph{required edges} and $\F$ is a set of \emph{forbidden edges}. We then say that a graph $\G$ \emph{encodes} the background knowledge $\K=(\R,\F)$ if $\G$ contains all edges in $\R$ and no edges from $\F$. If $\C$ is a PDAG, then $\K$ and $\C$ are \emph{consistent} if and only if there exists a DAG in the set represented by $\C$, which encodes  $\K$. Background knowledge $\K=(\R,\F)$ is  added to a consistent PDAG $\C$ by orienting  specific undirected edges in $\C$  through enforcing those in $\R$ and ruling out orientations in $\F$.
It has been shown that if we then apply orientation rules 1-4 (see Figure \ref{fig:meeksrules} in Appendix \ref{app:meekrules}) we obtain a maximally informative graph \cite{meek1995}. Note that now rule 4 is needed to ensure completeness. We refer to the resulting graph  as a \emph{maximally oriented partially directed acyclic graph} (MPDAG) \cite{perkovic2017}. An MPDAG represents a \emph{restricted} equivalence class.

In this paper, we are concerned with \emph{tiered} background knowledge as given, say, by a partial temporal ordering of the variables, e.g., due to a cohort study design: 
\begin{definition}[Tiered ordering]
    Let $\V$ be a set of nodes of size $p$ and let $T\in\mathbb{N}$, $T \leq p$. A   tiered ordering of the nodes in $\mathbf{V}$ is a map $\tau: \mathbf{V}\mapsto \{ 1,\ldots ,T\}^p$ that assigns each node $V\in\mathbf{V}$ to a unique tier $t\in\{ 1,\ldots ,T\}$. 
\end{definition}

If there are multiple orderings, $\tau_1$ and $\tau_2$, to be compared we say that $\tau_1$ is \emph{finer} than $\tau_2$ (and $\tau_2$ is \emph{coarser} than $\tau_1$) if for every $V_i,V_j\in\V$: $\tau_2(V_i)<\tau_2(V_j)\Rightarrow\tau_1(V_i)<\tau_1(V_j)$. 

A tiered ordering entails background knowledge in the following way.

\begin{definition}[Tiered background knowledge]
Let $\tau$ be a tiered ordering of the node set $\V$,  the corresponding tiered background knowledge $\K_{\tau}$  is then defined by  $\R_{\tau}=\emptyset$ and  $\F_{\tau}=\{ V_i \leftarrow V_j\mid \tau(V_i)<\tau (V_j); V_i,V_j\in\V \}$.
\end{definition}

In previous work it was shown how tiered background knowledge can be combined with  a consistent CPDAG yielding a restricted equivalence class represented by a \emph{tiered} MPDAG with desirable properties \cite{bang2023we}. Example \ref{ex:tiers} illustrates how tiered background knowledge improves the informativeness by orienting all  \emph{cross-tier} edges.

\begin{definition}[Cross-tier edge]
Let $\G=(\V,\E)$ be a PDAG and $\tau$ a tiered ordering of $\V$. An edge of any type $\{V_i\any V_j\}\in\E$ is a cross-tier edge (relative to $\tau$) if $\tau(V_i)<\tau(V_j)$.
\end{definition}

\begin{example}\label{ex:tiers}
The DAG $A\rightarrow B \rightarrow C$ (1) is Markov equivalent to  $A\leftarrow B \leftarrow C$ (2) and $A\leftarrow B \rightarrow C$ (3), and their equivalence class is represented by the CPDAG $A - B - C$. Assume that the variables $X_A$, $X_B$ and $X_C$ represented by the nodes  are measured at three different time points. This naturally imposes tiered background knowledge: With  $\tau (A) <\tau(B) <\tau(C)$, we have $\K_\tau=\{\emptyset, \{A\leftarrow B, A\leftarrow C, B\leftarrow C\}\}$. This rules out the graphs (2) and (3) leaving only the graph (1).
\end{example}

\bigskip

The following more complex example shows that more than just the cross-tier edges are oriented as a consequence of tiered background knowledge.

\begin{figure}[!htbp]
\centering
\begin{tikzpicture}

\filldraw[silvergray] (7,2.5) rectangle +(2,-4.5)  ;
\filldraw[silvergray] (9.5,2.5) rectangle +(2,-4.5)  ;
\filldraw[silvergray] (12,2.5) rectangle +(2,-4.5)  ;

\filldraw[silvergray] (-1,2.5) rectangle +(7,-4.5)  ;

\node[align=center] (A') at (0,1.125) {\small \textsf{parental}\\ \textsf{education}};
\node (B') at (0,-.25) {\small \textsf{breastfeeding}};
\node (C') at (2.5,1.75) {\small \textsf{screen time}};
\node (D') at (2.5,0.5) {\small \textsf{sleep}};
\node (E') at (2.5,-0.75) {\small \textsf{well-being}};
\node[align=center] (F') at (5,1.125) {\small \textsf{physical}\\ \small \textsf{activity}};
\node (G') at (5,-.25) {\small \textsf{BMI}};

\node[align=center] (A) at (8,1.125) {\small \textsf{parental}\\ \textsf{education}};
\node (B) at (8,-.25) {\small \textsf{breastfeeding}};
\node (C) at (10.5,1.75) {\small \textsf{screen time}};
\node (D) at (10.5,0.5) {\small \textsf{sleep}};
\node (E) at (10.5,-0.75) {\small \textsf{well-being}};
\node[align=center] (F) at (13,1.125) {\small \textsf{physical}\\ \small \textsf{activity}};
\node (G) at (13,-.25) {\small \textsf{BMI}};

\node (t1) at (8, -1.675) {\textsf{Tier 1}};
\node (t2) at (10.5, -1.675) {\textsf{Tier 2}};
\node (t3) at (13, -1.675) {\textsf{Tier 3}};

\node (t4) at (2.5, -1.675) {\textsf{Tier 1}};

\tikzset{undir/.style = {-, line width = 1pt}}
\tikzset{dir/.style = {->, -{To[length=5.5, width=6.5]}, line width = 1pt}}
\draw[dir]
(B') edge [bend right] [myblue] (E')
(D') edge [myblue] (E')
(A) edge [bend left] [mypurple] (C)
(B) edge [bend right] [myblue] (E)
(C) edge [myorange] (D)
(D) edge [myblue] (E)
(C) edge [bend left] [mypurple] (F)
(F) edge [myorange] (G)
;
\draw[undir]
(A') edge (B')
(A') edge [bend left] (C')
(C') edge (D')
(C') edge [bend left] (F')
(G') edge (F')
(A) edge (B)
;

\end{tikzpicture}
\caption{Toy example of a children's cohort study. Left: CPDAG. The edges are oriented because they form a \myblue{\textbf{v-structure}}, in this particular graph no orientations follow from \myorange{\textbf{Meek's orientation rules}}. 
Right: Tiered MPDAG. Edges are oriented either because they are part of a \myblue{\textbf{v-structure}}, they are \mypurple{\textbf{cross-tier edges}}, or they follow from \myorange{\textbf{Meek's orientation rules}}. }
\label{fig:alg}
\end{figure}
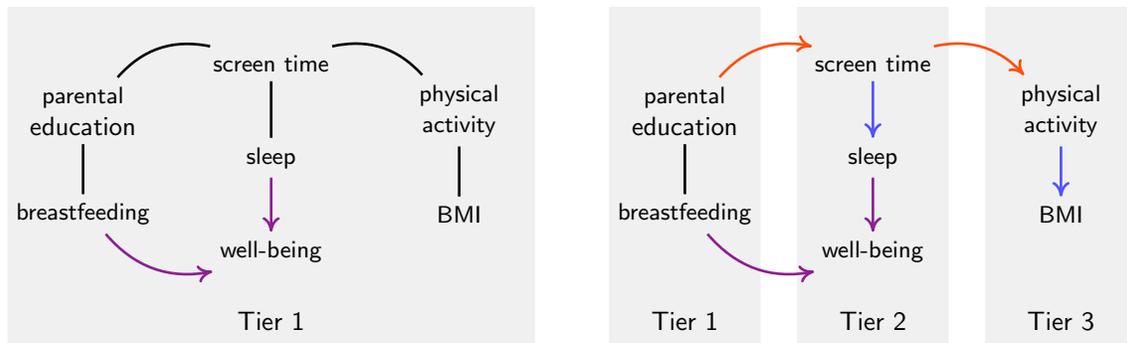

\begin{example}
\label{ex:alg}
    Consider the nodes in DAG $\D$ from Figure \ref{fig:dag} (a) and their time structure

    \begin{itemize}[align = left]
        \item[\textbf{Early life} (tier 1)] \textsf{parental education}, \textsf{breastfeeding} 
        \item[\textbf{Childhood} (tier 2)] \textsf{screen time}, \textsf{sleep}, \textsf{well-being} 
        \item[\textbf{Adolescence} (tier 3)] \textsf{physical activity}, \textsf{BMI} 
    \end{itemize}
   
    The  corresponding CPDAG and  tiered MPDAG  are given in Figure \ref{fig:alg}. With tiered background knowledge two cross-tier edges are oriented, and with Meek's rules two more edges can be oriented. In total, six out of seven edges are directed in the tiered MPDAG, while only two of seven are directed in the CPDAG which uses no background information.
\end{example}

\bigskip

A first approach for combining background knowledge with a discovery algorithm is to impose it  post hoc onto the output. We define the following naive tiered PC (naive tPC) algorithm.

\subsubsection*{The naive tPC algorithm}

\begin{itemize}[align=left]
    \item[Step 1:] Estimate a CPDAG $\C$ using the PC or LMPC-stable algorithm.
    \item[Step 2:] For each adjacent pair of nodes $V_i$ and $V_j$ in $\C$ with $\tau(V_i)<\tau(V_j)$, replace any cross-tier edge $V_i\any V_j$ with $V_i\rightarrow V_j$ to obtain $\C'$. 
    \item[Step 3:] In addition to Meek's rules 1-3, Meek's 4th rule is applied  to $\C'$ as well.
\end{itemize}

The oracle version of  the naive tPC algorithm is sound and complete (this follows directly from Meek's work \cite{meek1995}). 
However, the sample version of the naive tPC algorithm has a major flaw: The  estimated CPDAG of step 1 is not guaranteed to be consistent with the tiered background knowledge. While we can impose inconsistent background knowledge by `brute force' as in step 2, where edge directions may be reversed, it appears more sensible to ensure consistency throughout the algorithm. Moreover, if the background knowledge is correct, exploiting it early on should reduce statistical errors and decrease the required conditional independence tests, as the next Example \ref{ex:nonancestor} illustrates.

\begin{example}\label{ex:nonancestor}
    Assume that we are given three nodes $A$, $B$ and $C$, and an ordering $\tau$ with $\tau(A)=\tau(B)<\tau(C)$ and $\K_\tau=\{\emptyset,\{ A\leftarrow C, B\leftarrow C\}\}$, and that we have to estimate the CPDAG over $A,B,C$. Then, we do not know whether there are any paths between $A$ and $B$, $B$ and $C$, and $A$ and $C$, or whether these paths are causal. It is only known that $C$ cannot be an ancestor of $A$ or $B$; if there is a path between $A$ and $B$ that goes through $C$ it must contain a collider at $C$ and is therefore blocked. Hence, to decide whether $A$ and $B$ are adjacent we need not carry out any statistical tests that condition on $C$.
\end{example}

\section{The tPC algorithm}\label{sec:alg}

We now introduce a way how to ensure consistency  with the tiered background knowledge throughout the sample versions of the search algorithms while avoiding any unnecessary conditional independence tests. Below we describe how the phases of either the basic PC or the LMPC-stable algorithm  are to be altered. We loosely refer to this as `the tPC algorithm'; when specifically applied to the basic PC algorithm we obtain the `basic tPC' algorithm and when applied to the LMPC-stable algorithm we obtain the `tLPMC-stable' algorithm. Due to its better finite sample properties, we mostly  focus on the tLMPC-stable algorithm of which formal details are provided as Algorithm \ref{alg:tpc} in Appendix \ref{app:algs}.

\subsubsection*{The tPC algorithm}

\begin{itemize}
    \item[(II)] Only test whether $A$ and $B$ should be adjacent given separating sets that are either in the same or earlier tiers as $A$. 
    \item[(IIIa)] 
    A pattern  $A-B-C$ ($A$ and $C$ non-adjacent) is not considered a potential v-structure if $\tau(B)<\tau(A)$ or $\tau(B)<\tau(C)$; all other cases are traversed. When deciding whether $A-B-C$ is a v-structure, only separating sets that are in the same or previous tiers as $A$ or $C$ are considered.
    \item[(IIIb)] Additional orientation of cross-tier edges: Any undirected edge $A-B$ with $\tau(A)<\tau(B)$ is oriented as $A\rightarrow B$.
    \item[(IV)] In addition to Meek's rules 1-3, Meek's 4th rule is applied as well.
\end{itemize}

The modifications of (II) and (III) reflect that variables in the future cannot be direct causes of earlier ones, and are thus not needed for obtaining independence.  The modification of (IIIa) takes into account that if $B$ is in a later tier than $A$ and $C$, then this clearly forms a collider at $B$ and no independence test is performed for the orientation, as  illustrated in Example \ref{ex:nonancestor}. In addition, (IIIa) ensures that no v-structures contradicting the tiered ordering are created. 
Step (IIIb) includes an additional step where  cross-tier edges are oriented as a direct consequence of forbidden edges, see  illustration in Example \ref{ex:tiers}. In (IV) the additional 4th rule is needed to ensure that the output is maximally informative, i.e.\ includes any additional orientations that are consequences of the tiered ordering and acyclicity (c.f. Meek \cite{meek1995}).
Note that in the first round of (II) with separating sets of size zero, the same number of tests are carried out by the sample versions of tPC as by the PC algorithm; in the next round with separating sets of size one, tPC typically performs fewer tests as it avoids conditioning on the future (see Figures \ref{fig:n_tests} and \ref{fig:n_tests_sparse} in Appendix \ref{app:ntestnedges}). As the following rounds depend on the updated adjacencies, we cannot say definitely that the tPC algorithm has fewer tests in these later rounds, but it still omits separating sets in the future.

\subsection{Soundness and completeness}

The following result establishes soundness and completeness of the oracle version of the tPC algorithm.

\begin{proposition}\label{prop:soundandcomplete}
Let $\D=(\V,\E)$ be a DAG  and assume that the distribution over the random variables represented by $\V$ is faithful to $\D$. Then, given tiered background knowledge  $\K_{\tau}$  encoded by $\D$ and oracle knowledge of the conditional independencies, the tPC algorithm is sound and complete.
\end{proposition}

The proof of Proposition \ref{prop:soundandcomplete} can be found in Appendix \ref{app:proofs} for the tLMPC-stable algorithm (Algorithm \ref{alg:tpc}).

\bigskip

This result means that the oracle versions of the  tPC algorithm output the unique MPDAG representing the independence model and the tiered background knowledge, i.e. separating sets `in the future' can indeed be ignored as in phases (II) and (IIIa). However, the oracle version of the naive tPC algorithm is also sound and complete, and we now turn to the properties under finite samples. We have the following proposition.

\begin{proposition}\label{prop:consistent}
    Let $\K_\tau$ be the tiered background knowledge used by the tPC algorithm. Under any input set of conditional independencies, the partially directed graph obtained at the end of  phase (IIIa) of the tPC algorithm is consistent with $\K_\tau$.
\end{proposition}

Unlike the sample version of the naive tPC algorithm, the above states that the sample versions of the basic tPC or tLPMC-stable algorithm are consistent with the tiered background knowledge before using it to orient the cross-tier edges. 

Proposition \ref{prop:consistent} only concerns the skeleton and the v-structures. In phase (IIIb), background knowledge adds more information to the estimated graph, and potentially creates more v-structures. Note that it is important to apply Meek's rules only in phase (IV) after (IIIb). If we were to switch (IIIb) and (IV) of the tPC algorithm, the sample version would  not guarantee the consistency with the tiered background knowledge of  all  orientations following  the application of Meek's rules. Hence,  we should orient cross-tier edges  before applying Meek's rules in order to avoid unnecessary errors, as illustrated in the following example.

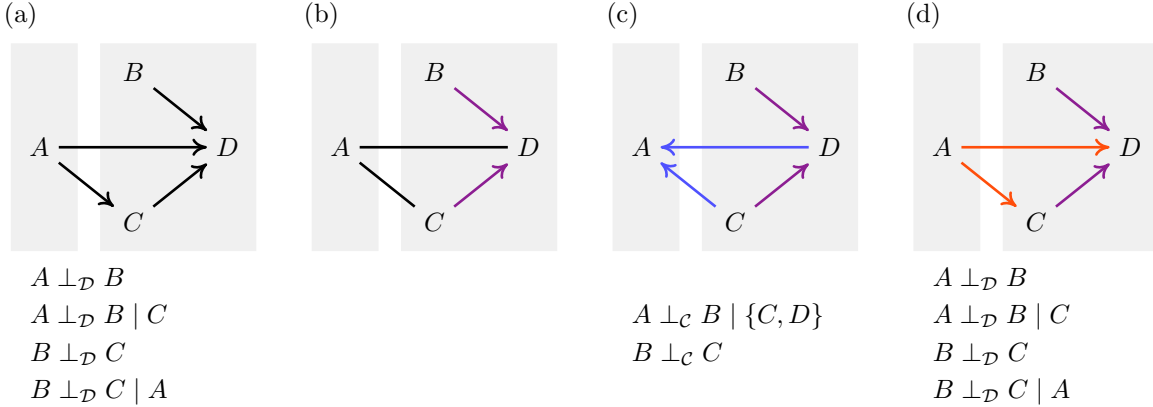
\begin{figure}[!htbp]
\centering
\begin{tikzpicture}

\filldraw[silvergray] (-0.375, 1.375) rectangle +(0.875,-2.75);
\filldraw[silvergray] (0.8125, 1.375) rectangle +(2.0625,-2.75);

\filldraw[silvergray] (3.625, 1.375) rectangle +(0.875,-2.75);
\filldraw[silvergray] (4.8125, 1.375) rectangle +(2.0625,-2.75);

\filldraw[silvergray] (7.625, 1.375) rectangle +(0.875,-2.75);
\filldraw[silvergray] (8.8125, 1.375) rectangle +(2.0625,-2.75);

\filldraw[silvergray] (11.625, 1.375) rectangle +(0.875,-2.75);
\filldraw[silvergray] (12.8125, 1.375) rectangle +(2.0625,-2.75);

\node (A1) at (0,0) {$A$};
\node (B1) at (1.25,1) {$B$};
\node (C1) at (1.25,-1) {$C$};
\node (D1) at (2.5,0) {$D$};

\node (A2) at (4,0) {$A$};
\node (B2) at (5.25,1) {$B$};
\node (C2) at (5.25,-1) {$C$};
\node (D2) at (6.5,0) {$D$};

\node (A3) at (8,0) {$A$};
\node (B3) at (9.25,1) {$B$};
\node (C3) at (9.25,-1) {$C$};
\node (D3) at (10.5,0) {$D$};

\node (A4) at (12,0) {$A$};
\node (B4) at (13.25,1) {$B$};
\node (C4) at (13.25,-1) {$C$};
\node (D4) at (14.5,0) {$D$};

\tikzset{undir/.style = {-, line width = 1pt}}
\tikzset{dir/.style = {->, -{To[length=5.5, width=6.5]}, line width = 1pt}}
\draw[dir]
(A1) edge (C1)
(A1) edge (D1)
(B1) edge (D1)
(C1) edge (D1)

(B2) edge [myblue] (D2)
(C2) edge [myblue] (D2)

(B3) edge [myblue] (D3)
(C3) edge [myblue] (D3)
(D3) edge [myorange](A3)
(C3) edge [myorange] (A3)

(B4) edge [myblue] (D4)
(C4) edge [myblue] (D4)
(A4) edge [mypurple] (D4)
(A4) edge [mypurple] (C4)
;
\draw[undir]
(A2) edge (C2)
(A2) edge (D2)
;

\node (a) at (-0.25, 1.75) {(a)};
\node (b) at (3.75, 1.75) {(b)};
\node (c) at (7.75, 1.75) {(c)};
\node (d) at (11.75, 1.75) {(d)};

\node[anchor=west] (x1) at (-0.25, -1.75) {$A\perp_\D B$};
\node[anchor=west] (x2) at (-0.25, -2.25) {$A\perp_\D B\mid C$};
\node[anchor=west] (x3) at (-0.25, -2.75) {$B\perp_\D C$};
\node[anchor=west] (x4) at (-0.25, -3.25) {$B\perp_\D C\mid A$};

\node[anchor=west] (x2) at (7.75, -2.25) {$A\perp_{\C} B\mid \{C, D\}$};
\node[anchor=west] (x3) at (7.75, -2.75) {$B\perp_{\C} C$};

\node[anchor=west] (x1) at (11.75, -1.75) {$A\perp_\D B$};
\node[anchor=west] (x2) at (11.75, -2.25) {$A\perp_\D B\mid C$};
\node[anchor=west] (x3) at (11.75, -2.75) {$B\perp_\D C$};
\node[anchor=west] (x4) at (11.75, -3.25) {$B\perp_\D C\mid A$};

\end{tikzpicture}
\caption{Example of how tiered background knowledge can prevent statistical errors. Figure (a) depicts the true DAG $\D$ and the d-separations encoded by the graph. Figure (b) depicts an intermediate PDAG $\C'$ obtained from correct independence tests and one erroneous finding of $X_A\indep X_B\mid X_D$. Figure (c) depicts a CPDAG $\C$ obtained from applying Meeks's rules to (b) -- note that $\C$ does not represent the equivalence class of $\D$. Figure (d) depicts the tiered MPDAG estimated with tPC based on the erroneous independence test, combined with tiered background knowledge.}
\label{fig:consistent}
\end{figure}

\begin{example}\label{ex:consistent}
Consider the DAG $\D=(\V,\E)$ in Figure \ref{fig:consistent} (a) and the tiered ordering $\tau$ of the nodes $A$, $B$, $C$ and $D$ with $\tau(A)<\tau(B)=\tau(C)=\tau(D)$. Assume that the distribution of the variables is faithful to $\D$, then we will expect to observe the following independencies: $X_A\indep X_B$, $X_A\indep X_B\mid X_C$, $X_B\indep X_C$ and $X_B\indep X_C\mid X_A$. 
Suppose that due to some statistical error, the observed distribution in addition to the correct independencies also yielded $X_A\indep X_B\mid X_D$. Then, the tPC algorithm would still recover the correct skeleton of $\D$ as well as the v-structure $B\rightarrow D\leftarrow C$. In this setup, when determining whether $A\any D\any B$ is a v-structure, the tPC algorithm only performs one conditional independence test, which is exactly the test of whether $X_A$ and  $X_B$ are independent given $X_D$: The potential separating sets consists of all nodes that are adjacent to $A$ and belong to the same or earlier tier as $A$, and all nodes that are adjacent to $B$ and belong to the same or earlier tier as $B$ (see Algorithm \ref{alg:vstructures} in Appendix \ref{app:algs}). The only node that satisfies this is $D$, i.e., the potential separating set $(C,D)$ is not  visited. With the statistical error, the tPC algorithm would then determine that $A\any D\any B$ could not be a v-structure, since $D$ is in all visited sets separating $A$ and $B$, and the edge  $A-D$ would  remain undirected. The resulting PDAG after phase (IIIa)  is depicted in Figure \ref{fig:consistent} (b). Note that this graph  does not represent the equivalence class of $\D$, but it is still consistent with $\tau(A)<\tau(B)=\tau(C)=\tau(D)$. 
    
Suppose now that we switch phases (IIIb) and (IV). Then, under the above erroneous independence, we would obtain the graph in Figure \ref{fig:consistent} (b) after the v-structure phase (IIIa), and  the graph in Figure \ref{fig:consistent} (c) after applying (IV) Meek's rules. Clearly, the two new edge orientations violate the tiered ordering, and in addition, the graph does not encode the correct independencies. Since early  errors in the algorithm might propagate through the following steps, in particular through Meek's rules, as illustrated here, it is important to exploit background knowledge as early as possible and only apply Meek's rules at the end.

Returning to the tPC algorithm as proposed here, phase (IIIb) would orient cross-tier edges and  obtain the graph in Figure \ref{fig:consistent} (d). This graph is consistent with the tiered background knowledge, and we have in fact recovered the full DAG. Note that even though phase (IIIa) could not recover the v-structure $A\rightarrow D\leftarrow B$ due to a statistical error, the tiered background knowledge recovers it in the next step. Also, without background knowledge we would not be able to orient $A\rightarrow C$, which is not identifiable  from independencies alone. On the other hand, $A\rightarrow D$ would  be identifiable from oracle knowledge of the independence model but not with the statistical error. In this particular example, exploiting background knowledge allows us to orient both of these undirected edges. In Section \ref{sec:simstudy} we illustrate via simulation that the tPC algorithm typically results in fewer errors. 
\end{example}

\subsection{Order independence}\label{sec:stability}

As mentioned in Section \ref{sec:discovery}, statistical errors may result in the output of the sample version of the basic PC algorithm being dependent on the order in which the independence tests are performed, and thus on the input sequence of the variables. This order dependence is eliminated by the LMPC-stable algorithm.  Here, we establish formally that the tLMPC-stable  algorithm is also order independent.

\begin{proposition}\label{prop:stable}
The output of the tLMPC-stable algorithm does not depend on the sequence in which the variables are visited.
\end{proposition}

The proof of Proposition \ref{prop:stable} can be found in Appendix \ref{app:proofs}.

\bigskip

As discussed in Section \ref{sec:discovery}, the output of the sample version of the LMPC-stable algorithm may contain bidirected edges, and the same is the case for the tLMPC-stable algorithm. However, the latter can be expected to have fewer bidirected edges since some of the conflicts might be prevented or resolved by  the tiered background knowledge. We illustrate this empirically in Section \ref {sec:simstudy}. This means that even in cases where LMPC-stable algorithm does not output a valid CPDAG, it might still be possible to obtain a valid tiered MPDAG with the tLMPC-stable algorithm. 

In the remainder we only consider the LMPC-stable or tLMPC-stable  algorithms when we refer to the (t)PC algorithms.

\subsection{Informativeness and robustness}

In the oracle version of the PC algorithm, the benefit of incorporating (tiered) background knowledge is that we cannot lose, and typically gain, informativeness by possibly orienting some of the undirected edges. Surprisingly,  the same  is not valid for the sample version: The restricted equivalence class output by the tPC algorithm can actually be larger than that of the PC algorithm when using finite samples. Hence, there might not be a gain in informativeness. Instead, using tiered background knowledge offers robustness by reducing  errors, as explained earlier, and in this way, we expect the output to be at least as or more reliable than the output obtained without background knowledge. In the example below, we illustrate how incorporating tiered background knowledge in the PC algorithm can increase the estimated equivalence class.

\begin{figure}[!htbp]
\centering
\begin{tikzpicture}

\node (a1) at (0, 1.5) {$A$};
\node (b1) at (1.5, 1.5) {$B$};
\node (c1) at (0, 0) {$C$};
\node (d1) at (1.5, 0) {$D$};

\node (s1) at (1,-1) {$A\perp_{\D} D\mid \{C,D\}$};

\node (a2) at (3.75, 1.5) {$A$};
\node (b2) at (5.25, 1.5) {$B$};
\node (c2) at (3.75, 0) {$C$};
\node (d2) at (5.25, 0) {$D$};

\node (s2) at (4.75,-1) {$A\perp_{\C} D\mid \{C,D\}$};

\node (a3) at (7.5, 1.5) {$A$};
\node (b3) at (9, 1.5) {$B$};
\node (c3) at (7.5, 0) {$C$};
\node (d3) at (9, 0) {$D$};

\node (s3) at (8.5,-1) {$A\perp_{\C'} D\mid \{C,D\}$};
\node (s4) at (8.3,-1.5) {$B\perp_{\C'} C\mid \{D\}$};

\node (a4) at (11.25, 1.5) {$A$};
\node (b4) at (12.75, 1.5) {$B$};
\node (c4) at (11.25, 0) {$C$};
\node (d4) at (12.75, 0) {$D$};

\node (s5) at (12.25,-1) {$A\perp_{\G} D\mid \{C,D\}$};

\node (t1) at (-0.75, 1.75) {(a)};
\node (t2) at (3, 1.75) {(b)};
\node (t3) at (6.75, 1.75) {(c)};
\node (t4) at (10.5, 1.75) {(d)};

\tikzset{undir/.style = {-, line width = 0.5pt}}
\tikzset{dir/.style = {->, -{To[length=5.5, width=6.5]}, line width = 0.5pt}}
\draw[dir]
(c1) edge (a1)
(b1) edge (a1)
(c1) edge (d1)
(b1) edge (d1)
(c1) edge (b1)

(c4) edge (d4)
(b4) edge (d4)

(c3) edge (a3)
(b3) edge (a3)
;
\draw[undir]
(c2) edge (a2)
(b2) edge (a2)
(c2) edge (d2)
(b2) edge (d2)
(c2) edge (b2)

(c4) edge (a4)
(b4) edge (a4)
(c4) edge (b4)

(c3) edge (d3)
(b3) edge (d3)
;

\end{tikzpicture}
\caption{Four graphs and their corresponding independence models. (a) True DAG ($\D$), (b) true CPDAG of $\D$ ($\C$), (c) estimated CPDAG of $\D$ ($\C'$) with erroneous $B\indep C\mid\{ D\}$, (d) estimated tiered MPDAG of $\D$  with background knowledge $\tau(A)=\tau(B)=\tau(C)<\tau(D)$ ($\G$).}
\label{fig:class_size}
\end{figure}
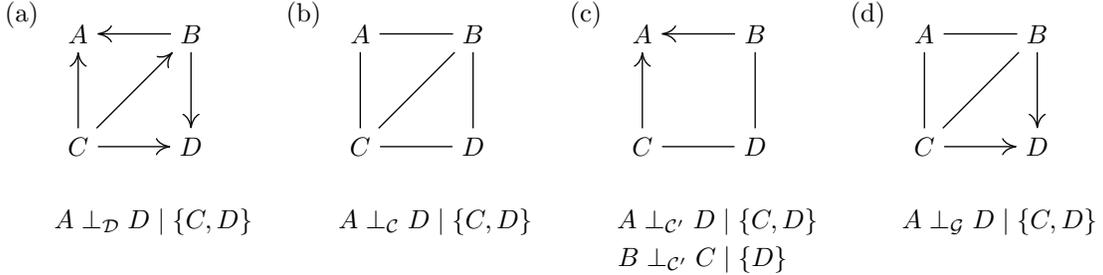

\begin{example}\label{ex:class_size}
Consider the DAG $\D$ in Figure \ref{fig:class_size}(a). Assuming the Markov properties, the only conditional independence relation encoded by $\D$ is $X_A\indep X_D\mid\mathbf{X}_{\{ B,C\}}$ and the CPDAG in Figure \ref{fig:class_size}(b) represents an equivalence class of ten DAGs. If we run the PC algorithm with finite data on  $X_A, X_B, X_C$ and $X_D$ and erroneously find that $X_B\indep X_C\mid\{ X_D\}$, then the skeleton phase (II)  removes the edge $B-C$. If $X_B\not\indep X_C\mid\mathbf{X}_{\{ A, D\}}$ is still correctly recovered from the data, then combined with the erroneous independence this implies that $C\rightarrow A\leftarrow B$ is found to be a v-structure; no additional edges are oriented, and the PC algorithm outputs the incorrect CPDAG in Figure \ref{fig:class_size}(c)  representing an equivalence class of three DAGs.

Instead, incorporating  the tiered ordering $\tau$ with $\tau(A)=\tau(B)=\tau(C)<\tau(D)$ and using the tPC algorithm, avoids  testing the conditional independence between $C$ and $B$ given $D$. Hence, the edge $C-B$ is retained. Moreover, the structure  $C\rightarrow D\leftarrow B$ is recovered in the cross-tier phase (IIIb). The tPC algorithm  outputs the estimated MPDAG in Figure \ref{fig:class_size}(c), which represents a restricted equivalence class of six graphs. Thus, the output of the sample version of the tPC algorithm in this example is a larger equivalence class than that of the sample version of the PC algorithm. Note, however, that the output of the sample version of the tPC algorithm is, in this example, still more informative than the true CPDAG (Figure \ref{fig:class_size}(b)).
\end{example}

\subsection{Statistical consistency}

The PC algorithm is statistically consistent for high-dimensional settings in the sense that the estimated CPDAG converges in probability to the true CPDAG under some regularity conditions \cite{kalisch2007estimating}. The authors assume Gaussian data, and allow for the number of nodes to grow as a function of the sample size under the assumption of a particular type of sparseness. Using slightly different assumptions, consistency can be shown for a broader distributional class with Gaussian copula \cite{harris2013pc}. The consistency results directly extend to the LMPC-stable  algorithm \cite{colombo2014}, and it is straightforward to argue that the tPC algorithm inherits the consistency property as well. In order to prove consistency, the crucial phases of the algorithm are the estimation of the skeleton and  of the v-structures. The  rest follows from the tiered ordering and Meek's rules, and only depends on the quality of skeleton and v-structure estimation. Consistency is shown by bounding the error rates from independence testing, which follows from the consistency of the estimated partial correlations. Since the tPC algorithm can be expected to decrease the number of tests carried out, the number of errors will still be bounded, and one can use assumptions similar to those by Kalisch and Bühlmann \cite{kalisch2007estimating} under which the tPC algorithm will be consistent. 

\section{Simulation study} \label{sec:simstudy}

In this section we conduct a simulation study to demonstrate how the tiered background knowledge improves graph estimation. 
We use the precision and recall for specific characteristics of the estimated graphs and compare them in a variety of settings.
First, we evaluate the estimation of the equivalence classes by considering adjacencies and v-structures of the estimated graphs. 
Second, we focus on the recovery of the causal information  by considering ancestral relations  in the graphs. Lastly, we compute the proportion of bidirected edges, and by this we demonstrate the reduction in conflicts. 

We  compare different levels of  background knowledge: None, partial, and detailed. No background knowledge corresponds to running the regular LMPC-stable algorithm, partial background knowledge corresponds to knowing a coarse ordering of the variables, and detailed background knowledge corresponds to knowing a finer ordering of the variables. In addition, we compare the tLMPC-stable algorithm to  the naive tPC algorithm  of  Section \ref{sec:backgroundknowledge}. 
As explained earlier, this algorithm does not use background knowledge to construct the equivalence class, it only imposes the background knowledge in a final step by orienting undirected and ambiguous edges according to the tiered ordering, as well as reversing any directed edges that are contradicting the background knowledge. By comparing the tLMPC-stable algorithm to the naive tPC algorithm, we demonstrate the importance of incorporating background knowledge as early as possible in the causal discovery algorithm. 

\subsection{Setup}

The simulation study is performed using \textsf{R} version 4.3.0. We generate DAGs with either 10, 20 or 40 nodes from an  Erdös-Rényi model. The DAGs have an adjacency probability of 0.2 (sparse graphs) or 0.4 (dense graphs). This corresponds to an expected number of neighbours of either 1.8, 3.8 or 7.8 (sparse graphs) or 3.6, 7.6 or 15.6 (dense graphs). Based on the generated DAGs, we simulate Gaussian data, where the error terms were all $\mathcal{N}(0,\sigma^2)$ distributed, where $\sigma$ is randomly drawn from $\textnormal{Unif}([0.5, 1.25])$, and the coefficients are randomly drawn from $\textnormal{Unif}([-1, -0.1]\cup [0.1,1])$. We generate datasets of sizes 100, 1,000 and 10,000. For the conditional independence tests, we use a  partial correlation test as in  the \texttt{gaussCItest} function from the \texttt{pcalg} package.  Variables are considered conditionally dependent if the test yields a value below a threshold alpha; here we use an alpha value of either 0.01 or 0.1. This setup amounts to 108 estimated MPDAGs. Each setting is repeated 1000 times. We estimate MPDAGs using the tLMPC-stable algorithm and the naive tPC algorithm, as described in Sections \ref{sec:theory} and \ref{sec:alg}. Both algorithms are equal in the case with no background knowledge, but for partial or detailed background knowledge, the estimated MPDAGs  differ. We evaluate the precision and recall of adjacencies, v-structures, ancestral relations, and possible ancestral relations (for details see Appendix \ref{app:performance}). Precision and recall are calculated using the true DAGs as references. In the case of possible ancestors, we use the true MPDAGs as references.  

For each simulated DAG, we partition the variables into five tiers. This means that for larger graphs, the tiers are larger, while for smaller graphs, knowing the full ordering will provide information on a large proportion of the edges. No background knowledge corresponds to a $\tau_1$ that assigns every node to tier 1. Detailed background knowledge corresponds to a $\tau_5$ assigning each node to a tier between 1 and 5. Partial background knowledge corresponds to a $\tau_2$ that is coarser than $\tau_5$ and finer than $\tau_1$: Here, we will assume that two tiers are known, but the percentage of nodes contained in the first tier varies between 20\% ($\tau_2^A$), 40\% ($\tau_2^B$), 60\% ($\tau_2^C$), and 80\% ($\tau_2^D$), these are all consistent with $\tau_5$. See Figure \ref{fig:sim_taus} in Appendix \ref{app:simorderings} for a visualisation of the different types of background knowledge. In the simulation study, we compare MPDAGs constructed using $\tau_1$ (no background knowledge) with $\tau_2^A,\tau_2^B,\tau_2^C$ or $\tau_2^D$ (randomly chosen), and $\tau_5$.

\subsection{Results}

In this section we include of the results of the main analysis obtained for graphs with 20 nodes and a sample size of 1,000. The remaining results can be found in Figures \ref{fig:sim_adj_rec_full} to \ref{fig:sim_conflicts_full} in Appendix \ref{app:mainanalysis}. The results of the analysis of the naive tPC algorithm are included in Figures \ref{fig:sim_vstruct_rec_full_naive} to  \ref{fig:sim_conflicts_full_naive} in Appendix \ref{app:naiveresults}. In addition to these two analyses, we  explored the number of tests performed at each round (see Figures \ref{fig:n_tests} and\ref{fig:n_tests_sparse} in Appendix \ref{app:ntestnedges}).

\begin{figure}[!htbp]
    \centering
    \includegraphics{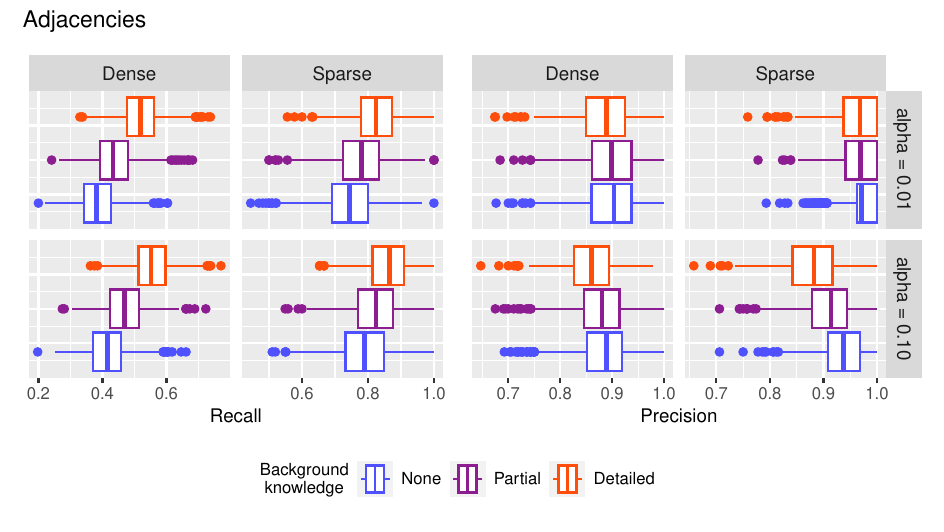}
    \caption{Recall (left two columns) and precision (right two columns) of adjacencies from 1,000 simulations in dense/sparse settings with  20 nodes, $\alpha\in \{0.01;0.1\}$ and sample size of 1,000. 
    For results on adjacencies under other settings  see Figure \ref{fig:sim_adj_rec_full} and Figure \ref{fig:sim_adj_prec_full} in Appendix \ref{app:mainanalysis}.
    }
    \label{fig:sim_adj_small}
\end{figure}

Figure \ref{fig:sim_adj_small} shows the precision and recall of the estimation of the skeleton, i.e. it assesses the quality of detected presence/absence of edges. The recall clearly improves when more detailed background knowledge is included. Thus, leveraging tiered background knowledge enables us to recover more correct adjacencies. Similar results can be found for other settings (see Figure \ref{fig:sim_adj_rec_full}). 
The improved recall comes at no or little loss in precision. This implies that with tiered background knowledge, the same proportion or a slightly higher of identified adjacencies are incorrect. This can be explained by the tPC algorithm  performing fewer, and consequently different, independence tests when using more detailed background knowledge (see Figure \ref{fig:sim_edgetests} in Appendix \ref{app:ntestnedges}), which tends to more edges being retained (see Figure \ref{fig:sim_nedges} in Appendix \ref{app:ntestnedges}). A similar pattern is found for other settings in Figure \ref{fig:sim_adj_prec_full}.  In summary, the improved recall of adjacencies at little cost in precision when using the tiered ordering is remarkable because it does not even make use of orienting cross-tier edges, it only exploits that we do not need to consider separating sets in the future.

\begin{figure}[!htbp]
    \centering
    \includegraphics{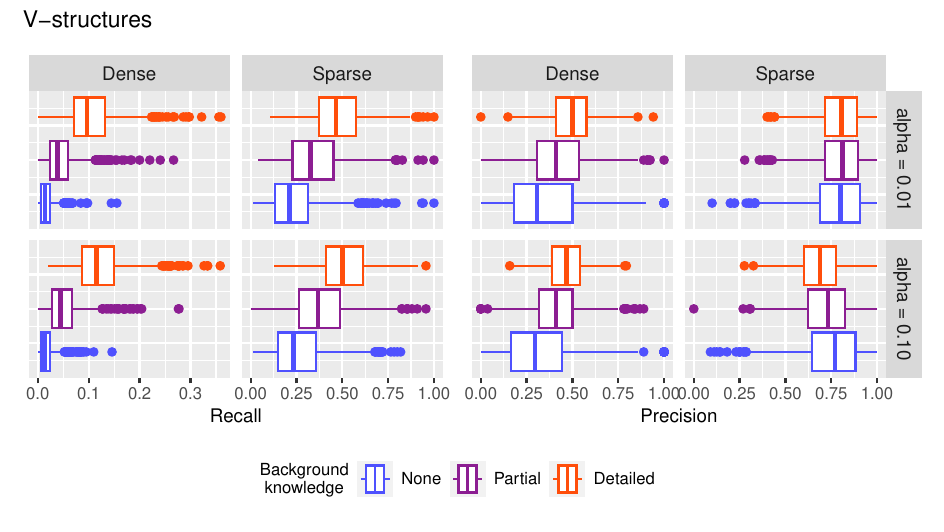}
    \caption{Recall (left two columns) and precision (right two columns) of v-structures from 1,000 simulations in dense/sparse settings with  20 nodes, $\alpha\in \{0.01;0.1\}$ and sample size of 1,000. 
    For results on v-structures under other settings  see Figure \ref{fig:sim_vstruct_rec_full} and Figure \ref{fig:sim_vstruct_prec_full} in Appendix \ref{app:mainanalysis}.
    }
    \label{fig:sim_vstruct_small}
\end{figure}

In Figure \ref{fig:sim_vstruct_small} we find that including more background knowledge clearly improves the recall of v-structures. For dense graphs, including background knowledge also improves the precision, but for sparse graphs and large alpha we find a slight tendency to  a lower precision. It should be noted that the estimation of v-structures requires a locally correct estimation of the skeleton: In order for $\langle A, B, C\rangle$ to be considered as a possible a v-structure, we need $A$ and $B$, and $B$ and $C$ to be adjacent, with $A$ and $C$ not adjacent.

\begin{figure}[!htbp]
    \centering
    \includegraphics{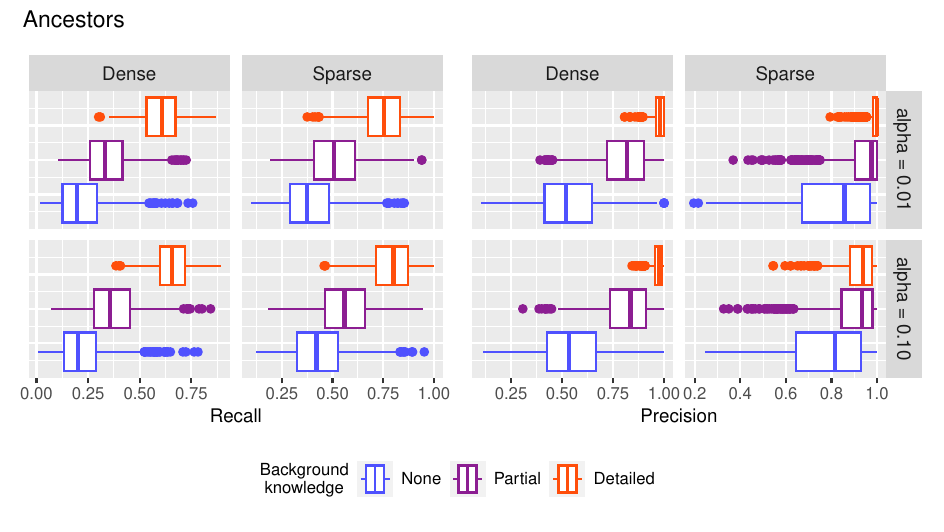}
    \caption{Recall (left two columns) and precision (right two columns) of ancestors from 1,000 simulations in dense/sparse settings with  20 nodes, $\alpha\in \{0.01;0.1\}$ and sample size of 1,000. 
    For results on ancestors under other settings  see Figure \ref{fig:sim_an_rec_full} and Figure \ref{fig:sim_an_prec_full} in Appendix \ref{app:mainanalysis}.
    }
    \label{fig:sim_an_small}
\end{figure}

Figure \ref{fig:sim_an_small} depicts the precision and recall of the estimated ancestral relations (including parent-child relations), i.e. this reflects the correctness of the directed paths and thus of (indirect) causal relations among the 20 nodes. Unsurprisingly, both clearly improve as the orientation of cross-tier edges using tiered background knowledge is guaranteed to be correct. With the tiered ordering we are thus able to recover more correct directed paths, which increases recall. At the same time, we  disallow paths in the opposite direction, which limits the proportion of incorrect directed paths, increasing the precision. Similar results are found for all other settings (see Figure \ref{fig:sim_an_rec_full} and Figure \ref{fig:sim_an_prec_full} in Appendix \ref{app:mainanalysis}).

\begin{figure}[!htbp]
    \centering
    \includegraphics{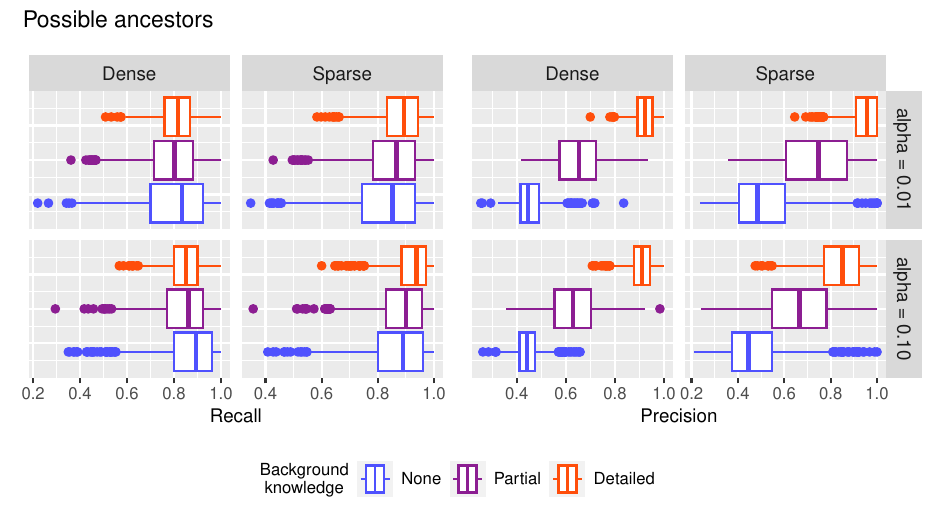}
    \caption{Recall (left two columns) and precision (right two columns) of possible ancestors from 1,000 simulations in dense/sparse settings with  20 nodes, $\alpha\in \{0.01;0.1\}$ and sample size of 1,000. 
    For results on possible ancestors under other settings  see Figure \ref{fig:sim_possan_rec_full} and Figure \ref{fig:sim_possan_prec_full} in Appendix \ref{app:mainanalysis}.
    }
    \label{fig:sim_possan_small}
\end{figure}

The recall and precision of the estimated possible ancestral relations (i.e. partially directed paths among the nodes) are shown in Figure \ref{fig:sim_possan_small}. 
Here, recall exhibits only marginal improvement with tiered knowledge in some settings, and Figure \ref{fig:sim_possan_rec_full} shows that for some graphs we get a slightly worsened recall. Comparing to the results for the ancestral relations above, possible ancestors is a much weaker property, and graphs with very few directed edges can have many possible ancestral relations. On the other hand, we find that the precision improves when including tiered background knowledge, not only in Figure \ref{fig:sim_possan_small} but also for other settings (Figure \ref{fig:sim_possan_prec_full}).

\begin{figure}[!htbp]
    \centering
    \includegraphics{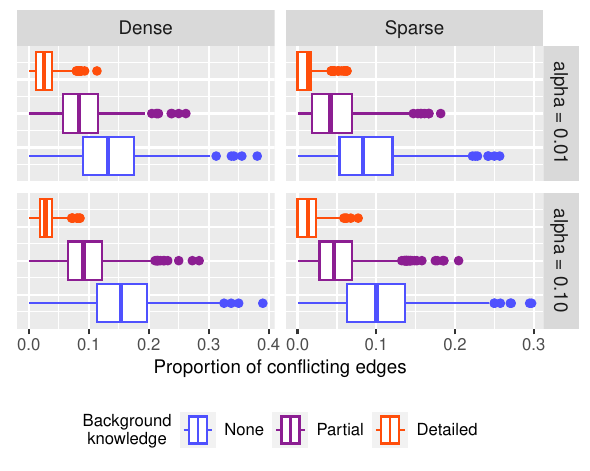}
    \caption{Proportions of bidirected edges from 1,000 simulations in dense/sparse settings with  20 nodes, $\alpha\in \{0.01;0.1\}$ and sample size of 1,000. 
    For results on bidirected under other settings  see Figure \ref{fig:sim_conflicts_full} in Appendix \ref{app:mainanalysis}.
    }
    \label{fig:sim_conflicts_small}
\end{figure}

Importantly, Figure \ref{fig:sim_conflicts_small} (and  \ref{fig:sim_conflicts_full} in Appendix \ref{app:mainanalysis}) depict the proportions of conflicting (bidirected) edges. These decrease as more background knowledge is included for all settings. Incorporating background knowledge enables the recovery of more of the true v-structures (Figure \ref{fig:sim_vstruct_rec_full}) and the appropriate orientation of cross-tier edges, overruling conflicting information in the data.

For the naive tPC algorithm (Figures \ref{fig:sim_vstruct_rec_full_naive} to \ref{fig:sim_conflicts_full_naive} in Appendix \ref{app:naiveresults}), we see similar trends for the recovery of v-structures and ancestral relations as for the tLMPC-stable algorithm. 
However, the improvement is much less pronounced than  for the tLMPC-stable algorithm.
Importantly, regarding the recovery of the skeleton, the naive tPC necessarily gives the same result as the tPC algorithm with no background knowledge (Figure \ref{fig:sim_adj_small}) since the naive tPC algorithm does not make use of the background knowledge for estimating the skeleton.

\section{Data application}\label{sec:application}

As an example of how the tPC algorithmn performs in practice, we  use it for estimating the causal structure of a subset of the data from the IDEFICS/I.Family study \cite{ahrens2017}, which is a prospective cohort study set up to explore the aetiology of nutrition related diseases among a cohort of children followed from early life through adolescence. The variables are measured in three  waves several years apart, and this provides an initial, unambiguous tiered ordering of the variables. Moreover, the variables within the first tier can be  subdivided into five tiers using expert knowledge. We will consider three tiered orderings, denoted by $\tau_1$, $\tau_2$ and $\tau_3$. Here, $\tau_1$ corresponds to having no background knowledge (one tier), $\tau_2$ corresponds to having knowledge of the order in which the variables have been measured (partial background knowledge, three tiers), and $\tau_3$ extends $\tau_2$ with expert knowledge on the early life and sociodemographic factors (detailed background knowledge, seven tiers). See Table \ref{tab:dataexmaple}  in Appendix \ref{app:dataordering} for details.

\subsection{Methods}

We use the dataset provided by Foraita et al.\ \cite{foraita2022}, where  we select one of their ten imputations of missing values for our analysis. The original aim was to answer specific research questions concerning the aetiology of BMI. Here,  our focus is on evaluating the algorithm's performance. In particular, we aim to illustrate the benefits of incorporating tiered background knowledge and how different levels of detail improve the accuracy of graph estimation. For this reason, we restrict our analysis to data from Germany and a subset of the variables. We estimate three different MPDAGs using either $\tau_1$, $\tau_2$ and $\tau_3$ using \textsf{R} version 4.2.1 and the \texttt{tpc} function from the \textsf{R} package \texttt{tpc} \cite{witte2022b}, which is an extension to the \texttt{pcalg} package \cite{kalisch2012causal}. We use the following conditional independence tests: For continuous data a correlation tests using the \texttt{gaussCItest} function, and for discrete data  the $G^2$ test using the \texttt{disCItest} function, both from the \texttt{pcalg} package. For mixed variables a likelihood ratio test is performed using the \texttt{mixCItest} function from the \texttt{micd} package. This test assumes that the  variables follow a conditional Gaussian distribution, i.e. conditional on each combination of values of the discrete variables, the continuous variables are multivariate Gaussian \cite{lauritzen1989graphical, andrews2018scoring}.

\subsection{Results}

\begin{figure}[!htbp]
\begin{tikzpicture}[scale=0.92, squarednode/.style={rectangle, draw=gray!80, fill=white, thin, minimum size=5mm}]

\node[squarednode] (sex) at (0, 2.75) {\small \textsf{sex}};
\node[squarednode] (age) at (0, 1.75) {\small \textsf{age}};
\node[squarednode] (migrant) at (0, 0.75) {\small  \textsf{migrant}};
\node[squarednode] (income) at (0, -.25) {\small \textsf{income}};
\node[squarednode] (isced) at (0,-1.25) {\small \textsf{ISCED}};

\node[squarednode, align=center] (bage) at (2.5,0.75) {\small \textsf{mother's} \\ \small \textsf{age}};

\node[squarednode, align=center] (pregweek) at (5, 1.25) {\small \textsf{weeks} \\ \small \textsf{pregnant}};
\node[squarednode, align=center] (bweight) at (5, -.25) {\small \textsf{birth} \\ \small \textsf{weight}};

\node[squarednode] (bf) at (7.5, 2) {\small \textsf{breastfeeding}};
\node[squarednode, align=center] (formula) at (7.5, 0.75) {\small \textsf{formula}\\ \small \textsf{milk}};
\node[squarednode, align=center] (hdiet) at (7.5, -.75) {\small \textsf{household}\\ \small \textsf{diet}};

\node[squarednode] (school) at (10,5) {\small \textsf{school}};
\node[squarednode] (media0) at (10,4) {\small \textsf{media}};
\node[squarednode] (bmi0) at (10, 3) {\small \textsf{BMI}};
\node[squarednode, align=center] (mbmi0) at (10, 1.75) {\small \textsf{mother's}\\ \small \textsf{BMI}};
\node[squarednode, align=center] (pa0) at (10, 0.25) {\small \textsf{physical}\\ \small \textsf{activity}};
\node[squarednode] (sleep0) at (10, -1) {\small \textsf{sleep}};
\node[squarednode] (wb0) at (10, -2) {\small \textsf{well-being}};
\node[squarednode, align=center] (yhei0) at (10, -3.25) {\small \textsf{healthy}\\ \small \textsf{eating}};
\node[squarednode, align=center] (homa0) at (10, -4.75) {\small \textsf{insulin}\\\small \textsf{resistance}};

\node[squarednode] (media1) at (12.5,4) {\small \textsf{media}};
\node[squarednode] (bmi1) at (12.5, 3) {\small \textsf{BMI}};
\node[squarednode, align=center] (mbmi1) at (12.5, 1.75) {\small \textsf{mother's}\\ \small \textsf{BMI}};
\node[squarednode, align=center] (pa1) at (12.5, 0.25) {\small \textsf{physical}\\ \small \textsf{activity}};
\node[squarednode] (sleep1) at (12.5, -1) {\small \textsf{sleep}};
\node[squarednode] (wb1) at (12.5, -2) {\small \textsf{well-being}};
\node[squarednode, align=center] (yhei1) at (12.5, -3.25) {\small \textsf{healthy}\\ \small \textsf{eating}};
\node[squarednode, align=center] (homa1) at (12.5, -4.75) {\small \textsf{insulin}\\\small \textsf{resistance}};

\node[squarednode] (pub) at (15, 5) {\small \textsf{puberty}};
\node[squarednode] (media2) at (15,4) {\small \textsf{media}};
\node[squarednode] (bmi2) at (15, 3) {\small \textsf{BMI}};
\node[squarednode, align=center] (mbmi2) at (15, 1.75) {\small \textsf{mother's}\\ \small \textsf{BMI}};
\node[squarednode, align=center] (pa2) at (15, 0.25) {\small \textsf{physical}\\ \small \textsf{activity}};
\node[squarednode] (sleep2) at (15, -1) {\small \textsf{sleep}};
\node[squarednode] (wb2) at (15, -2) {\small \textsf{well-being}};
\node[squarednode, align=center] (yhei2) at (15, -3.25) {\small \textsf{healthy}\\ \small \textsf{eating}};
\node[squarednode, align=center] (homa2) at (15, -4.75) {\small \textsf{insulin}\\\small \textsf{resistance}};

\tikzset{undir/.style = {-, line width = 0.5pt}}
\tikzset{dir/.style = {->, -{To[length=5.5, width=6.5]}, line width = 0.5pt}}
\tikzset{bidir/.style = {{To[length=4.5, width=6]}-{To[length=4.5, width=6]}, line width = 0.5pt}}

\node (t1) at (0, -5.75) {\small \textsf{context}};
\node (t2) at (2.5, -5.75) {\small \textsf{early life 1}};
\node (t3) at (5, -5.75) {\small \textsf{early life 2}};
\node (t4) at (7.5, -5.75) {\small \textsf{early life 3}};
\node (t5) at (10, -5.75) {\small \textsf{baseline}};
\node (t6) at (12.5, -5.75) {\small \textsf{wave 1}};
\node (t7) at (15, -5.75) {\small \textsf{wave 2}};

 \begin{pgfonlayer}{bg}
 
\filldraw[silvergray] (-1, 6.5) rectangle +(2,-12.75);
\filldraw[silvergray] (1.5, 6.5) rectangle +(2,-12.75);
\filldraw[silvergray] (4, 6.5) rectangle +(2,-12.75);
\filldraw[silvergray] (6.5, 6.5) rectangle +(2,-12.75);
\filldraw[silvergray] (9, 6.5) rectangle +(2,-12.75);
\filldraw[silvergray] (11.5, 6.5) rectangle +(2,-12.75);
\filldraw[silvergray] (14, 6.5) rectangle +(2,-12.75);

\draw[undir]
(migrant) edge (income)
(migrant) edge [bend right] (isced)
(income) edge (isced)
(bmi0) edge (mbmi0)
(media0) edge [bend right] (yhei0)
(bmi0) edge [bend right] (homa0)
;

\draw[bidir]
(sleep0) edge [bend left = 80] (mbmi0)
(media2) edge (pub)
(pub) edge [bend right] (wb2)
(sleep2) edge (wb2)
;

\draw[dir]
(sex) edge [bend right] (bweight)
(sex) edge [bend left] (bmi1)
(sex) edge [bend left] (media2)
(age) edge [bend left] (school)
(age) edge [bend left] (media0)
(age) edge [bend right] (sleep1)
(age) edge [bend left] (pub)
(age) edge [bend right] (sleep2)
(age) edge [bend right] (homa1)
(income) edge [bend right] (wb1)
(isced) edge [bend right] (bage)
(isced) edge [bend left] (bf)
(bage) edge [bend right] (wb0)
(pregweek) edge (bweight)
(pregweek) edge [bend left] (bf)
(bweight) edge [bend left] (bmi0)
(bweight) edge [bend left] (mbmi0)
(bf) edge (formula)
(school) edge [bend left] (sleep0)
(school) edge  (yhei1)
(school) edge (media2)
(school) edge (sleep2)
(media0) edge [ultra thick] (media1)
(media0) edge (sleep1)
(media0) edge [ultra thick] [bend left] (media2)
(bmi0) edge [ultra thick] (bmi1)
(mbmi0) edge (bmi1)
(mbmi0) edge [ultra thick] (mbmi1)
(mbmi0) edge [ultra thick] [bend left] (mbmi2)
(pa0) edge [bend right] (wb0) 
(pa0) edge [ultra thick] (pa1)
(sleep0) edge [ultra thick] (sleep1)
(wb0) edge [ultra thick] (wb1)
(yhei0) edge [ultra thick] (yhei1)
(yhei0) edge [ultra thick] [bend left] (yhei2)
(yhei0) edge (wb0)
(homa0) edge [ultra thick] [bend left] (homa2)
(media1) edge [ultra thick] (media2)
(bmi1) edge [ultra thick] (bmi2)
(bmi1) edge (pub)
(mbmi1) edge [ultra thick] (mbmi2)
(pa1) edge [bend left] (wb1)
(pa1) edge [bend left] (yhei1)
(pa1) edge [ultra thick] (pa2)
(sleep1) edge [ultra thick] (sleep2)
(wb1) edge [ultra thick] (wb2)
(yhei1) edge [ultra thick] (yhei2)
(homa1) edge [bend left] (bmi1)
(homa1) edge [bend right] (bmi2)
(bmi2) edge [bend right] (homa2)
(sleep2) edge [bend right] (media2)
(yhei2) edge [bend right] (media2)
;
\end{pgfonlayer}
\end{tikzpicture}
\caption{Estimated graph of the IDEFICS/I.Family dataset using the tPC algorithm with detailed background knowledge ($\tau_3$). Bold edges: These are directed edges between repeated measurements, which we expect to find. 
}
\label{fig:mpdag_estimated}
\end{figure}
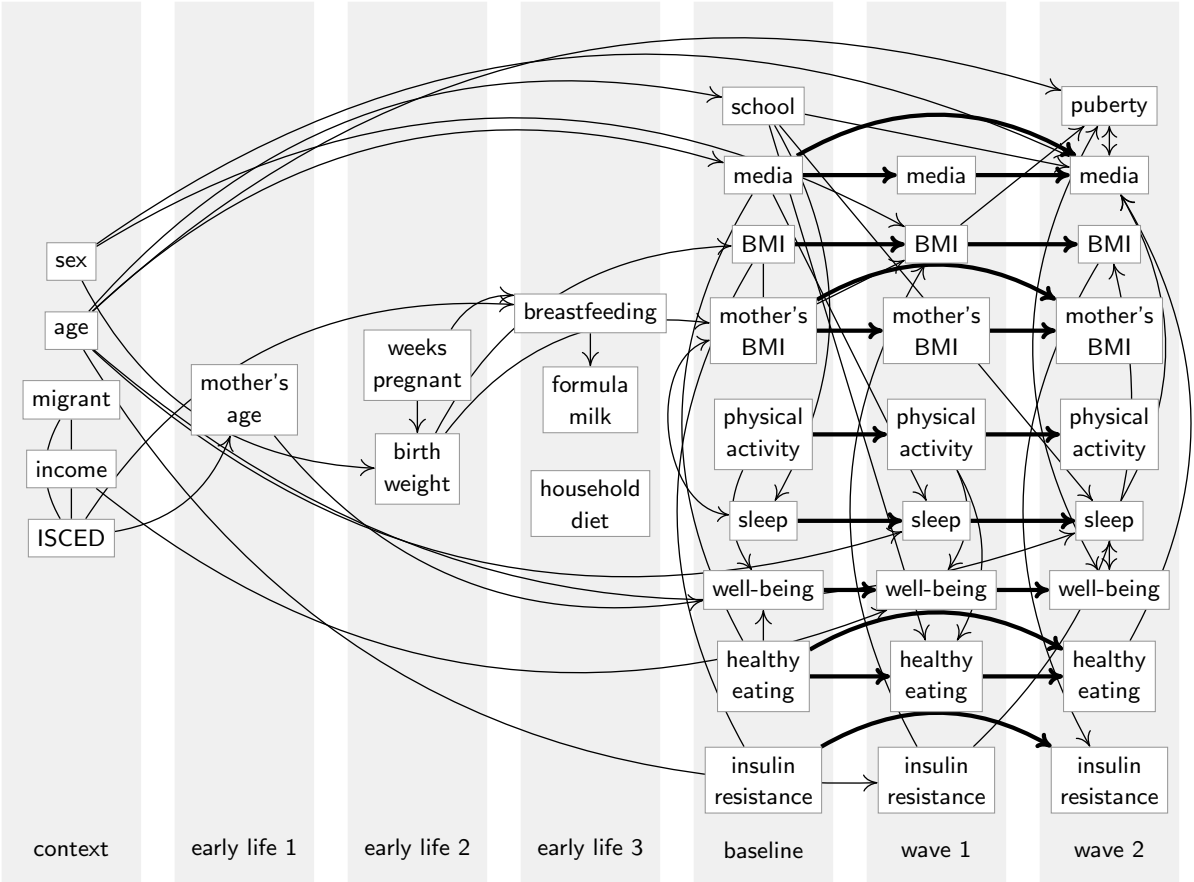

\begin{table}[!htbp]
    \centering
{\small
\begin{tabular}{| l  l | m{5.5em} | m{5.5em} | m{5.5em} | m{5.5em} | m{6em} |}

\hline
 & Ordering & Adjacencies & Directed edges & Bidirected edges & V-structures & Incorrect edges\\
\hline
\hline
\multirow{3}{5em}{Individual}  &  $\tau_3$ & 62 & 52 & 4 & 42 & -\\

& $\tau_2$ & 64 & 48 & 4 & 44 & 3\\

& $\tau_1$ & 60 & 48 & 6 & 24 & 16\\
\hline \hline
\multirow{4}{5em}{Shared} & $\tau_3$ \& $\tau_2$ & 61 & 41 & 0 & 30 & -\\

 & $\tau_3$ \& $\tau_1$ & 53 & 21 & 0 & 9 & -\\

& $\tau_2$ \& $\tau_1$ & 56 & 20 & 3 & 11 & 1\\

& all & 53 & 17 & 0 & 9 & -\\
\hline
\end{tabular}
}
    \caption{Individual and shared characteristics of the graphs  using  $\tau_1$ (no background knowledge), $\tau_2$ (partial background knowledge) and $\tau_3$ (detailed background knowledge).}
    \label{tab:descriptives}
\end{table}

The characteristics of the three estimated graphs are summarized in Table \ref{tab:descriptives}. All graphs have approximately the same density and share most of their adjacencies. In addition, the graphs with background knowledge ($\tau_2$ and $\tau_3$) share many of the same v-structures, and the equivalence classes represented by the two are closer compared to the graph estimated without using background knowledge ($\tau_1$). Unsurprisingly, the estimated graph using detailed background knowledge ($\tau_3$) has the largest number of directed edges, and the estimated graphs using background knowledge contain fewer bidirected (conflicting) edges and notably fewer edges contradicting the known time ordering. The graph estimated  without using background knowledge not only contains more undirected edges, it also contains numerous directed edges that contradict the known time order (incorrect edges). 

The estimated graphs can be found in Figure \ref{fig:mpdag_estimated}, and in Figures \ref{fig:est_cpdag} and \ref{fig:est_partial_mpdag} in Appendix \ref{app:dataresults}, and their respective adjacency matrices  can be found in Figures \ref{fig:est.graph.notiers}-\ref{fig:est.graph.altiers} in Appendix \ref{app:dataamat}. The estimated graphs of Figure \ref{fig:mpdag_estimated} (detailed background knowledge) and Figure \ref{fig:est_partial_mpdag} (partial background knowledge) share almost all of the edge directions, and most of the directed edges are between repeated measurements. These edges are expected, but not all of them could be found without using background knowledge (Figure \ref{fig:est_cpdag} in Appendix \ref{app:dataresults}). This strengthens our confidence in the graphs  estimated using background knowledge. With partial background knowledge, all undirected edges are in the first tier. Some of these edges could be oriented using detailed background knowledge which illustrates the usefulness of imposing a finer ordering especially  in the earliest tiers \cite{bang2023we}. 

It should be noted that neither of the estimated graphs in Figure \ref{fig:mpdag_estimated} and Figures \ref{fig:est_partial_mpdag} and \ref{fig:est_cpdag} are CPDAGs or MPDAGs. This happens because of the way the algorithm deals with conflicting or ambiguous information in the data. For example, in Figure \ref{fig:mpdag_estimated} the undirected edge \textsf{BMI}$-$\textsf{insulin resistance} is a part of an ambiguous triple: Based on data, the algorithm cannot decide whether \textsf{birth weight}$\any$\textsf{BMI}$\any$\textsf{insulin resistance} constitutes a v-structure. While  \textsf{birth weight}$\rightarrow$\textsf{BMI} is implied by the time order, it is still uncertain whether \textsf{BMI}$\rightarrow$\textsf{insulin resistance} or \textsf{BMI}$\leftarrow$\textsf{insulin resistance}. Hence, we obtain \textsf{birth weight}$\rightarrow$\textsf{BMI}$-$\textsf{insulin resistance}; this structure cannot occur in a CPDAG or MPDAG, and this implies that the estimated graph in fact is not an MPDAG and does not represent a (restricted) equivalence class.

\section{Discussion}\label{sec:discussion}

In this paper we  demonstrated that the practical performance of causal discovery, specifically of the PC algorithm, can be much improved by efficiently exploiting temporal background knowledge, which is often available and unambiguous. 
We proved that the tPC algorithm is sound and complete, and that the sample version of the tLMPC-stable algorithm is order independent. The simulation study and data application illustrated the importance of including background knowledge at an early stage of the causal discovery algorithm, and showed how background knowledge can improve the accuracy of the estimated equivalence class over and above imposing the orientation of cross-tier edges.

Specifically, the simulation study suggested that exploiting tiered background knowledge offers a higher edge recall at almost no cost of precision; since the lack of an edge is a stronger assumption than the presence, this small cost should be of little concern. The result  highlights the importance of including background knowledge even when estimating only the skeleton. Moreover, the simulation study showed an overall increase in precision as well as recall of (possible) ancestral relations, implying that we may have more confidence in the causal conclusions drawn from graphs that are estimated using background knowledge. Recovering the correct edges is helpful but not necessary for recovering ancestral relations; hence, tiered background knowledge improves the graph estimation in these two separate aspects. Moreover, we found that the naive, post hoc, inclusion of background knowledge offered less improvement of the estimated graphs, emphasising the importance of efficiently incorporating background knowledge. 
The application to data from the IDEFICS/I.Family study, illustrated the use of the tPC algorithm and how its output should be interpreted in practice. Here we found that the estimated graphs were more aligned with our prior expectations when incorporating background knowledge. In conclusion, we find that making use of tiered background knowledge greatly improves graph estimation. Since temporal background knowledge is oftentimes highly reliable, we argue that this strengthens the credibility of the estimated graphs obtained from causal discovery.

There are many directions in which our work could be extended. In future work we will relax the causal sufficiency assumption by investigating the so-called Fast Causal Inference (FCI) algorithm, which allows for latent variables \cite{spirtes1999fci}. Tiered background knowledge can be incorporated into the FCI algorithm in a similar fashion as in the PC algorithm and has been implemented, e.g., in the TETRAD software \cite{scheines1998} and R packages \cite{tfci, petersen2023}; see also the application in medical research \cite{lee2022causal}. 
To our knowledge, there has been no thorough analysis of the finite sample properties of the  tiered FCI algorithm in a similar fashion as the tPC algorithm in this paper. Moreover, in the oracle case, we expect the tiered FCI algorithm to be sound, but  not necessarily complete because of the much more complex orientation rules which are themselves not complete for general background knowledge \cite{zhang2008completeness}. With tiered background knowledge and the additional restriction of no cross-tier latent confounding, completeness of the FCI has previously been shown \cite{andrews2020}.

Moreover, it is desirable to allow for more general types of background knowledge to be combined with data driven methods for estimating causal DAGs. 
Previous work has addressed, e.g., interventional background knowledge \cite{eberhardt2008almost, hauser2012characterization, hauser2014two, squires2020active}, or ancestral background knowledge \cite{fang2022representation}. However, in contrast to the tiered case, such background knowledge is not necessarily transitive and complete. These properties are fundamental to being able to incorporate the tiered structure in the algorithm while estimating the equivalence class. For other types of background knowledge  it cannot necessarily be ensured that the estimated equivalence class is consistent with that knowledge. Moreover,  general background knowledge may not be as unambiguous as a temporal ordering resulting in a further potential source of error. Therefore, many challenges remain to be tackled regarding entirely general approaches for combining background knowledge and causal discovery. 

\section*{Acknowledgements}
This project was funded by the Deutsche Forschungsgemeinschaft (DFG, German Research Foundation) – Project 281474342/GRK2224/2 and 256000800.

\bibliography{ref}

\pagebreak

\section*{Appendix}

\appendix

\section{Notation and terminology}

\subsection{Causal graphs}\label{app:theory}

A graph $\G=(\V, \E)$ consists of nodes $\V$ and edges $\E$. We consider edges that are either directed $V_1\rightarrow V_2$ ($V_1$ is a \emph{parent} of $V_2$ and $V_2$ is a \emph{child} of $V_1$) or undirected $V_1-V_2$ ($V_1$ and $V_2$ are \emph{neighbours}). In either case, we say that $V_1$ and $V_2$ are \emph{adjacent}.  $\G'=(\V',\E')$ is a \emph{subgraph} of $\G=(\V,\E)$ if $\V'\subseteq\V$ and $\E'\subseteq\E$. Let $\V'\subseteq\V$, then $\G_\V'=(\V',\E_{V'})$, where $\E_{\V'}\subseteq\E$ contains all the edges between the nodes in $\V'$, is the \emph{induced subgraph} of $\G$ over $\V'$.

A sequence of distinct, adjacent nodes $V_1 ,V_2,...,V_{K-1}, V_K$ in $\G$ constitutes a \emph{path} between $V_1$ and $V_K$ of length $K$ in $\G$, and $V_1$ and $V_K$ are the \emph{endpoint} nodes of the path. Let $\pi$ be a path between $V_1$ and $V_K$, if for every $1\leq j<K$, the edges is directed as $V_{j}\rightarrow V_{j+1}$ we call $\pi$ a \emph{directed path} from $V_1$ to $V_K$. $V_1$ is then called an \emph{ancestor} of $V_K$ and $V_K$ is a \emph{descendant} of $V_1$. If there is a directed path from $V_1$ to $V_K$ and a directed path from $V_K$ to $V_1$ then we call this a \emph{directed cycle}. Directed acyclic graphs (DAGs) consist of only directed edges and have no directed cycles. Undirected graphs only have undirected edges. Replacing every edge in a given DAG with an undirected edge results in the \emph{skeleton} of the DAG.

By $\pa{\G}{X}$ we refer to the set of parents of node $X$ in $\G$. Correspondingly, $\adj{\G}{X}$, $\de{\G}{X}$,  $\nd{\G}{X}$ are the sets of adjacent nodes, descendants and non-descendants of $X$ in $\G$, respectively. 

If a path $\pi$ contains the structure $V_i\rightarrow V_j\leftarrow V_k$ we call $V_j$ a \emph{collider} on $\pi$; any edge that is not a collider is called a \emph{non-collider}. If $V_i$ and $V_k$ are not adjacent we call $V_i\rightarrow V_j\leftarrow V_k$  a \emph{v-structure}.

V-structures encode a particular form of dependence:

\begin{definition}[blocking]
Let $\D$ be a DAG and let $\pi$ be a path from $V_i\in\V$ to $V_j\in\V$ in $\D$, and let $\textbf{S}\subseteq\V\backslash\{ X, Y\}$. If either
\begin{itemize}
    \item[(i)] there is a collider $V$ on $\pi$ such that $V\notin\mathbf{S}$ and for all $V'\in\de{\D}{V}$ $V'\notin\mathbf{S}$, or
    \item[(ii)] there is a non-collider on $\pi$ in $\mathbf{S}$
\end{itemize}
then $\pi$ is blocked by $\textbf{S}$.
\end{definition}

\begin{definition}[d-separation]\label{def:dsep}
Let $\D=(\V,\E)$ be a DAG. Two nodes $V_i\in\V$ and $V_j\in\V$ are d-separated by $S\subset\V$ in $\D$ if every path between $V_i$ and $V_j$ in $\D$ is blocked by $\textbf{S}$
\end{definition}

We denote $V_i$ and $V_j$ being d-separated by $\textbf{S}$ in $\D$ by $V_i\perp_{\D} V_j\mid\textbf{S}$. Two nodes are adjacent if and only if they cannot be d-separated by any subset of the remaining nodes (see \cite{lauritzen1996}).

\begin{definition}[Independence model]\label{def:indepmodel}
    The independence model $\I(\G)$ induced by a graph $\G$ is a collection of all separations in $\G$: $(\mathbf{A} \perp_\G\mathbf{B}\mid\mathbf{C})\in\I(\G)\Leftrightarrow\mathbf{A}$ and $\mathbf{B}$ are d-separated by $\mathbf{C}$ in $\G$.
\end{definition}

Two DAGs are \emph{Markov equivalent} if they induce the same independence model. A graphical criterion for determining Markov equivalence was provided by \cite{verma1990}: Two DAGs are Markov equivalent if and only if they have the same skeleton and v-structures. Hence, the skeleton and v-structures uniquely characterise the equivalence class.

Graphs represent random variables in the following way: Let $\V$  be a set of nodes and $\mathbf{X}_\mathbf{V}$ a set of random variables corresponding to the nodes in $\V$, then $X_V$ is  the random variable represented by $V\in\V$, and  $\mathbf{X}_{\V'}$ is the set of random variables represented by $\V'\subseteq\V$. The \emph{Markov properties}, and \emph{faithfulness}, allow us to use use the graphs to reason about conditional independencies in a distribution over the random variables represented by the nodes, and vice versa:

\begin{definition}[Global Markov property]\label{def:globalmarkov}
    Let $\D=(\V,\E)$ be a DAG and let $P$ be a probability distribution over the random variables $\mathbf{X}_\mathbf{V}$ represented by $\V$. We say that $P$ obeys the global Markov property with respect to $\D$ if for any distinct nodes $V_i,V_j\in\V$ and set $\mathbf{S}\subseteq\V\backslash\{ V_i,V_j\}$
    
    \begin{align}
        V_i\perp_{\D} V_j\mid \mathbf{S}\Rightarrow X_{V_i}\indep X_{V_j}\mid \mathbf{X}_\mathbf{S}
    \end{align}

    \noindent where $\indep$ denotes (conditional) independence.
\end{definition}

It can be shown that the global Markov property is equivalent to the following property \cite{lauritzen1996}:

\begin{definition}[Local Markov property]\label{def:localmarkov}
    Let $\D=(\V,\E)$ be a DAG and let $P$ be a probability distribution over the random variables $\mathbf{X}_\V$ represented by $\V$. We say that $P$ obeys the local Markov property with respect to $\D$ if for any node $V\in\V$

    \begin{align}
        X_V\indep \mathbf{X}_{\nd{\D}{V}\mid\pa{\D}{V}}
    \end{align}
    
\noindent where $\indep$ denotes (conditional) independence.
\end{definition}

\begin{definition}[Faithfulness]
\label{def:faithfulness}
    Let $\D=(\V,\E)$ be a DAG and let $P$ be a probability distribution over the random variables $\mathbf{X}_\V$ represented by $\V$. We say that $P$ is faithful to $\D$ if for any distinct nodes $V_i,V_j\in\V$ and set $\mathbf{S}\subseteq\V$
    
    \begin{align}
        V_i\perp_{\D} V_j\mid \mathbf{S}\Leftarrow X_{V_i}\indep X_{V_j}\mid \mathbf{X}_\mathbf{S}
    \end{align}

    \noindent where $\indep$ denotes (conditional) independence.
\end{definition}

\subsection{Meek's rules}\label{app:meekrules}

\label{sec:meek}
\begin{figure}[!htbp]
\centering
\begin{tikzpicture}[state/.style={thick}, scale = 0.85]

\node (r1) at (-1.125,-2.25) {Rule 1};
\node (a1) at (-1.125,-2.75) {$\Downarrow$};
\node (r2) at (2.875,-2.25) {Rule 2};
\node (a2) at (2.875,-2.75) {$\Downarrow$};
\node (r3) at (6.875,-2.25) {Rule 3};
\node (a3) at (6.875,-2.75) {$\Downarrow$};
\node (r4) at (10.875,-2.25) {Rule 4};
\node (a4) at (10.875,-2.75) {$\Downarrow$};

\node (i) at (-2.5,-1) {(i)};
\node (i') at (-2.5,-4.25) {(i')};
\node (ii) at (1.5,-1) {(ii)};
\node (ii') at (1.5,-4.25) {(ii')};
\node (iii) at (5.5,-1) {(iii)};
\node (iii') at (5.5,-4.25) {(iii')};
\node (iv) at (9.5,-1) {(iv)};
\node (iv') at (9.5,-4.25) {(iv')};

\node[state] (A1) at (-2,0) {$A$};
\node[state] (B1) at (-0.25,0) {$B$};
\node[state] (C1) at (-0.25,-1.75) {$C$};

\node[state] (A1') at (-2,-3.25) {$A$};
\node[state] (B1') at (-0.25,-3.25) {$B$};
\node[state] (C1') at (-0.25,-5) {$C$};

\node[state] (A2) at (2,0) {$A$};
\node[state] (B2) at (3.75,0) {$B$};
\node[state] (C2) at (3.75,-1.75) {$C$};

\node[state] (A2') at (2,-3.25) {$A$};
\node[state] (B2') at (3.75,-3.25) {$B$};
\node[state] (C2') at (3.75,-5) {$C$};

\node[state] (A3) at (6,0) {$A$};
\node[state] (B3) at (7.75,0) {$B$};
\node[state] (C3) at (6,-1.75) {$C$};
\node[state] (D3) at (7.75,-1.75) {$D$};

\node[state] (A3') at (6,-3.25) {$A$};
\node[state] (B3') at (7.75,-3.25) {$B$};
\node[state] (C3') at (6,-5) {$C$};
\node[state] (D3') at (7.75,-5) {$D$};

\node[state] (A4) at (10,0) {$A$};
\node[state] (B4) at (11.75,0) {$B$};
\node[state] (C4) at (10,-1.75) {$C$};
\node[state] (D4) at (11.75,-1.75) {$D$};

\node[state] (A4') at (10,-3.25) {$A$};
\node[state] (B4') at (11.75,-3.25) {$B$};
\node[state] (C4') at (10,-5) {$C$};
\node[state] (D4') at (11.75,-5) {$D$};

\tikzset{dir/.style = {->, -{To[length=6, width=7]}, thick}}
\draw[dir]
(A1) edge (B1)
(A1') edge (B1')
(B1') edge [dashed] (C1')
(A2) edge (B2)
(B2) edge (C2)
(A2') edge (B2')
(B2') edge (C2')
(A2') edge [dashed] (C2')
(B3) edge (D3)
(C3) edge (D3)
(B3') edge (D3')
(C3') edge (D3')
(A3') edge [dashed] (D3')
(A4) edge (B4)
(B4) edge (D4)
(A4') edge (B4')
(B4') edge (D4')
(C4') edge [dashed] (D4')
; 
\tikzset{undir/.style = {-,  thick}}
\draw[undir]
(B1) edge (C1)
(A2) edge (C2)
(A3) edge (B3)
(A3) edge (C3)
(A3) edge (D3)
(A3') edge (B3')
(A3') edge (C3')
(A4) edge (C4)
(B4) edge (C4)
(C4) edge (D4)
(A4') edge (C4')
(B4') edge (C4')
;                    
        
\end{tikzpicture}

\caption{Meek's orientation rules \cite{meek1995}. If (i), (ii), (iii) or (iv) occur as an induced subgraph of some PDAG, then orient them as (i'), (ii'), (iii') or (iv'), respectively.}
\label{fig:meeksrules}
\end{figure}
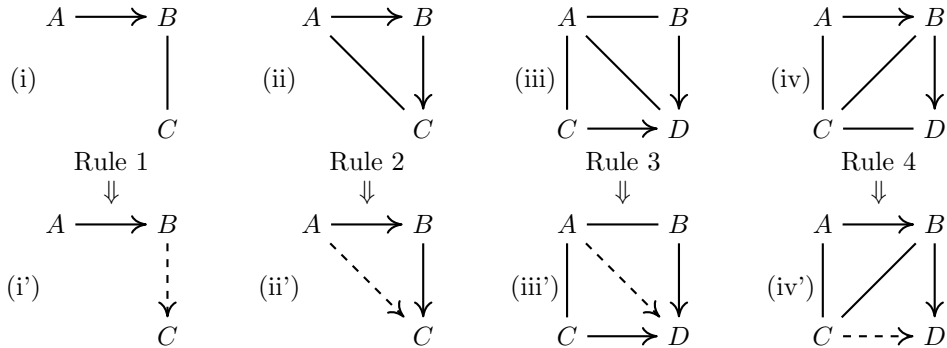

\pagebreak

\newgeometry{top = 2.8cm, bottom = 2.8cm, left = 3cm, right= 3cm}

\section{Algorithms}\label{app:algs}

\begin{algorithm}[!htbp]

\caption{tLMPC-stable algorithm}

\label{alg:tpc}

\Input{nodes $\V$, tiered ordering $\tau$, conditional independence oracle}

\Output{MPDAG $\G$}

\vspace{5pt} 

Skeleton phase: apply Algorithm \ref{alg:skeleton}\\

V-structure phase: apply Algorithm \ref{alg:vstructures}\\

Cross-tier phase: apply Algorithm \ref{alg:timeedges}\\

Meek's rules: apply Algorithm \ref{alg:meekrules}

\end{algorithm}

\begin{algorithm}[!htbp]

\caption{Skeleton phase (I-II)}

\label{alg:skeleton}

\Input{nodes $\V$, tiered ordering $\tau$, conditional independence oracle}

\Output{undirected graph $\mathcal{C}'$}

\vspace{5pt} 

Construct a complete undirected graph $\C$ over node set $\V$ \\
 
 $\C'=\C$\\
 
 $l=-1$

\Repeat{all ordered pairs of nodes $(V_i,V_j)$ adjacent in $\C'$ satisfy $|\mathrm{s}(V_i)\backslash\{ V_j\}|< l$}{

$l=l+1$

\For{all nodes $V_i$ in $\V$}{

$\mathrm{s}(V_i)=\{V\mid V\in\adj{\C'}{V_i} \text{ and } \tau(V)\leq \tau(V_i)\}$

} 

\Repeat{all ordered pairs of nodes $(V_i,V_j)$ with $V_j\in\adj{C'}{V_i}$ and $|\mathrm{s}(V_i)\backslash\{ V_j\}|\geq l$ have been chosen}{

select a new ordered pair of nodes $(V_i,V_j)$ with $V_j\in\adj{C'}{V_i}$ and $|\mathrm{s}(V_i)\backslash\{ V_j\}|\geq l$

\Repeat{edge $V_i- V_j$ is removed or all $\mathbf{S}\subseteq\mathrm{s}(V_i)\backslash\{ V_j\}$ with $|\mathbf{S}| = l$ have been chosen}{

choose a new $\mathbf{S}\subseteq\mathrm{s}(V_i)\backslash\{ V_j\}$ such that $|\mathbf{S}|=l$ 

\If{$X_{V_i}\indep X_{V_j}\mid\X_\mathbf{S}$}{

remove edge $V_i- V_j$ from $\C'$

} 

} 

} 

} 

\end{algorithm}

\begin{algorithm}[!htbp]

\caption{V-structure phase (IIIa)}

\label{alg:vstructures}

\Input{tiered ordering $\tau$, conditional independence oracle, undirected graph $\C'$}

\Output{PDAG $\C''$}

$\C''=\C'$\\

create empty lists $\textit{ambiguous}\_\textit{list}$ and $\textit{orient}\_\textit{list}$\\

\Repeat{all undirected unshielded triples $\langle V_i,V_j,V_k\rangle$ with $\max (\tau(V_i),\tau(V_k))\leq\tau(V_j)$ selected}{

select new undirected unshielded triple $\langle V_i,V_j,V_k\rangle$ with $\max (\tau(V_i),\tau(V_k))=\tau(V_j)$\\

$\mathrm{s}(V_i)=\{ V\mid V\in\adj{\C''}{V_i}\textnormal{ and } \tau(V)\leq\tau(V_i)\}$ and \\
$\mathrm{s}(V_k)=\{ V\mid V\in\adj{\C''}{V_k}\textnormal{ and } \tau(V)\leq\tau(V_k)\}$

$\textit{counter} = 0$

\For{all sets $\mathbf{S}$ such that $\mathbf{S}$ is a subset of $\mathrm{s}(V_i)$ or $\mathrm{s}(V_k)$ or both}{

\If{$X_{V_i}\indep X_{V_k}\mid \mathbf{X}_\mathbf{S}$}{

\eIf{$V_j\in\mathbf{S}$}{

$\textit{counter} = \textit{counter}-1$

}{

$\textit{counter} = \textit{counter}+1$

} 

} 

} 

\eIf{$\textit{counter}=0$}{

add $(V_i,V_j)$ to $\textit{ambiguous}\_\textit{list}$\\
add $(V_k,V_j)$ to $\textit{ambiguous}\_\textit{list}$

}{
\If{$\textit{counter}>0$}{

add $(V_i,V_j)$ to $\textit{orient}\_\textit{list}$\\
add $(V_k,V_j)$ to $\textit{orient}\_\textit{list}$

} 
} 

} 

\For{all pairs $(V_m,V_n)$ in $\textit{orient}\_\textit{list}$}{

\eIf{$V_m-V_n$ is in $\mathcal{C}''$}{

replace by $V_m\rightarrow V_n$

}{

\If{$V_m\leftarrow V_n$ is in $\mathcal{C}''$}{

replace by $V_m\leftrightarrow V_n$

} 

} 

} 

\end{algorithm}

\begin{algorithm}[!htbp]

\caption{Cross-tier phase (IIIb)}

\label{alg:timeedges}

\Input{tiered ordering $\tau$, PDAG $\C''$}

\Output{PDAG $\C'''$}

\vspace{5pt} 

$\C'''=\C''$

\For{all ordered pairs $(V_i, V_j)$ with $V_j\in\adj{\C'''}{V_i}$}{

\uIf{$\tau(V_i)<\tau(V_j)$}{

replace $V_i - V_j$ by $V_i\rightarrow V_j$

} 

} 

\end{algorithm}

\begin{algorithm}[!htbp]

\caption{Meek's rules (IV)}

\label{alg:meekrules}

\Input{PDAG $\C'''$}

\Output{MPDAG $\G$}

$\G = \C'''$\\

\Repeat{no further edges can be oriented}{

create empty list $\textit{rule1}\_\textit{list}$ and apply the following rule to as many undirected edges in $\G$ not listed in $\textit{ambiguous}\_\textit{list}$ as possible:\\

\vspace{2pt} 

$\textbf{Rule 1:}$ Add  $(V_i,V_j)$ to $\textit{rule1}\_\textit{list}$ if $\G$  contains an edge $V_k\rightarrow V_i$ with $V_k\not\in\adj{\G}{V_j}$.\\

\For{all pairs $(V_m,V_n)$ in $\textit{rule1}\_\textit{list}$}{

\eIf{$V_m-V_n$ is in $\G$}{

replace by $V_m\rightarrow V_n$

}{

\uIf{$V_m\leftarrow V_n$ is in $\G$}{

replace by $V_m\leftrightarrow V_n$

} 

} 

} 

create empty list $\textit{rule2}\_\textit{list}$ and apply the following rule to as many undirected edges in $\G$ not listed in $\textit{ambiguous}\_\textit{list}$ as possible:\\

\vspace{2pt} 

$\textbf{Rule 2:}$ Add  $(V_i,V_j)$ to $\textit{rule2}\_\textit{list}$ if $\G$  contains a path $V_i\rightarrow V_k\rightarrow V_j$ and $V_i\in\adj{\G}{V_j}$\\

\For{all pairs $(V_m,V_n)$ in $\textit{rule2}\_\textit{list}$}{

\eIf{$V_m-V_n$ is in $\G$}{

replace by $V_m\rightarrow V_n$

}{

\uIf{$V_m\leftarrow V_n$ is in $\G$}{

replace by $V_m\leftrightarrow V_n$

} 

} 

} 

create empty list $\textit{rule3}\_\textit{list}$ and apply the following rule to as many undirected edges in $\G$ not listed in $\textit{ambiguous}\_\textit{list}$ as possible:\\

\vspace{2pt} 

$\textbf{Rule 3:}$ Add  $(V_i,V_j)$ to $\textit{rule3}\_\textit{list}$ if $\G$  contains $V_k,V_m$ with $V_k\not\in\adj{\G}{V_m}$ and the edges $V_k-V_i$, $V_m-V_i$, $V_k\rightarrow V_j$ and $V_m\rightarrow V_j$.\\

\For{all pairs $(V_m,V_n)$ in $\textit{rule1}\_\textit{list}$}{

\eIf{$V_m-V_n$ is in $\G$}{

replace by $V_m\rightarrow V_n$

}{

\uIf{$V_m\leftarrow V_n$ is in $\G$}{

replace by $V_m\leftrightarrow V_n$

} 

} 

} 

create empty list $\textit{rule4}\_\textit{list}$ and apply the following rule to as many undirected edges in $\G$ not listed in $\textit{ambiguous}\_\textit{list}$ as possible:\\

\vspace{2pt} 

$\textbf{Rule 4:}$ Add  $(V_i,V_j)$ to $\textit{rule4}\_\textit{list}$ if $\G$  contains a paths $V_k\rightarrow V_m\rightarrow V_j$, $V_k-V_i$ and $V_m-V_i$, and $V_j\not\in\adj{\G}{V_k}$.\\

\For{all pairs $(V_m,V_n)$ in $\textit{rule4}\_\textit{list}$}{

\eIf{$V_m-V_n$ is in $\G$}{

replace by $V_m\rightarrow V_n$

}{

\uIf{$V_m\leftarrow V_n$ is in $\G$}{

replace by $V_m\leftrightarrow V_n$

} 

} 

} 

} 

\end{algorithm}

\restoregeometry

\clearpage

\section{Simulation study}\label{app:simulation}

\subsection{Performance measures}\label{app:performance}

\paragraph{Correct skeleton}

Let $\G$ be a graph, and let $\widehat{\G}$ be an estimate of $\G$. An edge $V_i \any V_j$ in $\widehat{\G}$ is a \emph{true positive} if $V_i$ and $V_j$ are adjacent in the true graph $\G$. Conversely, if $V_i$ and $V_j$ are not adjacent in $\G$, then $V_i \any V_j$ in $\widehat{\G}$ is a \emph{false positive}. Correspondingly, if there is no $V_i \any V_j$ in $\widehat{\G}$, but $V_i$ and $V_j$ are in fact adjacent in the true graph $\G$, then we call this a \emph{false negative}.

Given an estimated graph $\widehat{\G}$ and true graph $\G$ we then define $\mathrm{true.positives}(\widehat{\G},\G)$ as the set of true positives, $\mathrm{false.positives}(\widehat{\G},\G)$ as the set of false positives, and $\mathrm{false.negatives}(\widehat{\G},\G)$ as the set of false negatives. By $\vert \cdot \vert$ we denote the number of elements in a set.

We define the \emph{precision} relative to $\widehat{\G}$ and $\G$ as 

\begin{align}\label{def:precision}
    \mathrm{precision}(\widehat{\G},\G)=\dfrac{\vert\mathrm{true.positives}(\widehat{\G},\G)\vert}{\vert\mathrm{true.positives}(\widehat{\G},\G)\vert + \vert\mathrm{false.positives}(\widehat{\G},\G)\vert}
\end{align}

We define the \emph{recall} relative to $\widehat{\G}$ and $\G$ as 

\begin{align}\label{def:recall}
    \mathrm{recall}(\widehat{\G},\G)=\dfrac{\vert\mathrm{true.positives}(\widehat{\G},\G)\vert}{\vert\mathrm{true.positives}(\widehat{\G},\G)\vert + \vert\mathrm{false.negatives}(\widehat{\G},\G)\vert}
\end{align}

\paragraph{Correct v-structures}

We again consider $\G$ and $\widehat{\G}$ as above. A v-structure consisting of $\langle V_i, V_j, V_k\rangle$ in  $\widehat{\G}$ is a \emph{true positive} if the triple $\langle V_i, V_j, V_k\rangle$ constitutes a v-structure in the true graph $\G$. Conversely, if $\langle V_i, V_j, V_k\rangle$ does not constitute a v-structure in $\G$, then this is a \emph{false positive}. If $\langle V_i, V_j, V_k\rangle$ does not form a v-structure in $\widehat{G}$, but it is in fact a v-structure in $\G$, then this is a \emph{false negative}. By the definitions above, we are able to compute the precision and recall as described in equations (\ref{def:precision}) and (\ref{def:recall}) with respect to the v-structures.

\paragraph{Ancestral relations}

We again consider $\G$ and $\widehat{\G}$ as above. If $V_i\in\an{\widehat{\G}}{V_j}$ and $V_i\in\an{\G}{V_j}$ then this ancestral relation is a true positive. Conversely, if $V_i\in\an{\widehat{\G}}{V_j}$ and $V_i\not\in\an{\G}{V_j}$ then this ancestral relation is a false positive. Correspondingly, if $V_i\not\in\an{\widehat{\G}}{V_j}$ and $V_i\in\an{\G}{V_j}$, then this is a false negative. Using these definitions, we can compute the precision and recall with respect to the ancestral relations using (\ref{def:precision}) and (\ref{def:recall}).

We will extend this to \emph{possible ancestors}: We say that $V_i$ is a possible ancestor of $V_j$ in a graph $\G$ ($V_i\in\possan{\G}{V_j}$) if there is a path from $V_i$ in $V_j$, and there exists no path $V_i\any\ldots\any V_j$ with $V_k\leftarrow V_{k+1}$ for some $i\leq k<j$ in $\G$. This corresponds to what is called a b-possible ancestor in \cite{perkovic2017}, which is more strict than the usual definition on of possible ancestors. In general, PDAGs and MPDAGs might contain partially directed cycles, and we have to check for b-possible ancestors \cite{perkovic2017}, which is not the case for tiered MPDAGs \cite{bang2023we}. However, since the estimated graphs are not guaranteed to be valid tiered MPDAGs due to possible statistical errors, we will use the definition from \cite{perkovic2017} as it is more conservative.

We then say that if $V_i\in\possan{\widehat{\G}}{V_j}$ and $V_i\in\possan{\G}{V_j}$ then this possible ancestral relation is a true positive. Conversely, if $V_i\not\in\possan{\widehat{\G}}{V_j}$ and $V_i\in\possan{\G}{V_j}$ then this ancestral relation is a false positive. Correspondingly, if $V_i\not\in\possan{\widehat{\G}}{V_j}$ and $V_i\in\possan{\G}{V_j}$, then this is a false negative. As above, we can compute the precision and recall using equations (\ref{def:precision}) and (\ref{def:recall}) with respect to the possible ancestral relations.

\paragraph{Proportion of conflicting edges}

In Section \ref{sec:stability} we introduced an edge $V_i\any V_j$ in an estimated graph $\widehat{\G}$ as a conflicting edge if in the causal discovery procedure it was not possible to determine the orientation due to statistical errors, not statistical equivalence. In practice, the tPC algorithm encodes this in the estimated MPDAG by using a bidirected edge $V_i\leftrightarrow V_j$. Hence, we compute the proportion of conflicting edges in an estimated MPDAG $\widehat{\G}=(\V,\widehat{\E})$ in the following way

\begin{align}
    \mathrm{conflicting}(\widehat{\G})=\dfrac{\vert\{V_i\in\V \mid \exists V_j\in\V: \{V_i\leftrightarrow V_j\}\in\widehat{\E}\}\vert}{\vert\widehat{\E}\vert}
\end{align}

\vspace{10pt}

\subsection{Tiered orderings}\label{app:simorderings}

\begin{figure}[!htbp]
    \centering
\begin{tikzpicture}

\draw[dashed, thick] [lightblue] (1.75, 1.125) -- (1.75, -3.75);
\draw[dashed, thick] [lightblue] (4.25, 1.125) -- (4.25, -5);
\draw[dashed, thick] [lightblue] (6.75, 1.125) -- (6.75, -2.5);

\draw[dashed, thick] [lightblue] (6.75, -3.75) -- (6.75, -6.25);

\draw[dashed, thick] [lightblue] (9.25, 1.125) -- (9.25, -2.5);

\draw[dashed, thick] [lightblue] (9.25, -5) -- (9.25, -7.5);

\filldraw[silvergray] (-0.5,1) rectangle +(2,-2)  ;
\filldraw[silvergray] (2,1) rectangle +(2,-2)  ;
\filldraw[silvergray] (4.5,1) rectangle +(2,-2)  ;
\filldraw[silvergray] (7,1) rectangle +(2,-2)  ;
\filldraw[silvergray] (9.5,1) rectangle +(2,-2)  ;

\filldraw[lightblue] (-0.5, -1.25) rectangle +(2,-1) ;
\filldraw[lightblue] (2, -1.25) rectangle +(2,-1) ;
\filldraw[lightblue] (4.5, -1.25) rectangle +(2,-1) ;
\filldraw[lightblue] (7, -1.25) rectangle +(2,-1) ;
\filldraw[lightblue] (9.5, -1.25) rectangle +(2,-1) ;

\filldraw[lightblue] (-0.5, -2.5) rectangle +(2,-1) ;
\filldraw[lightblue] (2, -2.5) rectangle +(9.5,-1) ;

\filldraw[lightblue] (-0.5, -3.75) rectangle +(4.5,-1) ;
\filldraw[lightblue] (4.5, -3.75) rectangle +(7,-1) ;

\filldraw[lightblue] (-0.5, -5) rectangle +(7,-1) ;
\filldraw[lightblue] (7, -5) rectangle +(4.5,-1) ;

\filldraw[lightblue] (-0.5, -6.25) rectangle +(9.5,-1) ;
\filldraw[lightblue] (9.5, -6.25) rectangle +(2,-1) ;

\filldraw[lightblue] (-0.5, -7.5) rectangle +(12,-1) ;

\node (A) at (0.5, 0) {\textsf{Tier 1}};
\node (B) at (3, 0) {\textsf{Tier 2}};
\node (C) at (5.5, 0) {\textsf{Tier 3}};
\node (D) at (8, 0) {\textsf{Tier 4}};
\node (E) at (10.5, 0) {\textsf{Tier 5}};

\node (A) at (0.5,-1.75) {$\tau_5=1$};
\node (A) at (3,-1.75) {$\tau_5=2$};
\node (A) at (5.5,-1.75) {$\tau_5=3$};
\node (A) at (8,-1.75) {$\tau_5=4$};
\node (A) at (10.5,-1.75) {$\tau_5=5$};

\node (A) at (0.5,-3) {$\tau_2^A=1$};
\node (A) at (6.75,-3) {$\tau_2^A=2$};

\node (A) at (1.75,-4.25) {$\tau_2^B=1$};
\node (A) at (8,-4.25) {$\tau_2^B=2$};

\node (A) at (3,-5.5) {$\tau_2^C=1$};
\node (A) at (9.25,-5.5) {$\tau_2^C=2$};

\node (A) at (4.25,-6.75) {$\tau_2^D=1$};
\node (A) at (10.5,-6.75) {$\tau_2^D=2$};

\node (A) at (5.5,-8) {$\tau_1=1$};

\draw[decoration={brace, raise=5pt},decorate]
(11.5,1) -- node[right=8pt] {\textsf{True ordering}} (11.5,-1);

\draw[decoration={brace, raise=5pt},decorate]
(11.5,-1.25) -- node[align= center, right=8pt] {\textsf{Detailed background}\\ \textsf{knowledge}} (11.5,-2.25);

\draw[decoration={brace, raise=5pt},decorate]
(11.5,-2.5) -- node[align= center, right=8pt] {\textsf{Partial background}\\ \textsf{knowledge}} (11.5,-7.25);

\draw[decoration={brace, raise=5pt},decorate]
(11.5,-7.5) -- node[align= center, right=8pt] {\textsf{No background}\\ \textsf{knowledge}} (11.5,-8.5);

\end{tikzpicture}
    \caption{Tiered orderings considered in the simulation study. We compared the MPDAGs estimated using no background knowledge ($\tau_1$), partial background knowledge ($\tau_2^A$, $\tau_2^B$, $\tau_2^C$ or $\tau_2^D$) and detailed background knowledge ($\tau_5$). For each combination of parameters in the study, we constructed an MPDAG using $\tau_1$, an MPDAG using $\tau_5$, and an MPDAG using either $\tau_2^A$, $\tau_2^B$, $\tau_2^C$ or $\tau_2^D$ (chosen with equal probability).}
    \label{fig:sim_taus}
\end{figure}
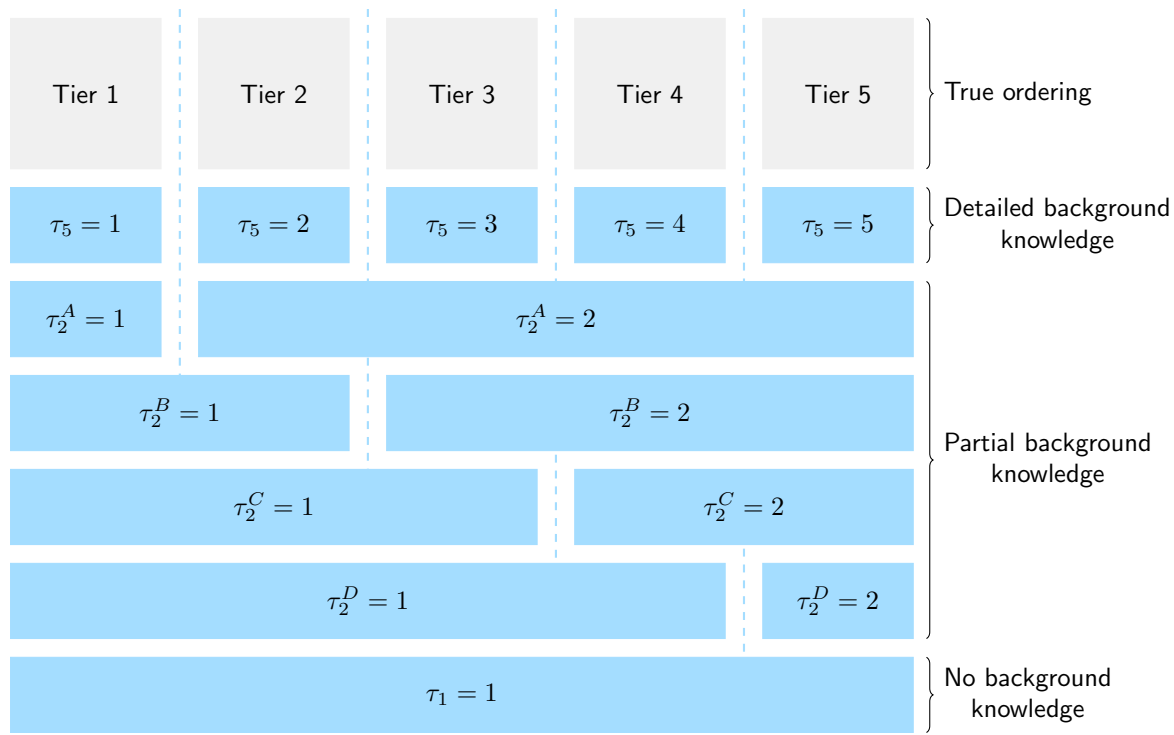

\newpage

\newgeometry{left=2cm, right = 2cm, top =3cm, bottom=3cm}

\subsection{Main analysis}\label{app:mainanalysis}

\begin{figure}[!htbp]
    \centering
    \includegraphics[scale = 0.975]{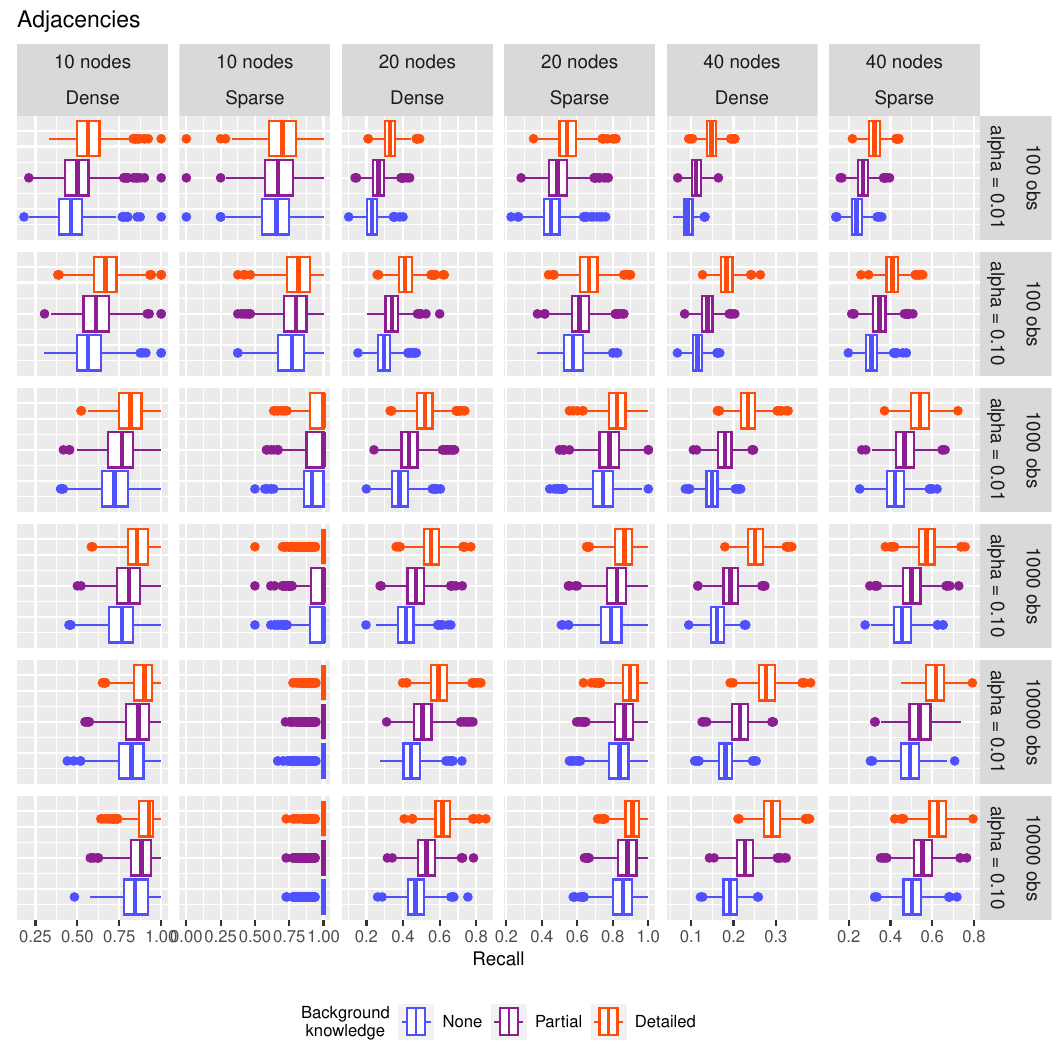}
    \caption{Recall of adjacencies from 1,000 simulations in dense/sparse settings with 10, 20 or 40 nodes, $\alpha\in \{0.01;0.1\}$ and a sample size of 100, 1,000, or 10,000, using the tLMPC-stable algorithm.}
    \label{fig:sim_adj_rec_full}
\end{figure}

\begin{figure}[!htbp]
    \centering
    \includegraphics[scale = 0.975]{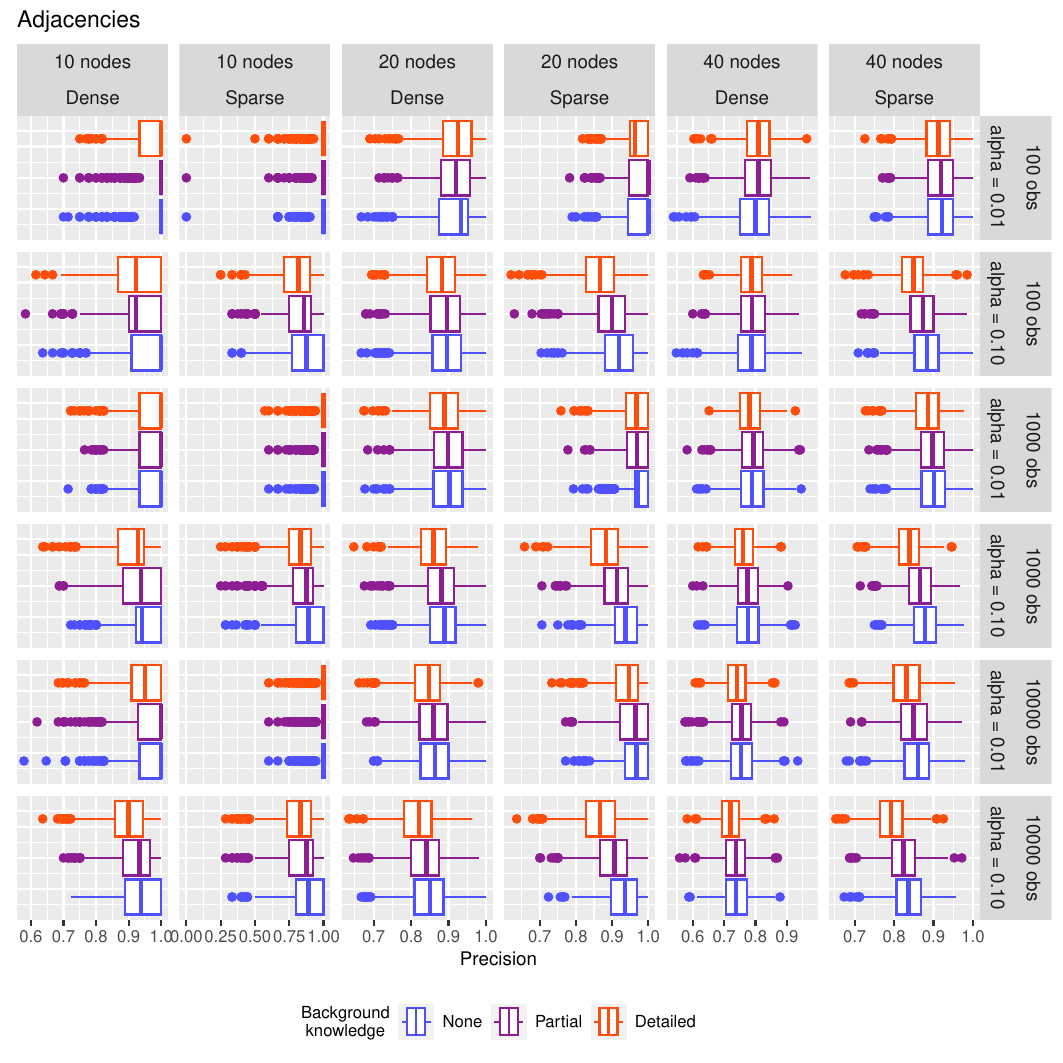}
    \caption{Precision of adjacencies from 1,000 simulations in dense/sparse settings with 10, 20 or 40 nodes, $\alpha\in \{0.01;0.1\}$ and a sample size of 100, 1,000, or 10,000, using the tLMPC-stable algorithm.}
    \label{fig:sim_adj_prec_full}
\end{figure}

\begin{figure}[!htbp]
    \centering
    \includegraphics[scale = 0.975]{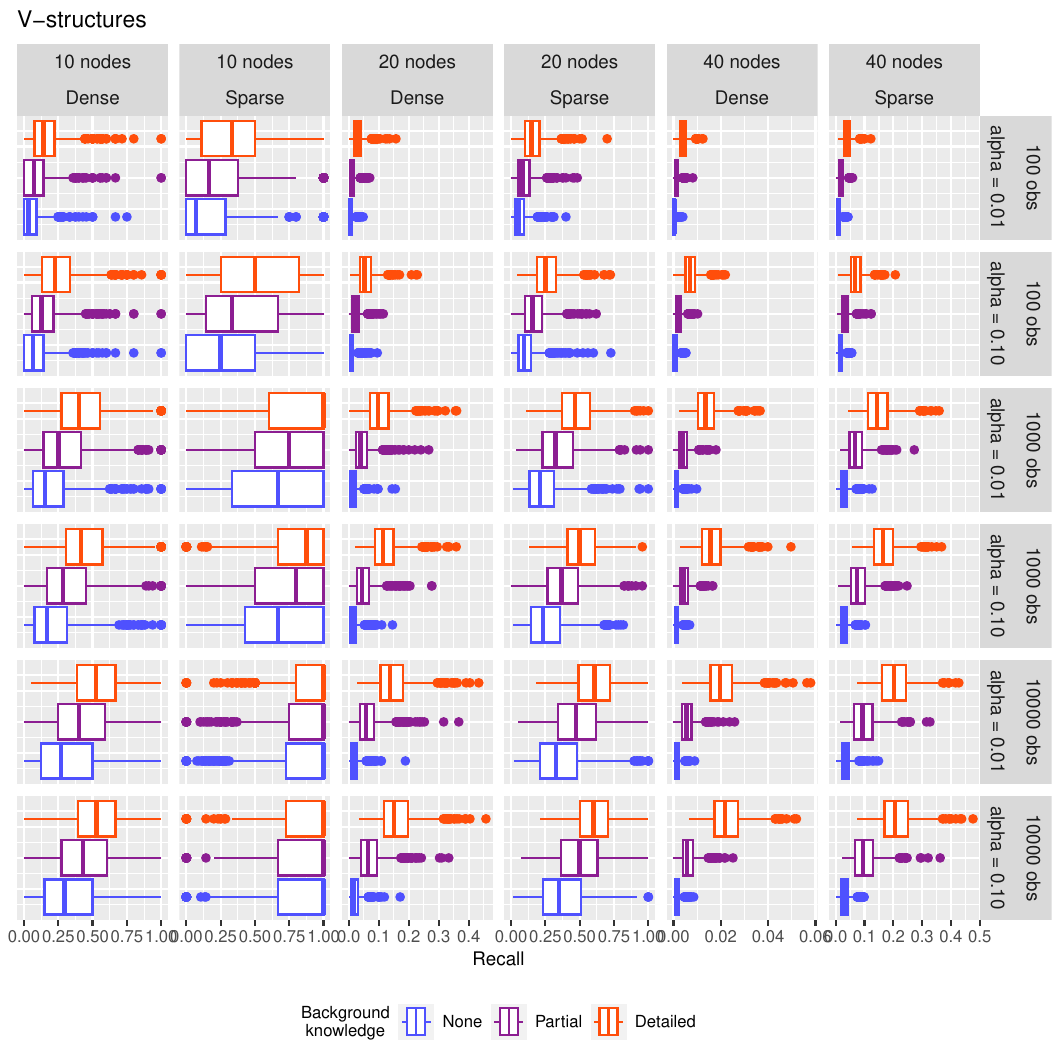}
    \caption{Recall of v-structures from 1,000 simulations in dense/sparse settings with 10, 20 or 40 nodes, $\alpha\in \{0.01;0.1\}$ and a sample size of 100, 1,000, or 10,000, using the tLMPC-stable algorithm.}
    \label{fig:sim_vstruct_rec_full}
\end{figure}

\begin{figure}[!htbp]
    \centering
    \includegraphics[scale = 0.975]{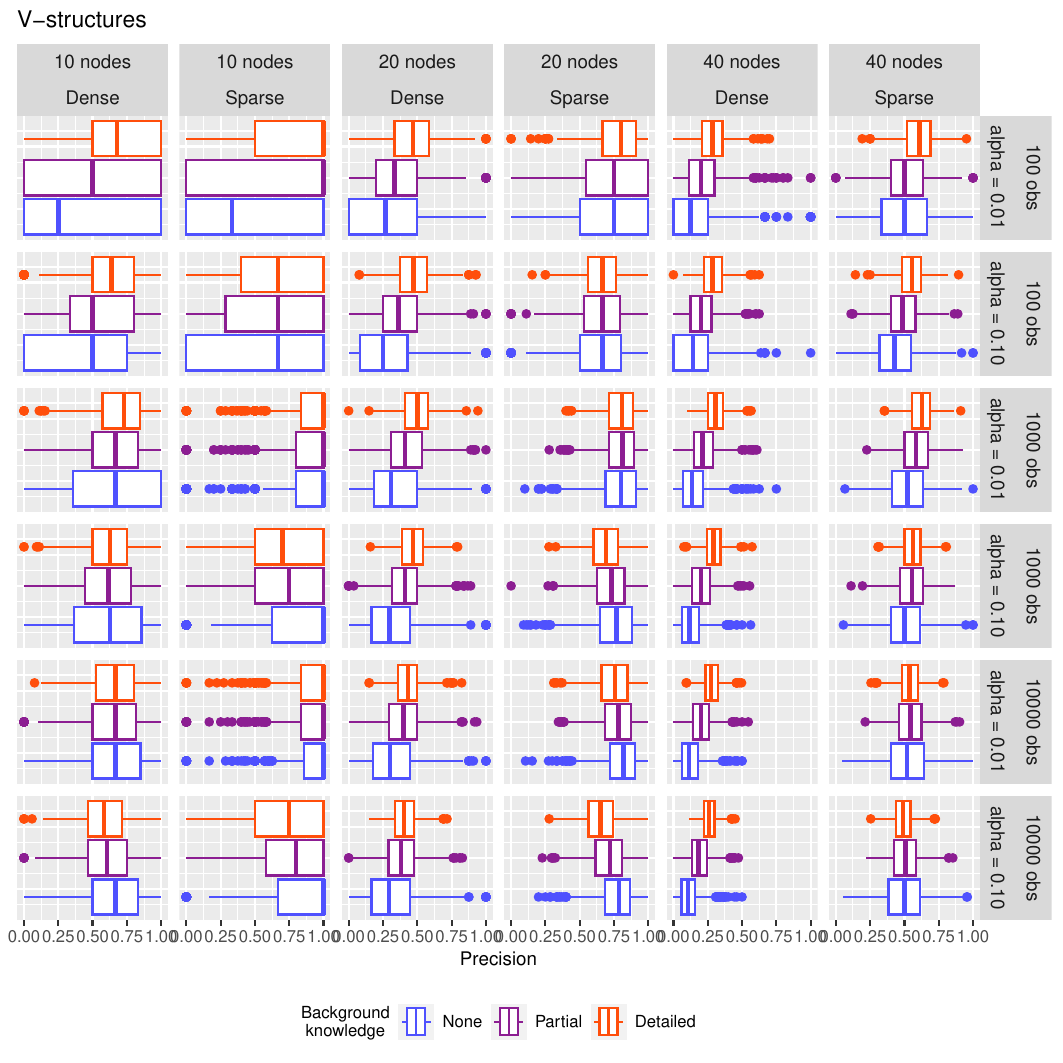}
    \caption{Precision of v-structures from 1,000 simulations in dense/sparse settings with 10, 20 or 40 nodes, $\alpha\in \{0.01;0.1\}$ and a sample size of 100, 1,000, or 10,000, using the tLMPC-stable algorithm.}
    \label{fig:sim_vstruct_prec_full}
\end{figure}

\begin{figure}[!htbp]
    \centering
    \includegraphics[scale = 0.975]{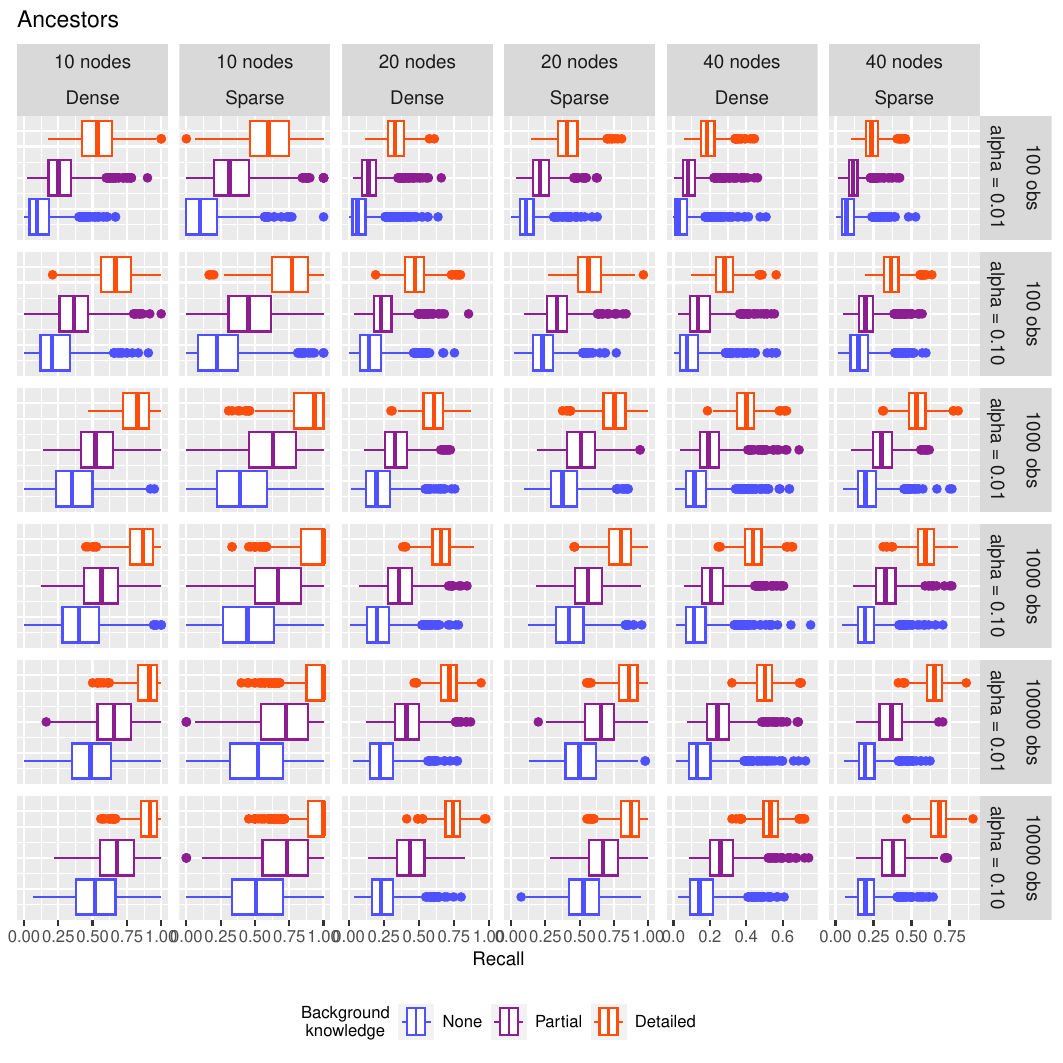}
    \caption{Recall of ancestral relations from 1,000 simulations in dense/sparse settings with 10, 20 or 40 nodes, $\alpha\in \{0.01;0.1\}$ and a sample size of 100, 1,000, or 10,000, using the tLMPC-stable algorithm.}
    \label{fig:sim_an_rec_full}
\end{figure}

\begin{figure}[!htbp]
    \centering
    \includegraphics[scale = 0.975]{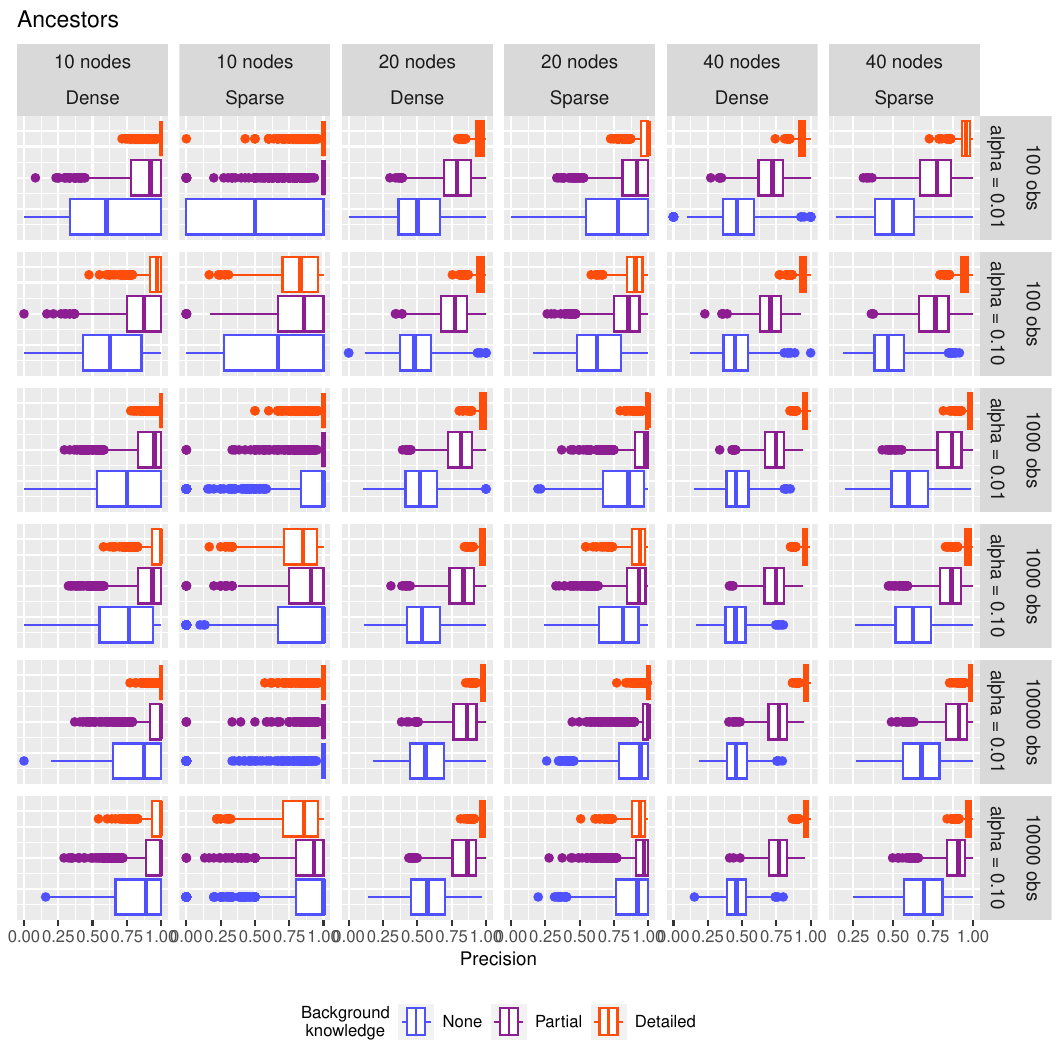}
    \caption{Precision of ancestral relations from 1,000 simulations in dense/sparse settings with 10, 20 or 40 nodes, $\alpha\in \{0.01;0.1\}$ and a sample size of 100, 1,000, or 10,000, using the tLMPC-stable algorithm.}
    \label{fig:sim_an_prec_full}
\end{figure}

\begin{figure}[!htbp]
    \centering
    \includegraphics[scale = 0.975]{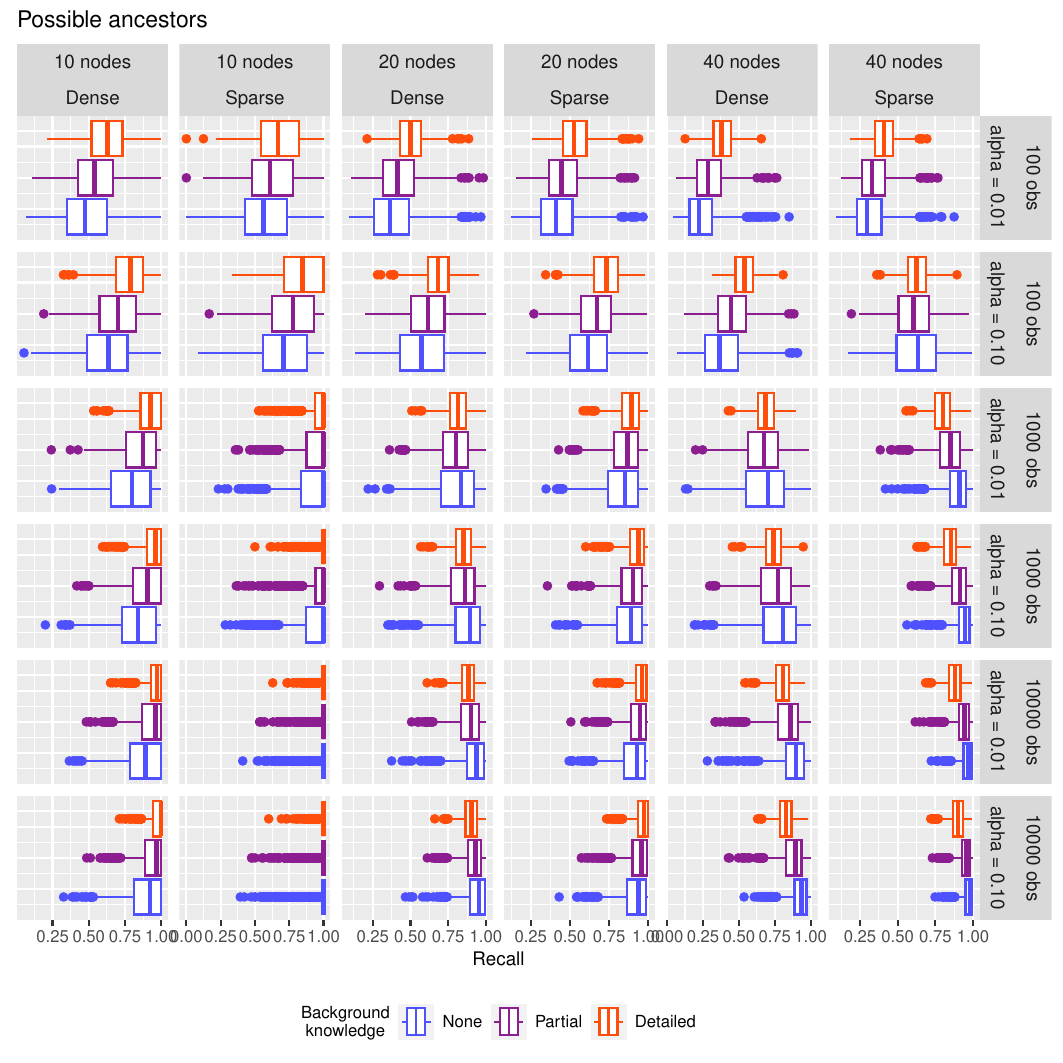}
    \caption{Recall of possible ancestral relations from 1,000 simulations in dense/sparse settings with 10, 20 or 40 nodes, $\alpha\in \{0.01;0.1\}$ and a sample size of 100, 1,000, or 10,000, using the tLMPC-stable algorithm.}
    \label{fig:sim_possan_rec_full}
\end{figure}

\begin{figure}[!htbp]
    \centering
    \includegraphics[scale = 0.975]{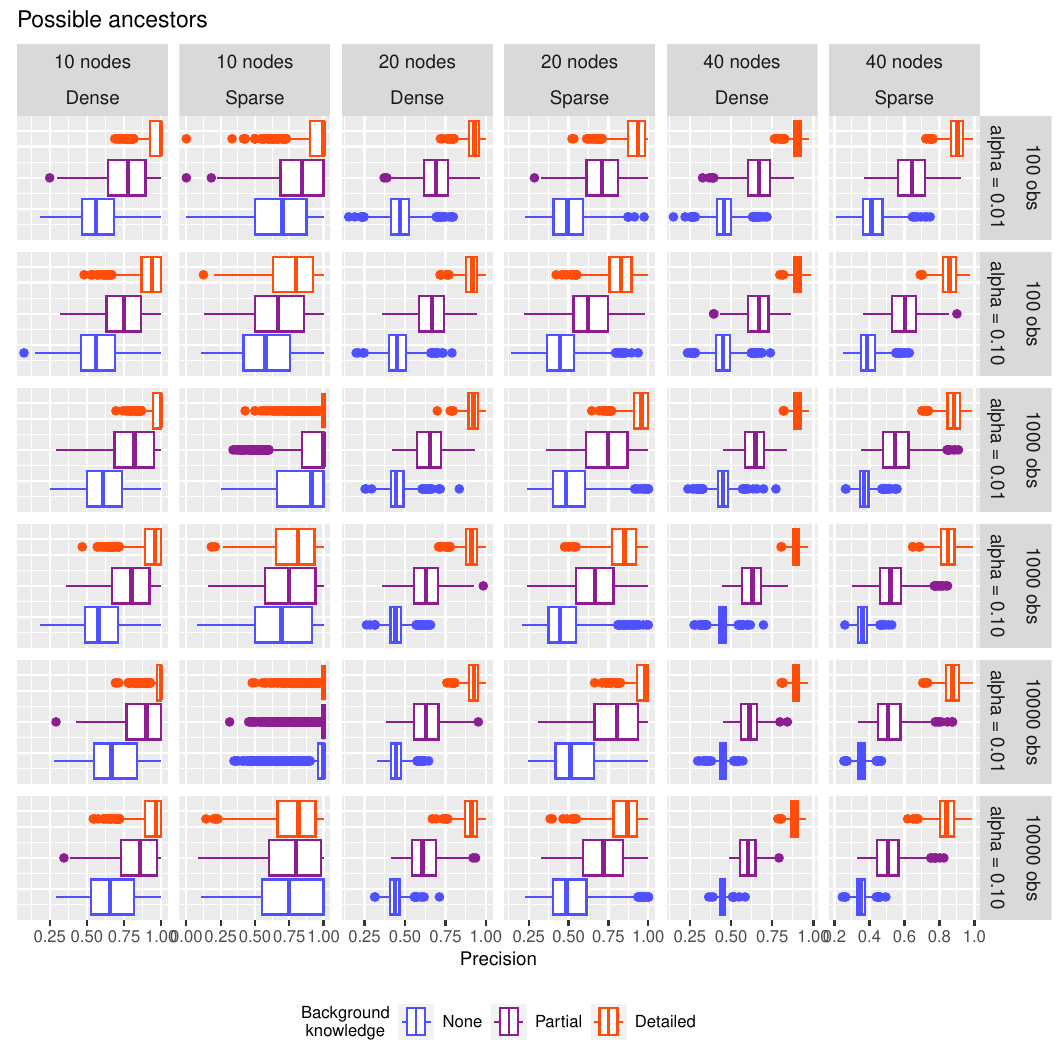}
    \caption{Precision of possible ancestral relations from 1,000 simulations in dense/sparse settings with 10, 20 or 40 nodes, $\alpha\in \{0.01;0.1\}$ and a sample size of 100, 1,000, or 10,000, using the tLMPC-stable algorithm.}
    \label{fig:sim_possan_prec_full}
\end{figure}

\begin{figure}[!htbp]
    \centering
    \includegraphics[scale = 0.975]{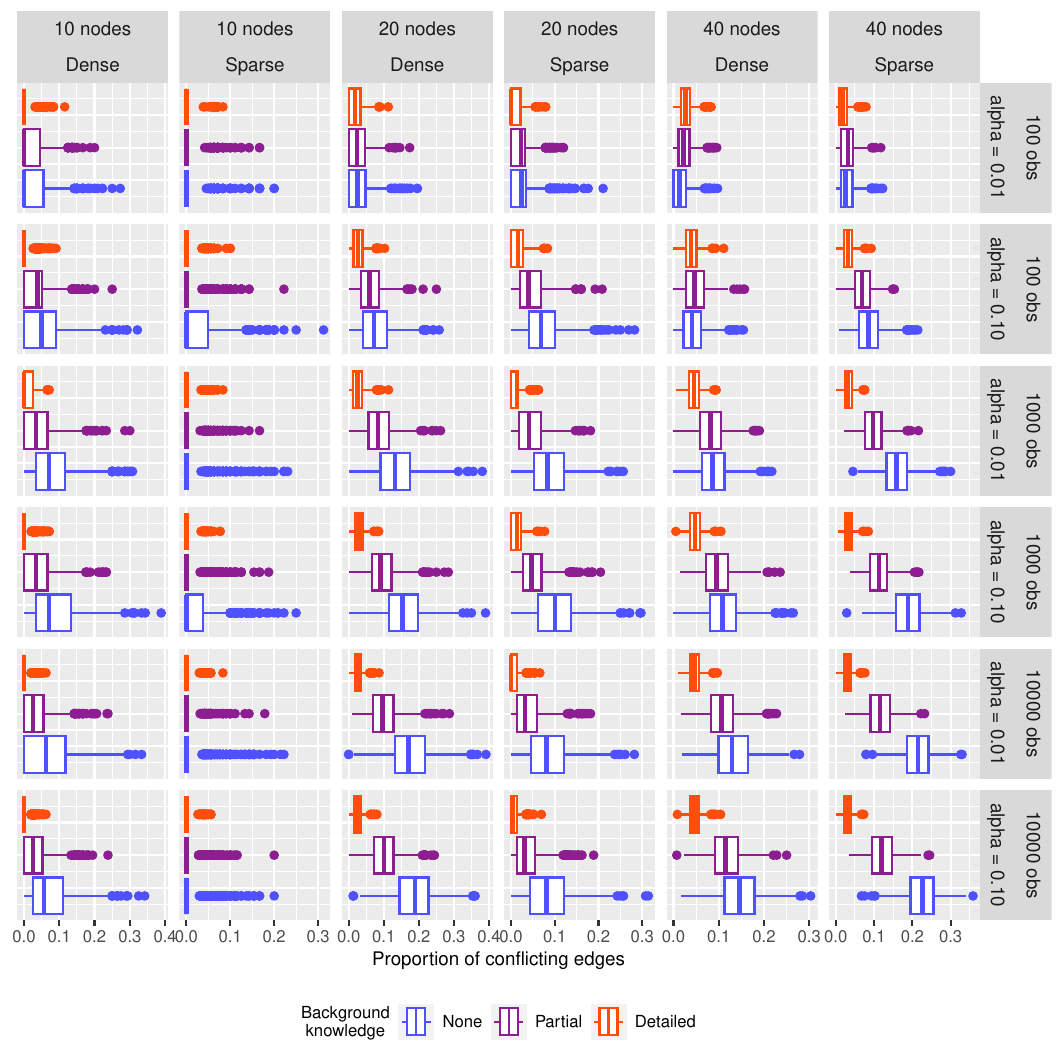}
    \caption{Proportion of bidirected edges from 1,000 simulations in dense/sparse settings with 10, 20 or 40 nodes, $\alpha\in \{0.01;0.1\}$ and a sample size of 100, 1,000, or 10,000, using the tLMPC-stable algorithm.}
    \label{fig:sim_conflicts_full}
\end{figure}

\newpage

\subsection{Naive tPC-algorithm}\label{app:naiveresults}

\begin{figure}[!htbp]
    \centering
    \includegraphics[scale = 0.975]{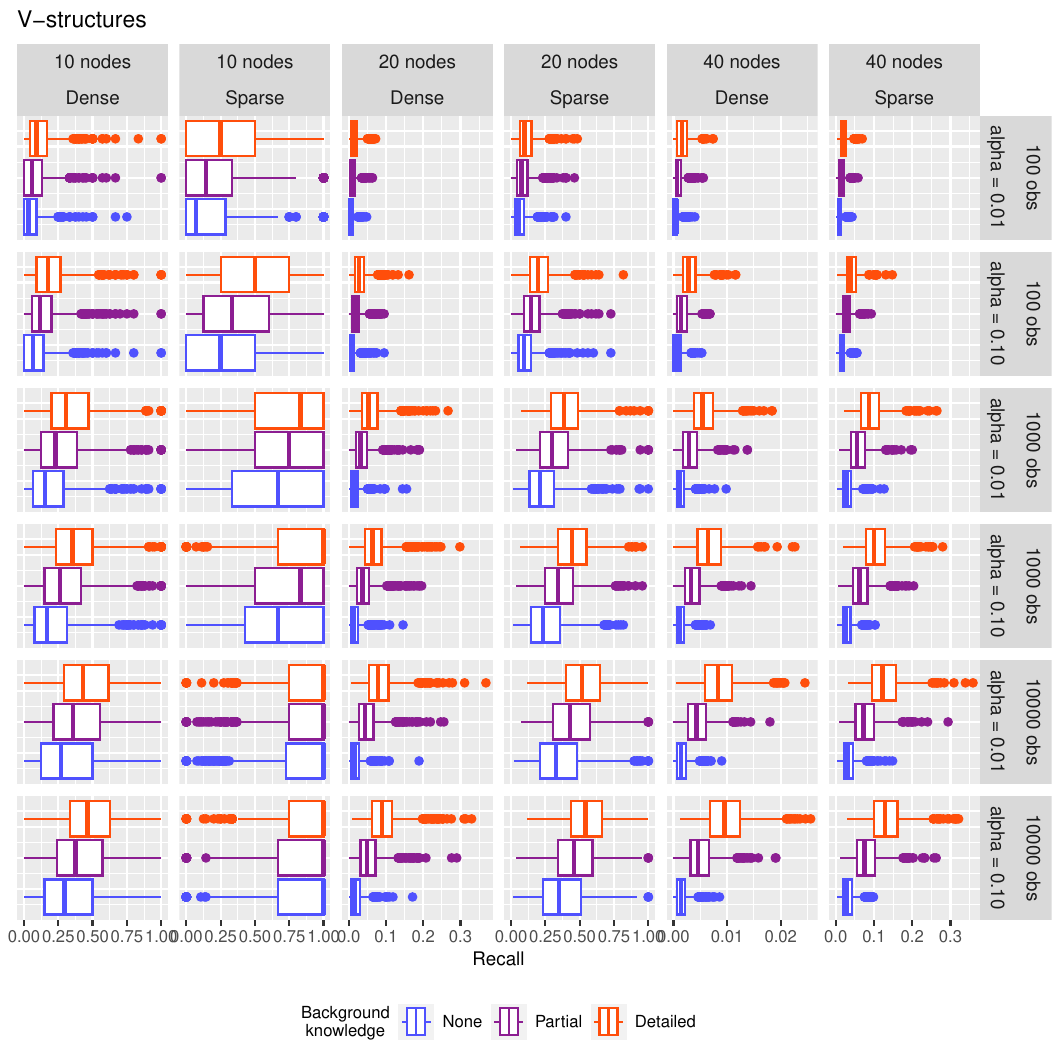}
    \caption{Recall of v-structures from 1,000 simulations in dense/sparse settings with 10, 20 or 40 nodes, $\alpha\in \{0.01;0.1\}$ and a sample size of 100, 1,000, or 10,000, using the naive tPC algorithm.}
    \label{fig:sim_vstruct_rec_full_naive}
\end{figure}

\begin{figure}[!htbp]
    \centering
    \includegraphics[scale = 0.975]{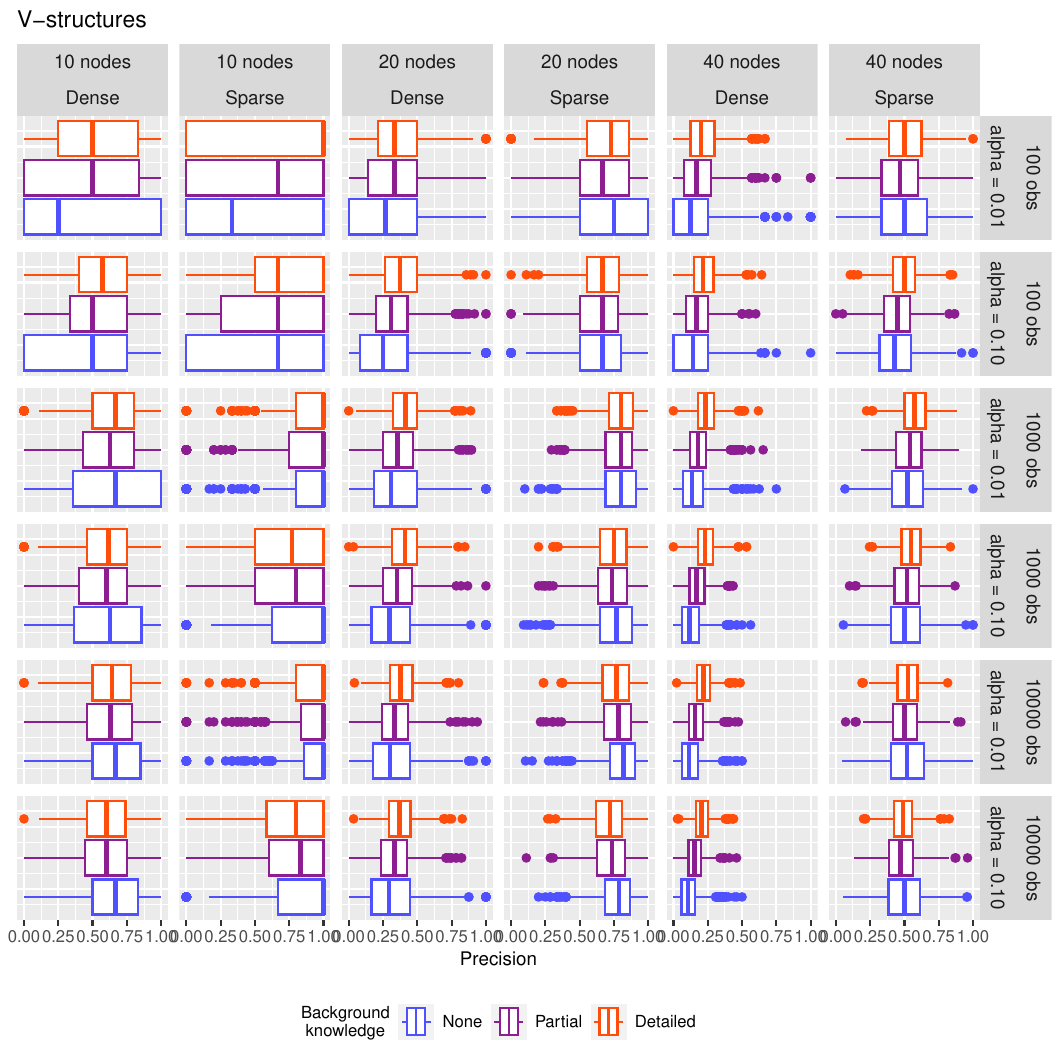}
    \caption{Precision of v-structures from 1,000 simulations in dense/sparse settings with 10, 20 or 40 nodes, $\alpha\in \{0.01;0.1\}$ and a sample size of 100, 1,000, or 10,000, using the naive tPC algorithm.}
    \label{fig:sim_vstruct_prec_full_naive}
\end{figure}

\begin{figure}[!htbp]
    \centering
    \includegraphics[scale = 0.975]{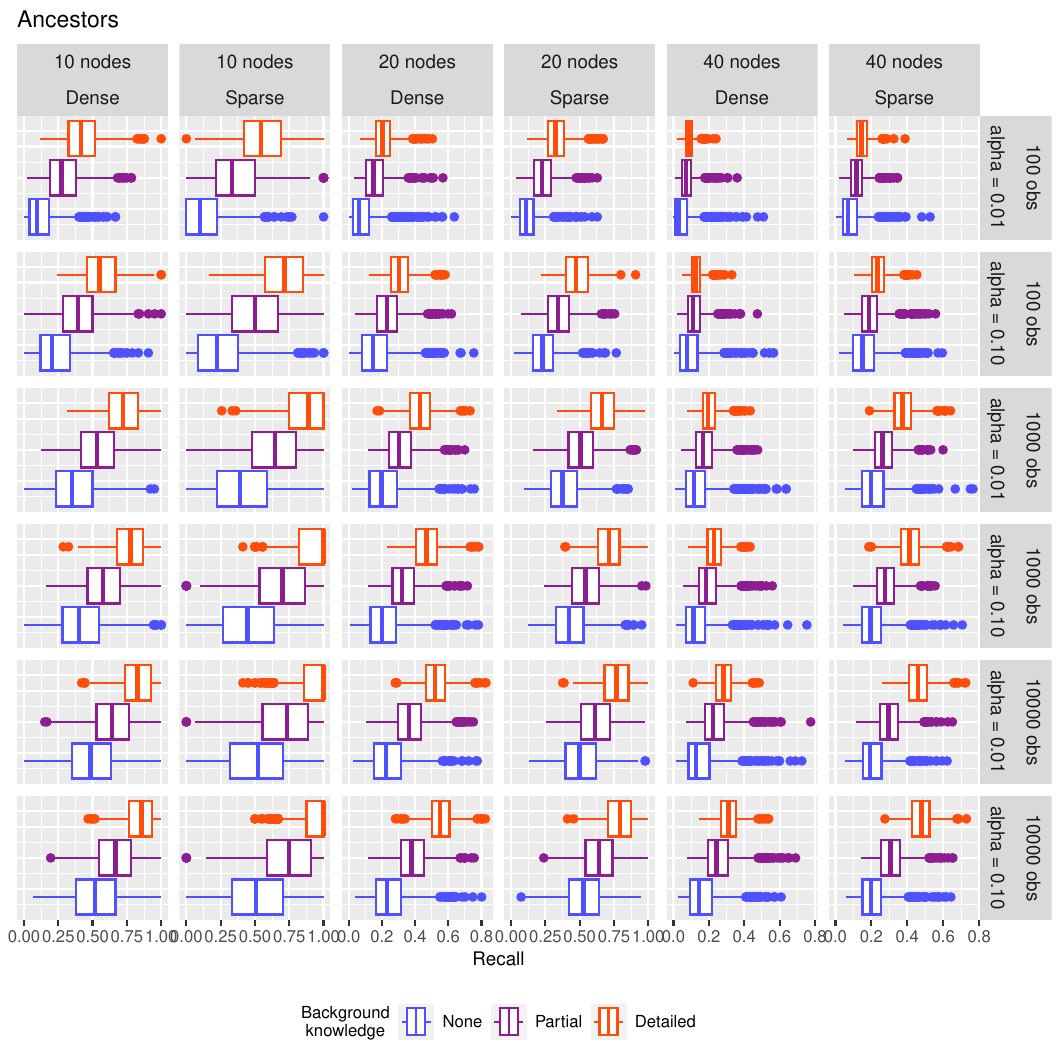}
    \caption{Recall of ancestral relations from 1,000 simulations in dense/sparse settings with 10, 20 or 40 nodes, $\alpha\in \{0.01;0.1\}$ and a sample size of 100, 1,000, or 10,000, using the naive tPC algorithm.}
    \label{fig:sim_an_rec_full_naive}
\end{figure}

\begin{figure}[!htbp]
    \centering
    \includegraphics[scale = 0.975]{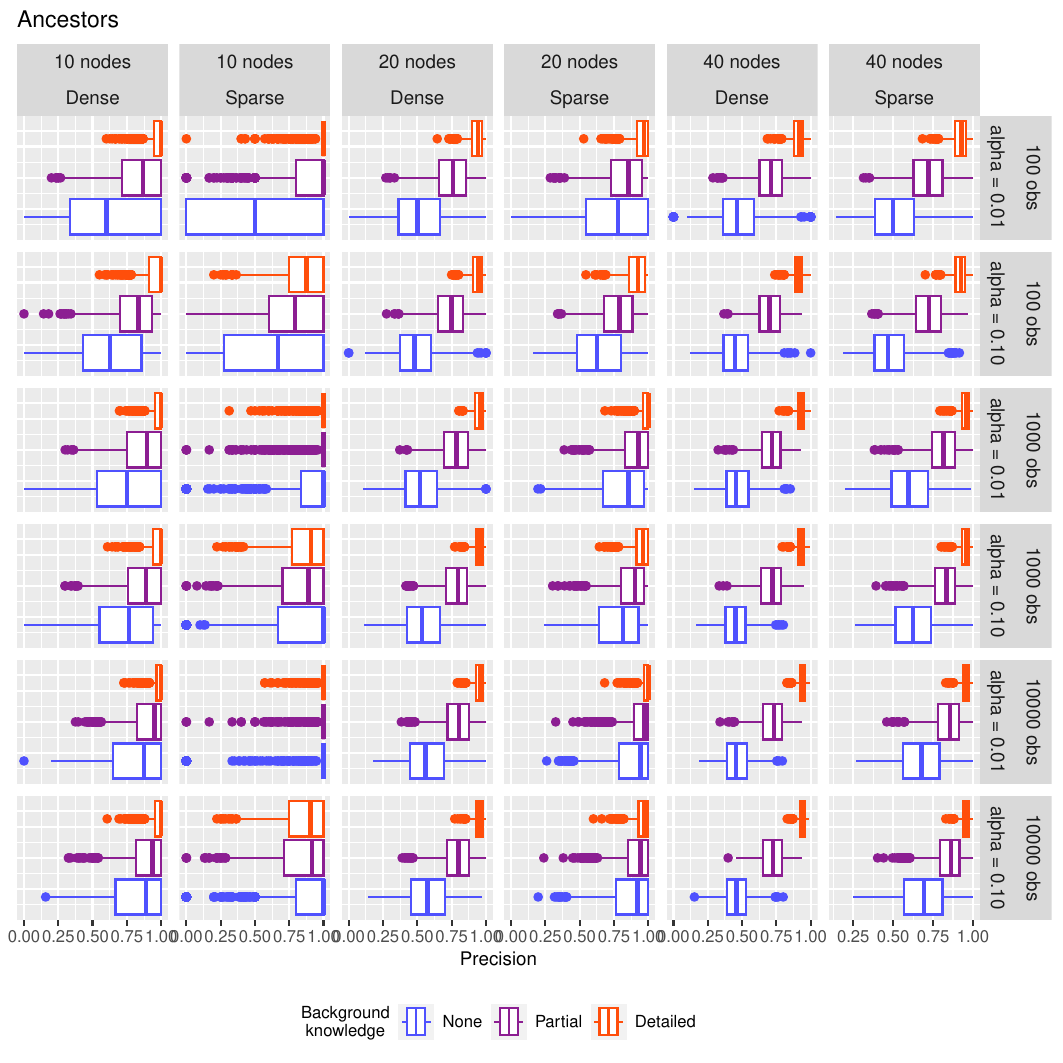}
    \caption{Precision of ancestral relations from 1,000 simulations in dense/sparse settings with 10, 20 or 40 nodes, $\alpha\in \{0.01;0.1\}$ and a sample size of 100, 1,000, or 10,000, using the naive tPC algorithm.}
    \label{fig:sim_an_prec_full_naive}
\end{figure}

\begin{figure}[!htbp]
    \centering
    \includegraphics[scale = 0.975]{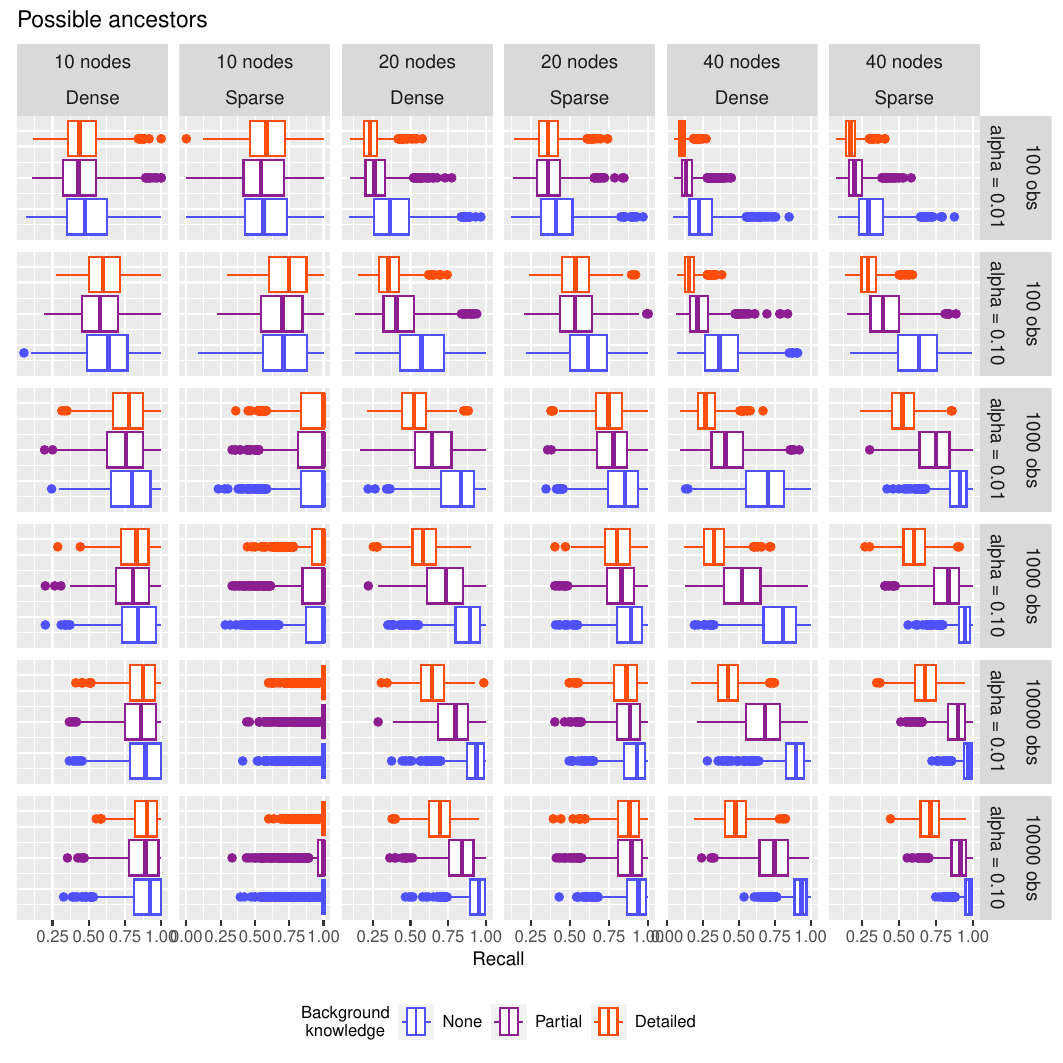}
    \caption{Recall of possible ancestral relations from 1,000 simulations in dense/sparse settings with 10, 20 or 40 nodes, $\alpha\in \{0.01;0.1\}$ and a sample size of 100, 1,000, or 10,000, using the naive tPC algorithm.}
\label{fig:sim_possan_rec_full_naive}
\end{figure}

\begin{figure}[!htbp]
    \centering
    \includegraphics[scale = 0.975]{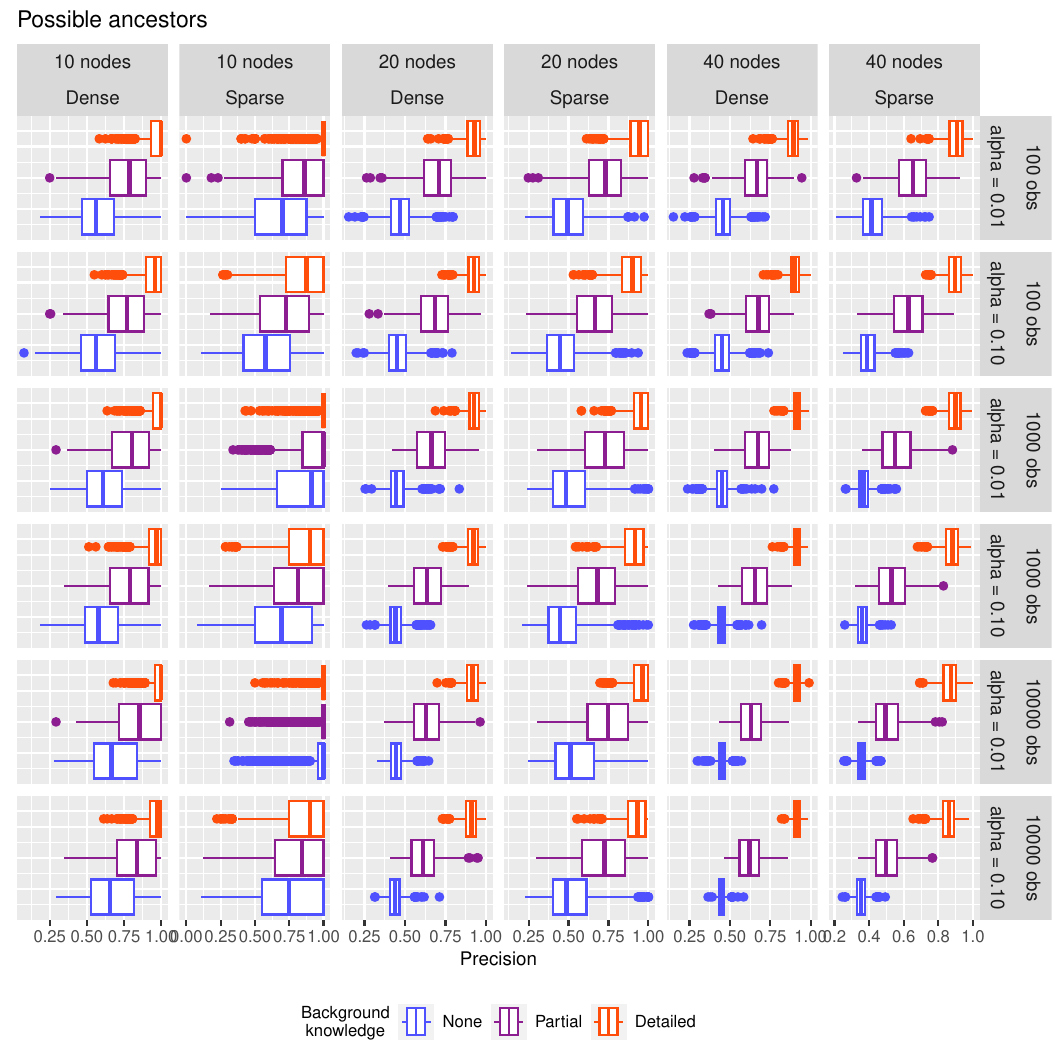}
    \caption{Precision of possible ancestral relations from 1,000 simulations in dense/sparse settings with 10, 20 or 40 nodes, $\alpha\in \{0.01;0.1\}$ and a sample size of 100, 1,000, or 10,000, using the naive tPC algorithm.}
    \label{fig:sim_possan_prec_full_naive}
\end{figure}

\begin{figure}[!htbp]
    \centering
    \includegraphics[scale = 0.975]{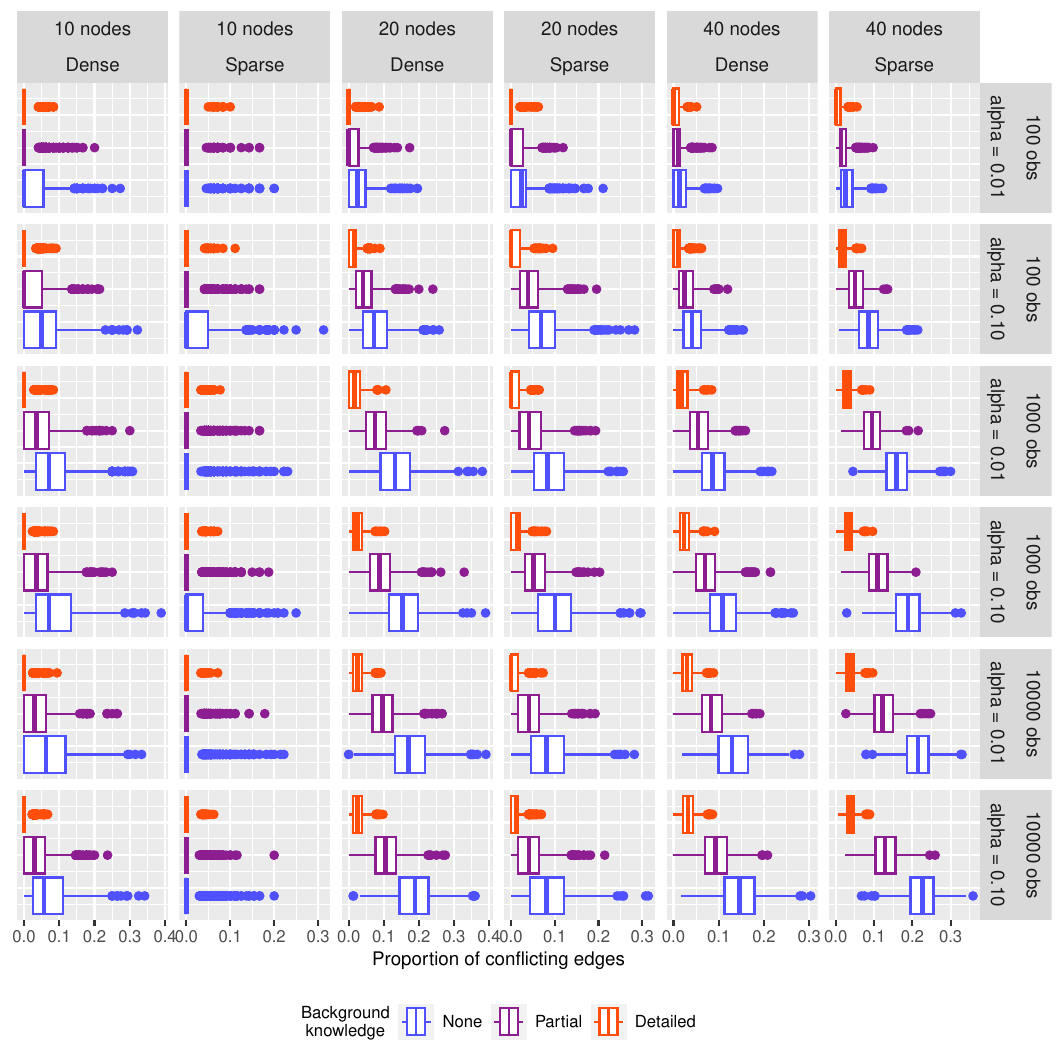}
    \caption{Proportion of bidirected edges from 1,000 simulations in dense/sparse settings with 10, 20 or 40 nodes, $\alpha\in \{0.01;0.1\}$ and a sample size of 100, 1,000, or 10,000, using the naive tPC algorithm.}
    \label{fig:sim_conflicts_full_naive}
\end{figure}

\newpage

\subsection{Number of estimated edges and conditional independence tests}\label{app:ntestnedges}

\begin{figure}[!htbp]
    \centering
    \includegraphics[scale = 0.975]{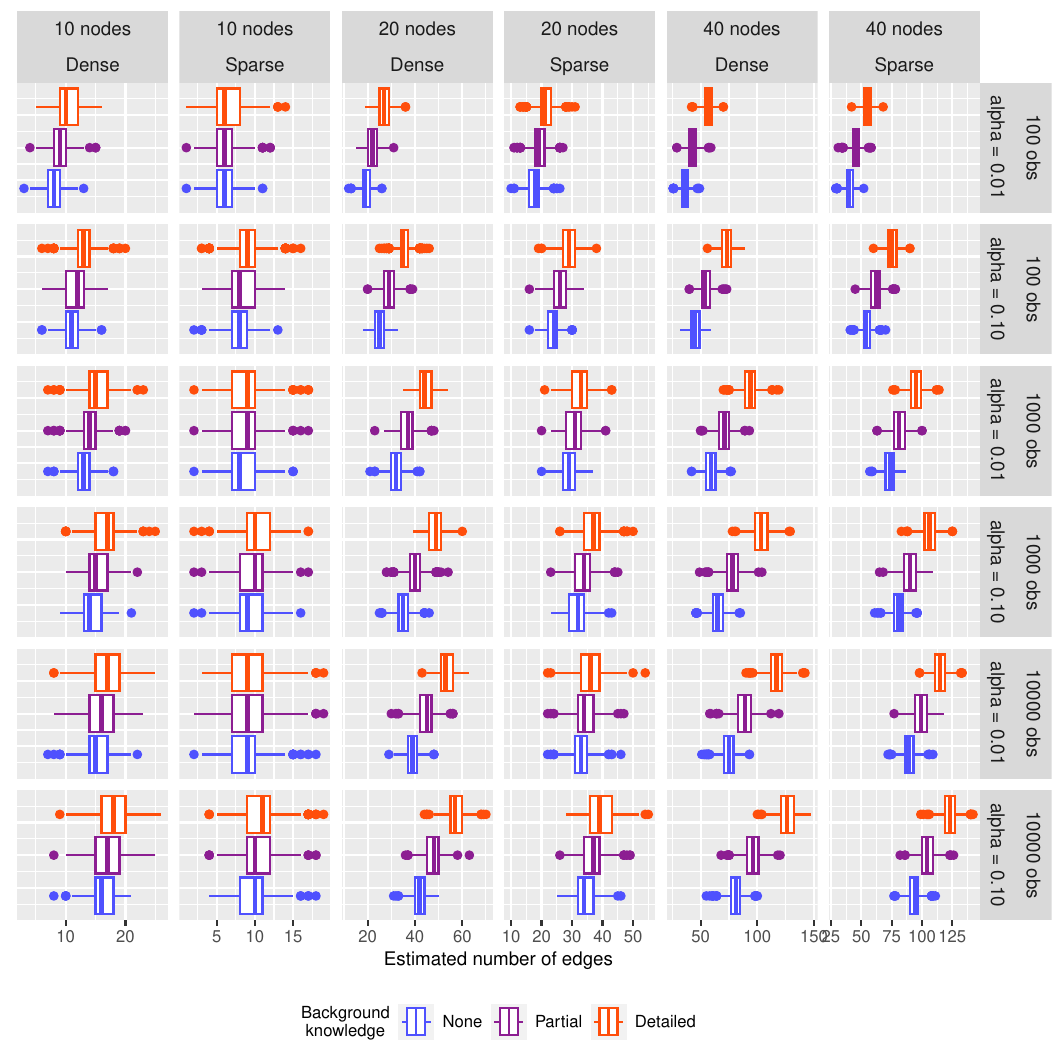}
    \caption{Number of estimated adjacencies ffrom 1,000 simulations in dense/sparse settings with 10, 20 or 40 nodes, $\alpha\in \{0.01;0.1\}$ and a sample size of 100, 1,000, or 10,000, using the tLMPC-stable algorithm}
    \label{fig:sim_nedges}
\end{figure}

\begin{figure}[!htbp]
    \centering
    \includegraphics[scale = 0.975]{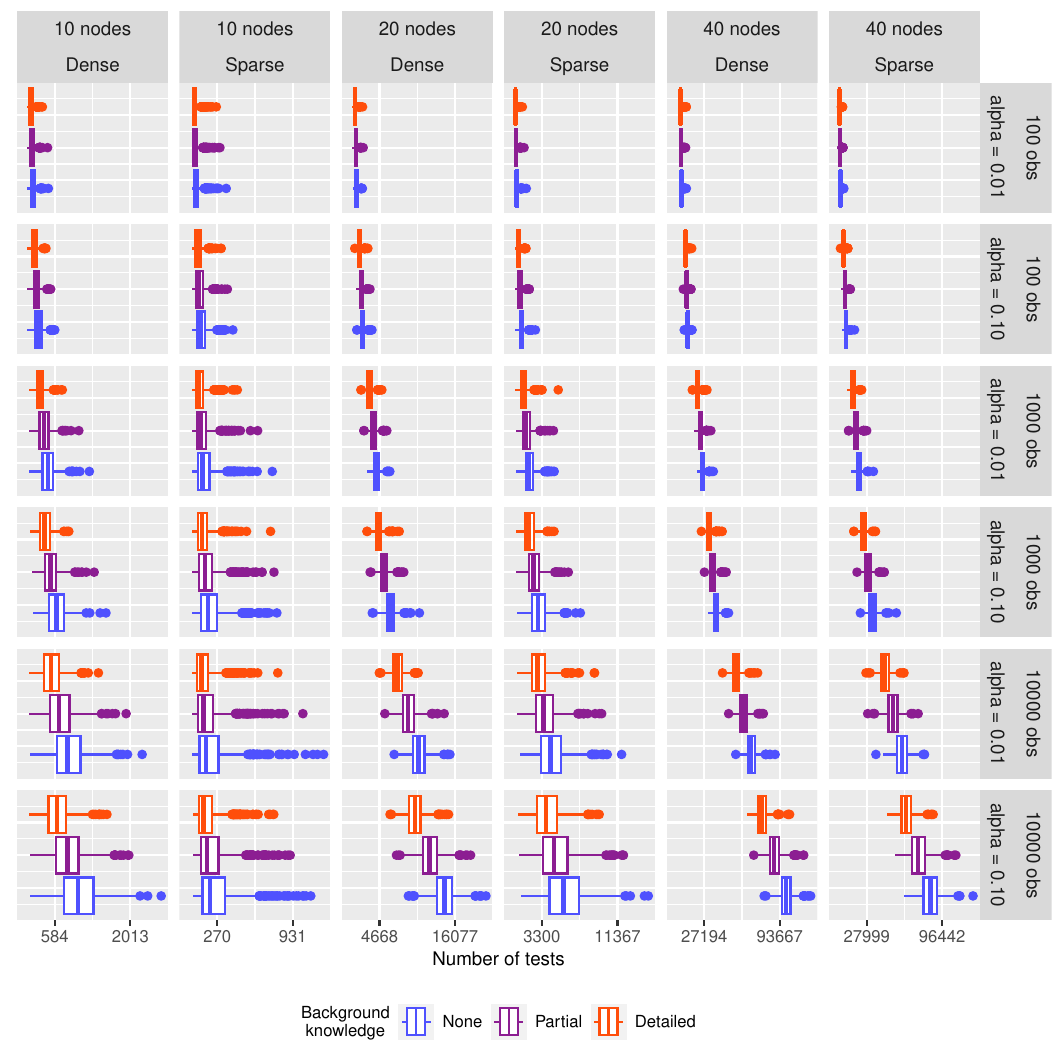}
    \caption{Total number of conditional independence tests from 1,000 simulations in dense/sparse settings with 10, 20 or 40 nodes, $\alpha\in \{0.01;0.1\}$ and a sample size of 100, 1,000, or 10,000, using the tLMPC-stable algorithm.}
    \label{fig:sim_edgetests}
\end{figure}
\restoregeometry
\begin{figure}
    \centering
    \includegraphics{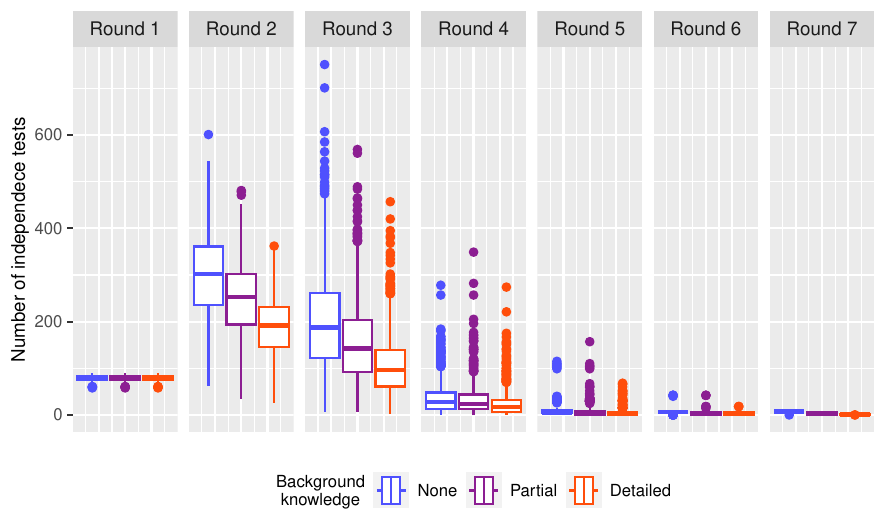}
    \caption{Total number of (conditional) independence tests performed each round of the tLMPC-stable algorithm. 1,000 simulations of dense graphs with 20 nodes and a sample size 1,000. The random variables follow a Gaussian distribution as in the main analysis.}
    \label{fig:n_tests}
\end{figure}

\begin{figure}
    \centering
    \includegraphics{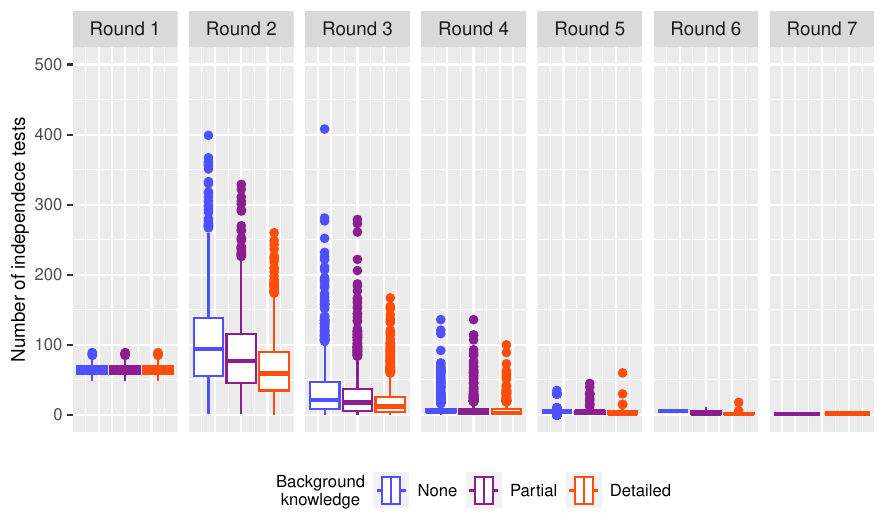}
    \caption{Total number of (conditional) independence tests performed each round of the tLMPC-stable algorithm. 1,000 simulations of sparse graphs with 20 nodes and a sample size 1,000. The random variables follow a Gaussian distribution as in the main analysis.}
    \label{fig:n_tests_sparse}
\end{figure}

\newpage

\newgeometry{top = 3cm, bottom = 3cm}

\section{Data application}\label{app:dataexample}

\subsection{Tiered orderings}\label{app:dataordering}

\begin{table}[!htbp]
    \centering
    \begin{tabular}{| l | c | c | c |}

    \hline
 & \multicolumn{3}{|c|}{Background knowledge} \\ 

  \hline
    
        Variable & None & Partial & Detailed  \\ \hline\hline
        Sex (B) & 1 & 1 & 1\\
        Age  (B) & 1 & 1 & 1\\
        Migrant (B) & 1 & 1 & 1\\
        Income (B) & 1 & 1 & 1\\
        ISCED (B) & 1 & 1 & 1\\
        Mother's age at birth (B) & 1 & 1 & 2\\
        Weeks pregnant (B) & 1 & 1 & 3\\
        Birth weight (B) & 1 & 1 & 3\\
        Breastfeeding (B) & 1 & 1 & 4\\
        Formula milk (B) & 1 & 1 & 4\\
        Eating household diet (B) & 1 & 1 & 4\\
        School (B) & 1 & 1 & 5\\
        Media consumption (B) & 1 & 1 & 6\\
        BMI (B) & 1 & 1 & 6\\
        Mother's BMI (B) & 1 & 1 & 6\\
        Physical activity (B) & 1 & 1 & 6\\
        Sleep (B) & 1 & 1 & 6\\
        Well-being (B) & 1 & 1 & 6\\
        Healthy eating (B) & 1 & 1 & 6\\
        Insulin resistance (B) & 1 & 1 & 6\\ 
        Media consumption (FU1) & 1 & 2 & 7\\
        BMI (FU1) & 1 & 2 & 7\\
        Mother's BMI (FU1) & 1 & 2 & 7\\
        Physical activity (FU1) & 1 & 2 & 7\\
        Sleep (FU1) & 1 & 2 & 7\\
        Well-being (FU1) & 1 & 2 & 7\\
        Healthy eating (FU1) & 1 & 2 & 7\\
        Insulin resistance (FU1) & 1 & 2 & 7\\ 
        Media consumption (FU2) & 1 & 3 & 8\\
        BMI (FU2) & 1 & 3 & 8\\
        Mother's BMI (FU2) & 1 & 3 & 8\\
        Physical activity (FU2) & 1 & 3 & 8\\
        Pubertal status (FU2) & 1 & 3 & 8 \\ 
        Sleep (FU2) & 1 & 3 & 8\\
        Well-being (FU2) & 1 & 3 & 8\\
        Healthy eating (FU2) & 1 & 3 & 8\\
        Insulin resistance (FU2) & 1 & 3 & 8 \\ \hline
    \end{tabular}
    \caption{Overview of the tiers used for estimating MPDAGs. The tiered ordering $\tau_1$ assigns every variable to the same tier, i.e. it does not contain any background knowledge and produces a CPDAG. The tiered ordering $\tau_2$ assigns the nodes to a tier according to the time ordering; i.e., it contains partial, but not detailed, background knowledge, and it produces an MPDAG. The ordering $\tau_3$ is the detailed ordering of the nodes as described in Figure \ref{tab:dataexmaple} and contain all available background knowledge; i.e. it produces an MPDAG.}
    \label{tab:dataexmaple}
\end{table}
\restoregeometry

\clearpage

\subsection{Estimated graphs}\label{app:dataresults}

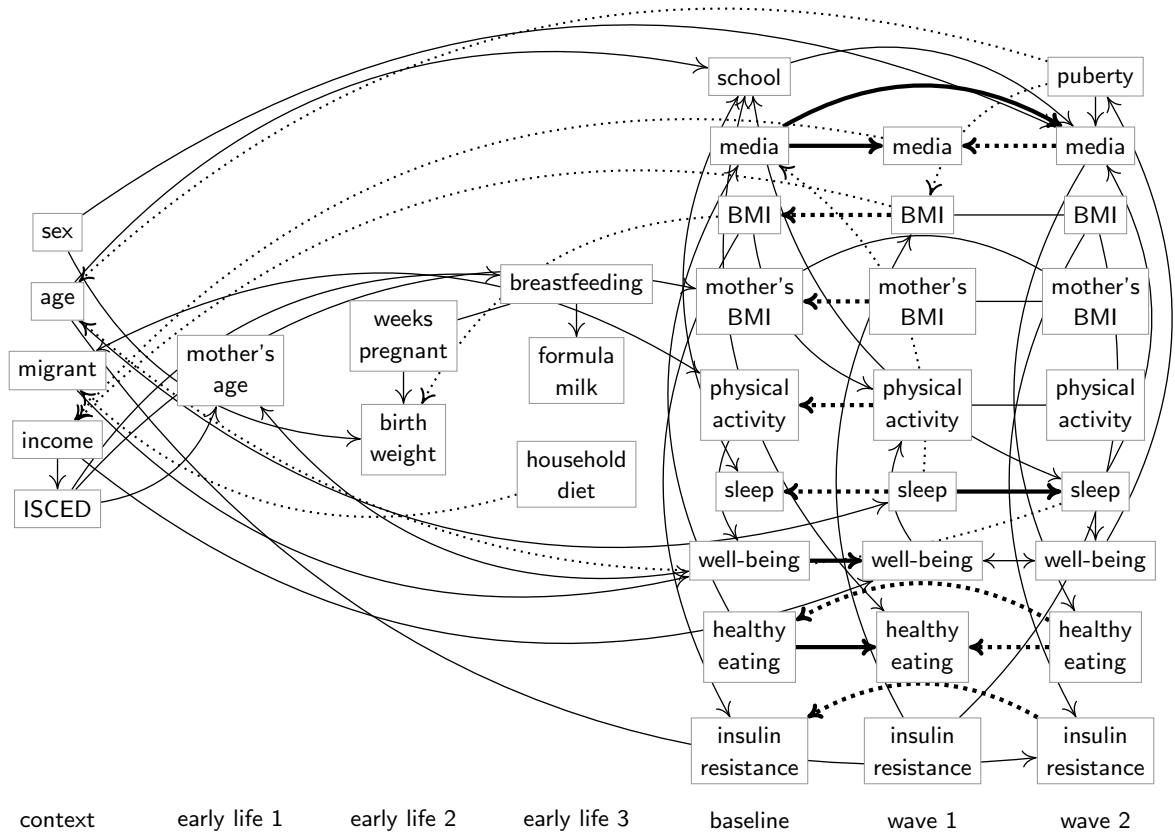
\begin{figure}[!htbp]
\begin{tikzpicture}[scale=0.92, squarednode/.style={rectangle, draw=gray!80, fill=white, thin, minimum size=5mm}]

\node[squarednode] (sex) at (0, 2.75) {\small \textsf{sex}};
\node[squarednode] (age) at (0, 1.75) {\small \textsf{age}};
\node[squarednode] (migrant) at (0, 0.75) {\small  \textsf{migrant}};
\node[squarednode] (income) at (0, -.25) {\small \textsf{income}};
\node[squarednode] (isced) at (0,-1.25) {\small \textsf{ISCED}};

\node[squarednode, align=center] (bage) at (2.5,0.75) {\small \textsf{mother's} \\ \small \textsf{age}};

\node[squarednode, align=center] (pregweek) at (5, 1.25) {\small \textsf{weeks} \\ \small \textsf{pregnant}};
\node[squarednode, align=center] (bweight) at (5, -.25) {\small \textsf{birth} \\ \small \textsf{weight}};

\node[squarednode] (bf) at (7.5, 2) {\small \textsf{breastfeeding}};
\node[squarednode, align=center] (formula) at (7.5, 0.75) {\small \textsf{formula}\\ \small \textsf{milk}};
\node[squarednode, align=center] (hdiet) at (7.5, -.75) {\small \textsf{household}\\ \small \textsf{diet}};

\node[squarednode] (school) at (10,5) {\small \textsf{school}};
\node[squarednode] (media0) at (10,4) {\small \textsf{media}};
\node[squarednode] (bmi0) at (10, 3) {\small \textsf{BMI}};
\node[squarednode, align=center] (mbmi0) at (10, 1.75) {\small \textsf{mother's}\\ \small \textsf{BMI}};
\node[squarednode, align=center] (pa0) at (10, 0.25) {\small \textsf{physical}\\ \small \textsf{activity}};
\node[squarednode] (sleep0) at (10, -1) {\small \textsf{sleep}};
\node[squarednode] (wb0) at (10, -2) {\small \textsf{well-being}};
\node[squarednode, align=center] (yhei0) at (10, -3.25) {\small \textsf{healthy}\\ \small \textsf{eating}};
\node[squarednode, align=center] (homa0) at (10, -4.75) {\small \textsf{insulin}\\\small \textsf{resistance}};

\node[squarednode] (media1) at (12.5,4) {\small \textsf{media}};
\node[squarednode] (bmi1) at (12.5, 3) {\small \textsf{BMI}};
\node[squarednode, align=center] (mbmi1) at (12.5, 1.75) {\small \textsf{mother's}\\ \small \textsf{BMI}};
\node[squarednode, align=center] (pa1) at (12.5, 0.25) {\small \textsf{physical}\\ \small \textsf{activity}};
\node[squarednode] (sleep1) at (12.5, -1) {\small \textsf{sleep}};
\node[squarednode] (wb1) at (12.5, -2) {\small \textsf{well-being}};
\node[squarednode, align=center] (yhei1) at (12.5, -3.25) {\small \textsf{healthy}\\ \small \textsf{eating}};
\node[squarednode, align=center] (homa1) at (12.5, -4.75) {\small \textsf{insulin}\\\small \textsf{resistance}};

\node[squarednode] (pub) at (15, 5) {\small \textsf{puberty}};
\node[squarednode] (media2) at (15,4) {\small \textsf{media}};
\node[squarednode] (bmi2) at (15, 3) {\small \textsf{BMI}};
\node[squarednode, align=center] (mbmi2) at (15, 1.75) {\small \textsf{mother's}\\ \small \textsf{BMI}};
\node[squarednode, align=center] (pa2) at (15, 0.25) {\small \textsf{physical}\\ \small \textsf{activity}};
\node[squarednode] (sleep2) at (15, -1) {\small \textsf{sleep}};
\node[squarednode] (wb2) at (15, -2) {\small \textsf{well-being}};
\node[squarednode, align=center] (yhei2) at (15, -3.25) {\small \textsf{healthy}\\ \small \textsf{eating}};
\node[squarednode, align=center] (homa2) at (15, -4.75) {\small \textsf{insulin}\\\small \textsf{resistance}};

\tikzset{undir/.style = {-, line width = 0.5pt}}
\tikzset{dir/.style = {->, -{To[length=5.5, width=6.5]}, line width = 0.5pt}}
\tikzset{bidir/.style = {{To[length=4.5, width=6]}-{To[length=4.5, width=6]}, line width = 0.5pt}}

\node (t1) at (0, -5.75) {\small \textsf{context}};
\node (t2) at (2.5, -5.75) {\small \textsf{early life 1}};
\node (t3) at (5, -5.75) {\small \textsf{early life 2}};
\node (t4) at (7.5, -5.75) {\small \textsf{early life 3}};
\node (t5) at (10, -5.75) {\small \textsf{baseline}};
\node (t6) at (12.5, -5.75) {\small \textsf{wave 1}};
\node (t7) at (15, -5.75) {\small \textsf{wave 2}};

 \begin{pgfonlayer}{bg}
\draw[undir]
(pregweek) edge  (bf)
(mbmi0) edge [bend left] (mbmi2)
(mbmi1) edge (mbmi2)
(bmi1) edge (bmi2)
(pa1) edge (pa2)
(homa1) edge [bend right] (bmi2) 
;

\draw[bidir]
(migrant) edge [bend left] (pa0)
(bage) edge [bend right] (wb0)
(school) edge [bend right] (sleep0)
(school) edge [bend right] (yhei1)
(school) edge [bend right] (sleep2)
(wb1) edge (wb2)
;

\draw[dir]
(sex) edge [bend right] (bweight)
(sex) edge [bend left] (media2)
(age) edge [bend left] (school)
(age) edge [bend right] (sleep1)
(age) edge [bend right] (homa2)
(migrant) edge [bend right] (wb0)
(income) edge (isced)
(income) edge [bend right] (wb1)
(isced) edge [bend right] (bage)
(isced) edge [bend left] (bf)
(isced) edge [bend left] (mbmi0)
(pregweek) edge (bweight)
(bf) edge (formula)
(hdiet) edge [bend left, dotted, thick] (migrant)
(school) edge [bend left] (media2)
(media0) edge [ultra thick] (media1)
(media0) edge [bend left, ultra thick] (media2)
(bmi0) edge [bend right, dotted, thick] (bweight)
(bmi0) edge [bend right] (homa0)
(bmi0) edge [bend right] (pa1)
(pa0) edge [bend right] (wb0)
(wb0) edge [ultra thick] (wb1)
(yhei0) edge [bend left] (media0)
(yhei0) edge [ultra thick] (yhei1)
(media1) edge [bend right, thick, dotted] (income)
(bmi1) edge [bend right, thick, dotted] (income)
(bmi1) edge [ultra thick, dotted] (bmi0)
(mbmi1) edge [ultra thick, dotted] (mbmi0)
(pa1) edge [ultra thick, dotted] (pa0)
(sleep1) edge [thick, dotted, bend right] (media0)
(sleep1) edge [ultra thick, dotted] (sleep0)
(sleep1) edge [ultra thick] (sleep2)
(wb1) edge [bend left] (pa1)
(homa1) edge [bend left] (bmi1)
(media2) edge [ultra thick, dotted] (media1)
(media2) edge [bend right] (yhei2)
(bmi2) edge [bend right] (homa2)
(pub) edge [bend right, thick, dotted] (age)
(pub) edge [bend right, thick, dotted] (bmi1)
(pub) edge (media2)
(sleep2) edge [bend left, thick, dotted] (age)
(sleep2) edge [bend right] (media2)
(sleep2) edge (wb2)
(wb2) edge [bend right] (pub)
(yhei2) edge [bend right, ultra thick, dotted] (yhei0)
(yhei2) edge [ultra thick, dotted] (yhei1)
(homa2) edge [bend right, ultra thick, dotted] (homa0)
;
\end{pgfonlayer}
\end{tikzpicture}
\caption{Graph estimated without background knowledge. Bold edges: These are directed edges between repeated measurements, which we expect to find. Dotted edges: These edges are implausible since they contradict the known time ordering.}
\label{fig:est_cpdag}
\end{figure}

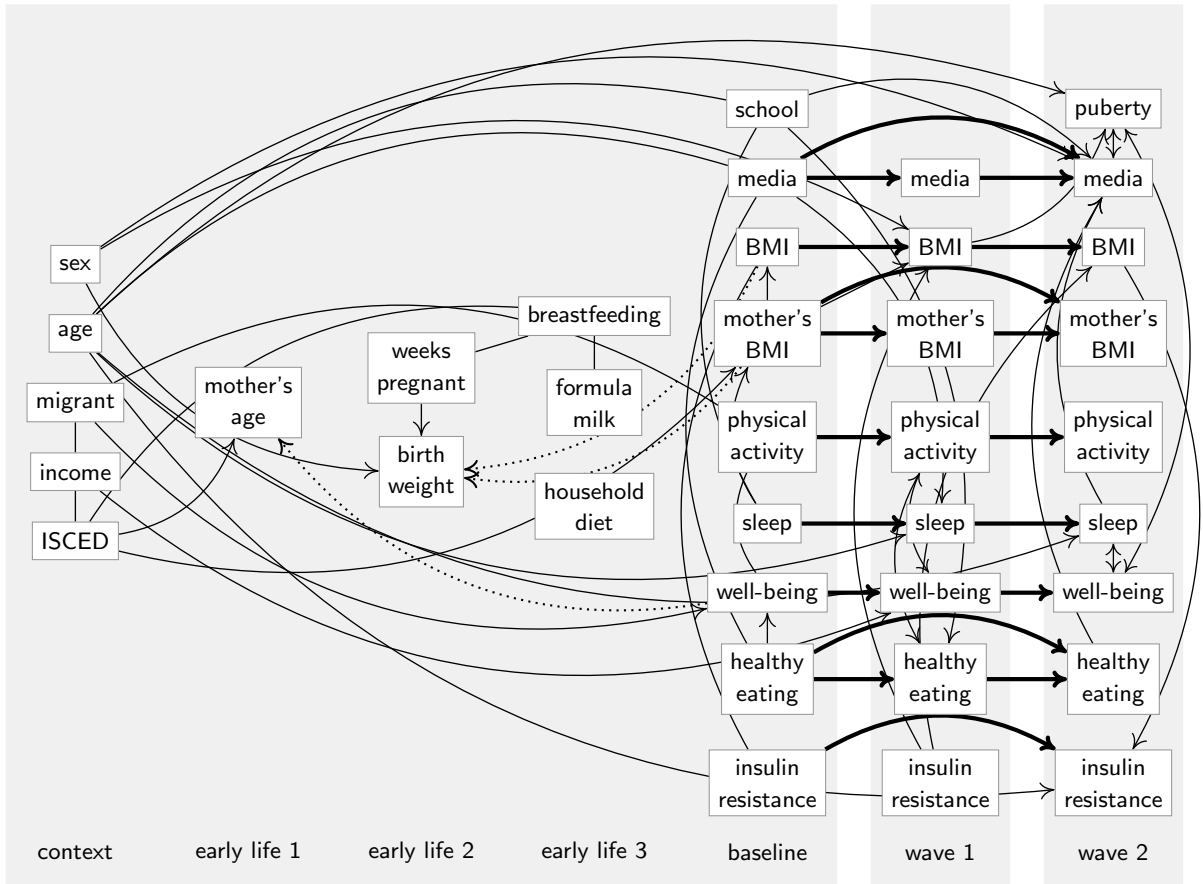
\begin{figure}[!htbp]
\begin{tikzpicture}[scale=0.92, squarednode/.style={rectangle, draw=gray!80, fill=white, thin, minimum size=5mm}]

\node[squarednode] (sex) at (0, 2.75) {\small \textsf{sex}};
\node[squarednode] (age) at (0, 1.75) {\small \textsf{age}};
\node[squarednode] (migrant) at (0, 0.75) {\small  \textsf{migrant}};
\node[squarednode] (income) at (0, -.25) {\small \textsf{income}};
\node[squarednode] (isced) at (0,-1.25) {\small \textsf{ISCED}};

\node[squarednode, align=center] (bage) at (2.5,0.75) {\small \textsf{mother's} \\ \small \textsf{age}};

\node[squarednode, align=center] (pregweek) at (5, 1.25) {\small \textsf{weeks} \\ \small \textsf{pregnant}};
\node[squarednode, align=center] (bweight) at (5, -.25) {\small \textsf{birth} \\ \small \textsf{weight}};

\node[squarednode] (bf) at (7.5, 2) {\small \textsf{breastfeeding}};
\node[squarednode, align=center] (formula) at (7.5, 0.75) {\small \textsf{formula}\\ \small \textsf{milk}};
\node[squarednode, align=center] (hdiet) at (7.5, -.75) {\small \textsf{household}\\ \small \textsf{diet}};

\node[squarednode] (school) at (10,5) {\small \textsf{school}};
\node[squarednode] (media0) at (10,4) {\small \textsf{media}};
\node[squarednode] (bmi0) at (10, 3) {\small \textsf{BMI}};
\node[squarednode, align=center] (mbmi0) at (10, 1.75) {\small \textsf{mother's}\\ \small \textsf{BMI}};
\node[squarednode, align=center] (pa0) at (10, 0.25) {\small \textsf{physical}\\ \small \textsf{activity}};
\node[squarednode] (sleep0) at (10, -1) {\small \textsf{sleep}};
\node[squarednode] (wb0) at (10, -2) {\small \textsf{well-being}};
\node[squarednode, align=center] (yhei0) at (10, -3.25) {\small \textsf{healthy}\\ \small \textsf{eating}};
\node[squarednode, align=center] (homa0) at (10, -4.75) {\small \textsf{insulin}\\\small \textsf{resistance}};

\node[squarednode] (media1) at (12.5,4) {\small \textsf{media}};
\node[squarednode] (bmi1) at (12.5, 3) {\small \textsf{BMI}};
\node[squarednode, align=center] (mbmi1) at (12.5, 1.75) {\small \textsf{mother's}\\ \small \textsf{BMI}};
\node[squarednode, align=center] (pa1) at (12.5, 0.25) {\small \textsf{physical}\\ \small \textsf{activity}};
\node[squarednode] (sleep1) at (12.5, -1) {\small \textsf{sleep}};
\node[squarednode] (wb1) at (12.5, -2) {\small \textsf{well-being}};
\node[squarednode, align=center] (yhei1) at (12.5, -3.25) {\small \textsf{healthy}\\ \small \textsf{eating}};
\node[squarednode, align=center] (homa1) at (12.5, -4.75) {\small \textsf{insulin}\\\small \textsf{resistance}};

\node[squarednode] (pub) at (15, 5) {\small \textsf{puberty}};
\node[squarednode] (media2) at (15,4) {\small \textsf{media}};
\node[squarednode] (bmi2) at (15, 3) {\small \textsf{BMI}};
\node[squarednode, align=center] (mbmi2) at (15, 1.75) {\small \textsf{mother's}\\ \small \textsf{BMI}};
\node[squarednode, align=center] (pa2) at (15, 0.25) {\small \textsf{physical}\\ \small \textsf{activity}};
\node[squarednode] (sleep2) at (15, -1) {\small \textsf{sleep}};
\node[squarednode] (wb2) at (15, -2) {\small \textsf{well-being}};
\node[squarednode, align=center] (yhei2) at (15, -3.25) {\small \textsf{healthy}\\ \small \textsf{eating}};
\node[squarednode, align=center] (homa2) at (15, -4.75) {\small \textsf{insulin}\\\small \textsf{resistance}};

\tikzset{undir/.style = {-, line width = 0.5pt}}
\tikzset{dir/.style = {->, -{To[length=5.5, width=6.5]}, line width = 0.5pt}}
\tikzset{bidir/.style = {{To[length=4.5, width=6]}-{To[length=4.5, width=6]}, line width = 0.5pt}}

\node (t1) at (0, -5.75) {\small \textsf{context}};
\node (t2) at (2.5, -5.75) {\small \textsf{early life 1}};
\node (t3) at (5, -5.75) {\small \textsf{early life 2}};
\node (t4) at (7.5, -5.75) {\small \textsf{early life 3}};
\node (t5) at (10, -5.75) {\small \textsf{baseline}};
\node (t6) at (12.5, -5.75) {\small \textsf{wave 1}};
\node (t7) at (15, -5.75) {\small \textsf{wave 2}};

 \begin{pgfonlayer}{bg}

\filldraw[silvergray] (-1, 6.5) rectangle +(12,-12.75);
\filldraw[silvergray] (11.5, 6.5) rectangle +(2,-12.75);
\filldraw[silvergray] (14, 6.5) rectangle +(2,-12.75);
 
\draw[undir]
(income) edge (migrant)
(isced) edge (income)
(bf) edge [bend right] (isced)
(bf) edge (pregweek)
(formula) edge (bf)
(school) edge [bend right] (age)
(media0) edge [bend right] (age)
(pa0) edge [bend right] (migrant)
(sleep0) edge [bend left] (school)
(wb0) edge [bend left] (pa0)
(yhei0) edge [bend left] (media0)
(homa0) edge [bend left] (bmi0)
;

\draw[bidir]
(wb1) edge [bend left] (pa1)
(pub) edge (media2)
(wb2) edge [bend right] (pub)
(wb2) edge (sleep2)
;

\draw[dir]
(sex) edge [bend right] (bweight)
(sex) edge [bend left] (bmi1)
(sex) edge [bend left] (media2)
(age) edge [bend right] (sleep1)
(age) edge [bend left] (pub)
(age) edge [bend right] (sleep2)
(age) edge [bend right] (homa2)
(migrant) edge [bend right] (wb0)
(income) edge [bend right] (wb1)
(isced) edge [bend right] (bage)
(isced) edge [bend right] (mbmi0)
(pregweek) edge (bweight)
(school) edge [bend left] (yhei1)
(school) edge [bend left] (media2)
(media0) edge [ultra thick] (media1)
(media0) edge [bend left] (sleep1)
(media0) edge [ultra thick, bend left] (media2)
(bmi0) edge [thick, dotted, bend left] (bweight)
(bmi0) edge [ultra thick] (bmi1)
(mbmi0) edge [thick, dotted, bend left] (bweight)
(mbmi0) edge (bmi0)
(mbmi0) edge (bmi1)
(mbmi0) edge [ultra thick] (mbmi1)
(mbmi0) edge [ultra thick, bend left] (mbmi2)
(pa0) edge [ultra thick] (pa1)
(sleep0) edge [bend left] (mbmi0)
(sleep0) edge [ultra thick] (sleep1)
(wb0) edge [bend left, dotted, thick]  (bage)
(wb0) edge [ultra thick] (wb1)
(yhei0) edge (wb0)
(yhei0) edge [ultra thick] (yhei1)
(yhei0) edge [ultra thick, bend left] (yhei2)
(homa0) edge [ultra thick, bend left] (homa2)
(media1) edge [ultra thick] (media2)
(bmi1) edge [ultra thick] (bmi2)
(bmi1) edge [bend right] (pub)
(mbmi1) edge [ultra thick] (mbmi2)
(pa1) edge [bend right] (yhei1)
(pa1) edge [ultra thick] (pa2)
(sleep1) edge [ultra thick] (sleep2)
(wb1) edge [ultra thick] (wb2)
(yhei1) edge [ultra thick] (yhei2)
(homa1) edge [bend left] (bmi1)
(homa1) edge [bend left] (bmi2)
(bmi2) edge [bend left] (homa2)
(sleep2) edge [bend left] (media2)
(yhei2) edge [bend left] (media2)
;
\end{pgfonlayer}
\end{tikzpicture}
\caption{Graph estimated with partial background knowledge. Bold edges: These are directed edges between repeated measurements, which we expect to find.
Dotted edges: These edges are implausible since they contradict the known time ordering.}
\label{fig:est_partial_mpdag}
\end{figure}

\newgeometry{left=2cm, right = 2cm, top = 2cm, bottom = 2cm}
\newpage

\subsection{Estimated adjacency matrices}\label{app:dataamat}

\begin{figure}[!htbp]
    \centering
    \includegraphics{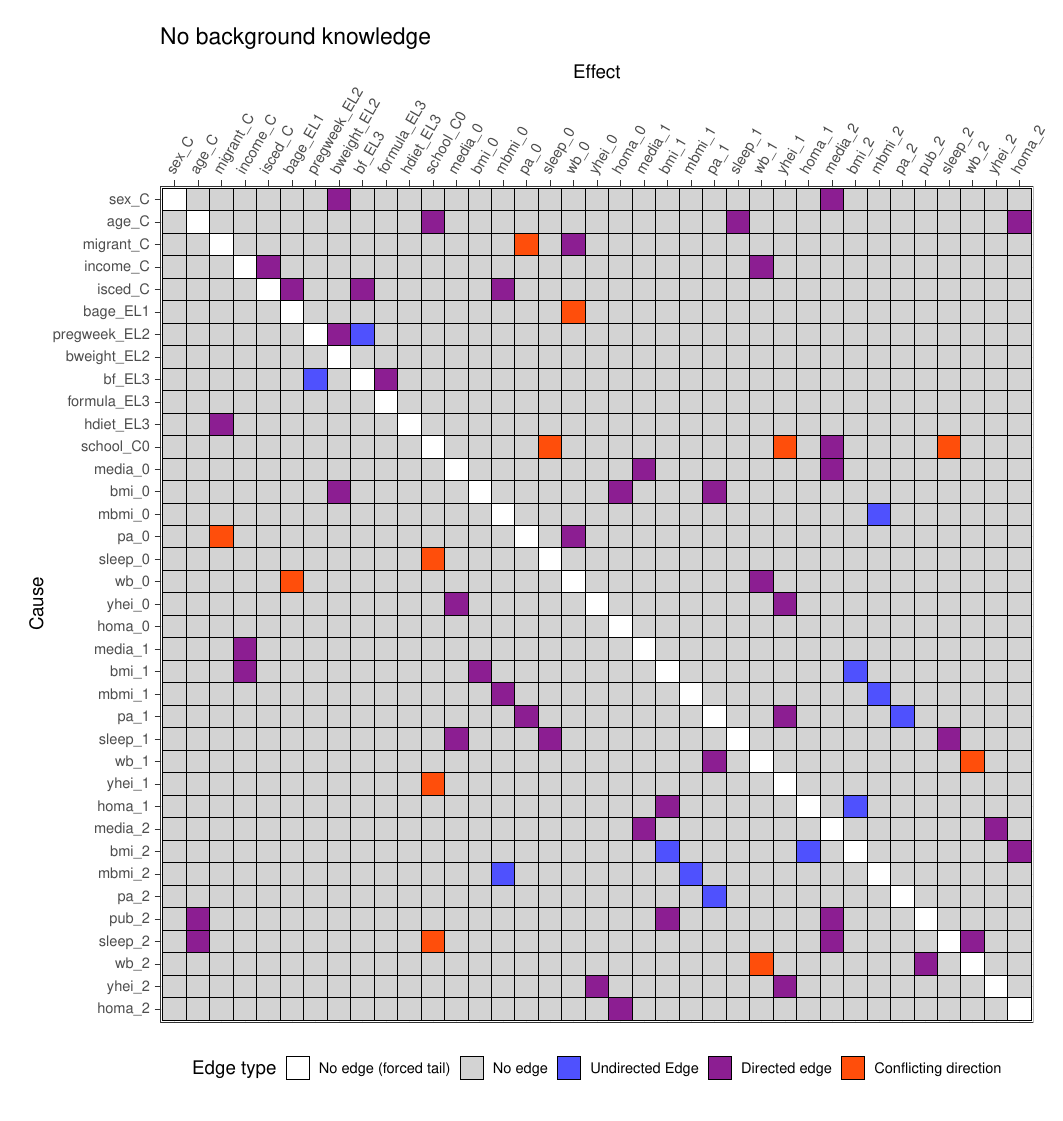}
    \caption{Adjacency matrix of estimated graph using ordering $\tau_1$ (no background knowledge). Edge gaps are only enforced to avoid self-loops, any other missing edges is due to conditional independence. Undirected edges means that there is no evidence based on the estimated independence model that the causation is one way or the other. Conflicting edges means that there is evidence of the edge going each way, i.e. the data contains conflicting evidence. Directed edges means that the data contains unambiguous evidence of the causal direction.}
    \label{fig:est.graph.notiers}
\end{figure}

\begin{figure}[!htbp]
    \centering
    \includegraphics{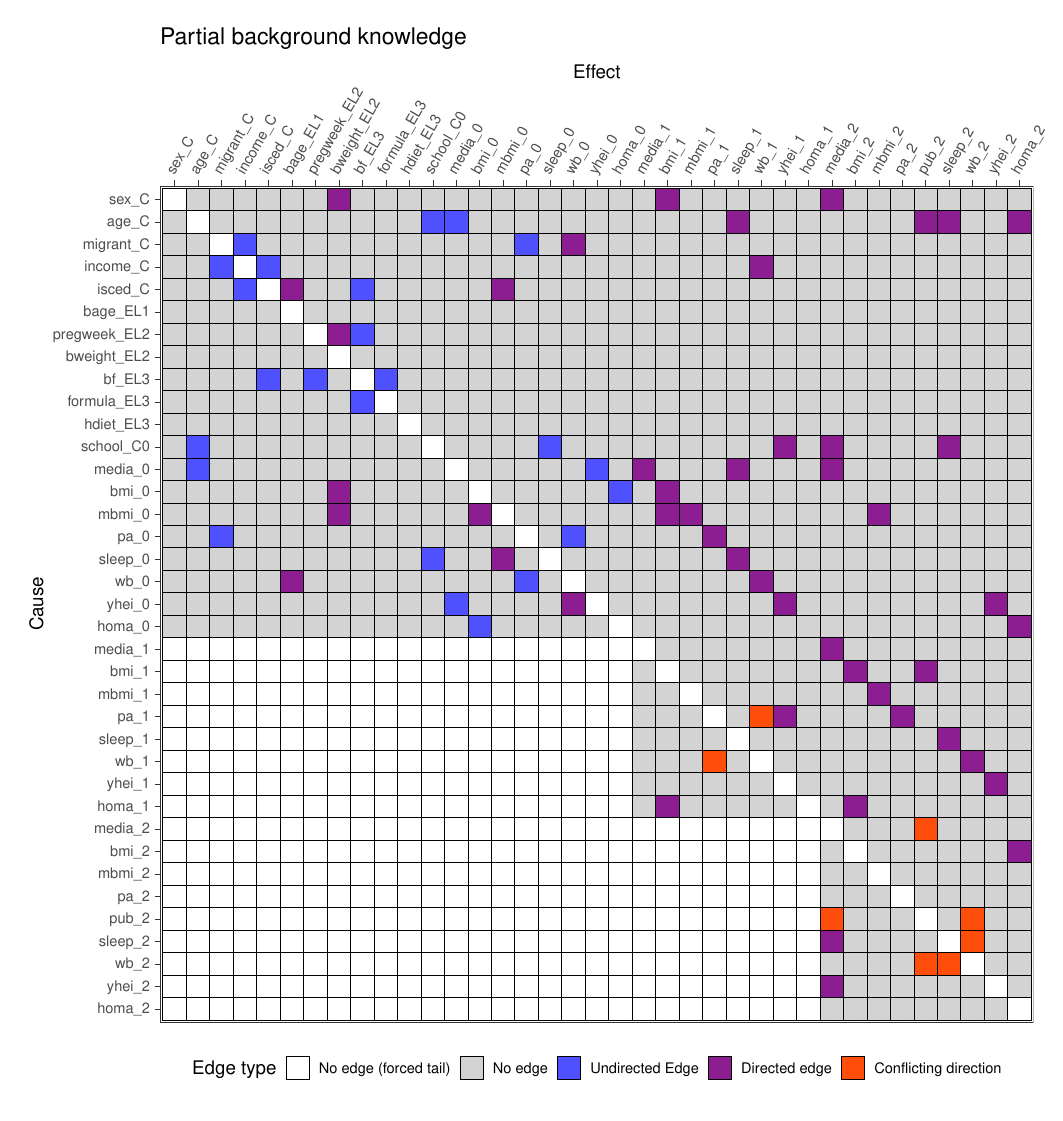}
    \caption{Adjacency matrix of estimated graph using ordering $\tau_2$ (partial background knowledge). Edge gaps are enforced either to avoid self-loops, or to avoid edges directed contrary to the flow of time. Undirected edges means that there is no evidence based on the estimated independence model that the causation is one way or the other. Conflicting edges means that there is evidence of the edge going each way, i.e. the data contains conflicting evidence. Directed edges means that the data contains unambiguous evidence of the causal direction.}
    \label{fig:est.graph.somtiers}
\end{figure}

\begin{figure}[!htbp]
    \centering
    \includegraphics{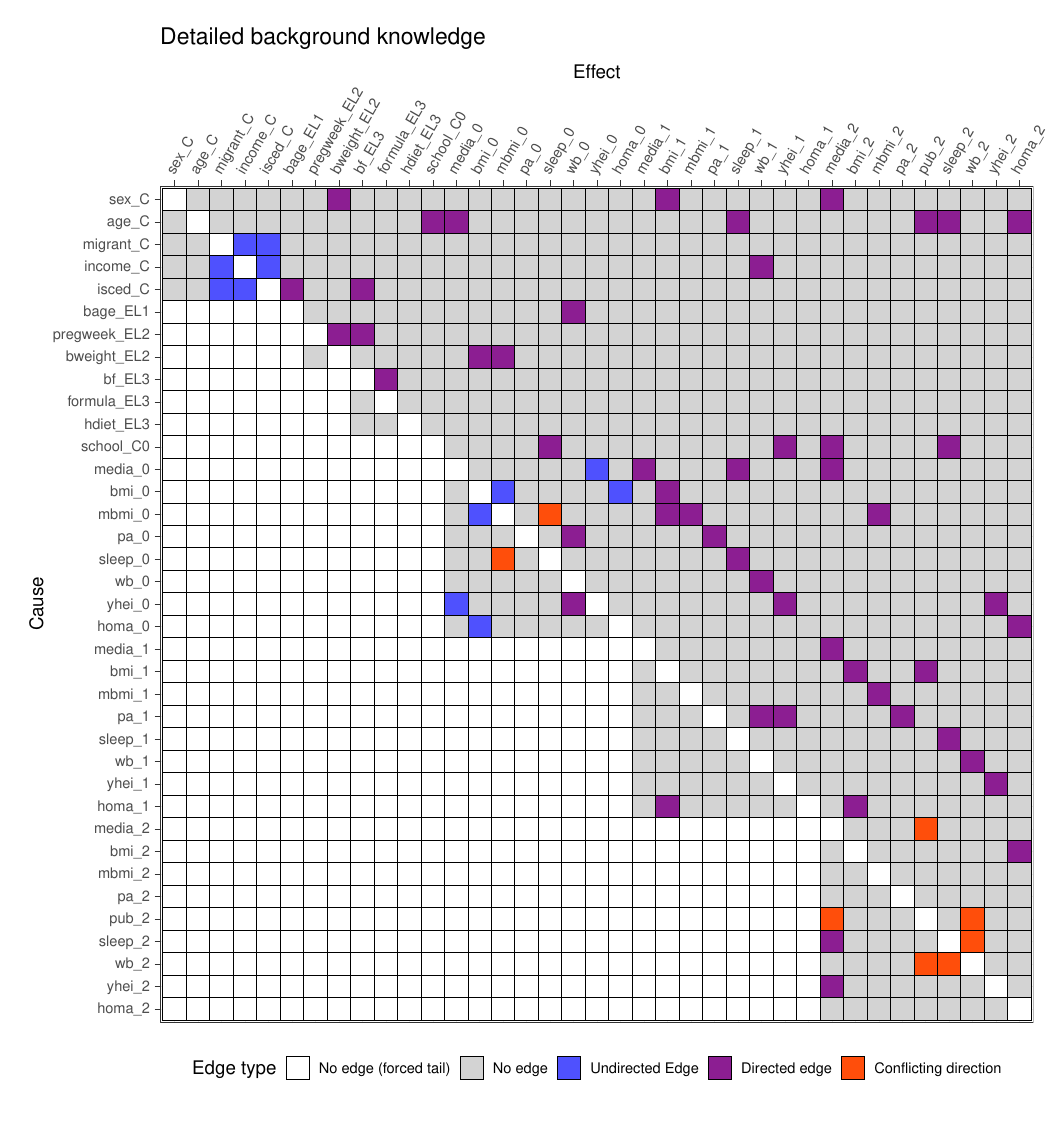}
    \caption{Adjacency matrix of estimated graph using ordering $\tau_3$ (detailed background knowledge). Edge gaps are enforced either to avoid self-loops, or to avoid edges directed contrary to the flow of time. Undirected edges means that there is no evidence based on the estimated independence model that the causation is one way or the other. Conflicting edges means that there is evidence of the edge going each way, i.e. the data contains conflicting evidence. Directed edges means that the data contains unambiguous evidence of the causal direction.}
    \label{fig:est.graph.altiers}
\end{figure}
\restoregeometry

\newpage

\section{Proofs}\label{app:proofs}

\subsection{Proof of Proposition \ref{prop:soundandcomplete}}

\begin{proof}
We will consider the phases I-IV (Algorithms 2, 3, 4 and 5) of the tLMPC-stable algorithm individually. 

(I-II) Let $\D$ be the true underlying DAG, then two nodes $V_i$ and $V_j$ are non-adjacent in $\C'$ if and only if they are non-adjacent in $\D$. The proof is similar to the proof of Theorem 5.1 in Spirtes et al. (2000).

``Only if'': Since the tPC algorithm starts with a fully connected graph, two nodes $V_i$ and $V_j$ are non-adjacent in $\C'$ only if $X_{V_i}$ and $X_{V_j}$ are conditionally  independent given some $\mathbf{X}_\mathbf{S}$ with $\mathbf{S}\subseteq\V\backslash\{ V_i,V_j\}$. Due to the faithfulness assumption, $V_i$ and $V_j$ must then be d-separated by $\mathbf{S}$ in $\D$.

``If'': Suppose now that $V_i$ and $V_j$ are not adjacent in $\D$, and assume without loss of generality that $V_j\in\nd{\D}{V_i}$. By the local Markov property, $X_{V_i}\indep X_{V_j}\mid \mathbf{X}_{\pa{\D}{V_i}}$. We established above that the tPC algorithm does not erroneously delete edges, so $\pa{\D}{V_i}\subset\adj{\C'}{V_i}$. Further, for every node $P\in\pa{\D}{V_i}$, $\tau(P)\leq \tau(V_i)$. This implies that the candidate conditional independence $X_{V_i}\indep X_{V_j}\mid \mathbf{X}_{\pa{\D}{V_i}}$ will be tested, and $V_i$ and $V_j$ are then not adjacent in $\C'$.

(IIIa): Consider now $\C''$  from Algorithm 3. If the tiered background knowledge were not taken into account, this phase would be identical to the second phase of the LMPC-stable algorithm, which was shown to be sound and complete \cite{colombo2014}. We show that modifications differentiating the tPC algorithm from the LMPC-stable algorithm do not affect which v-structures will be directed. For a triple $(V_i, V_j, V_k)$ that is a candidate v-structure the algorithm looks for sets $\mathbf{S}$ separating $V_i$ and $V_k$, and it then checks whether they contain $V_j$. The tiered background knowledge is included in two ways: (a) only triples with $\tau(V_j)=\max (\tau(V_i),\tau(V_k))$ will be considered and (b) candidate separating sets $\mathbf{S}$ for nodes $V_i$ and $V_k$ are not allowed to include any node $V$ with $\tau(V)>\tau(V_i)$.

(a) First, for any collider $V_j$ of $V_i$ and $V_k$ in $\G$ with $\tau(V_j)>\max (\tau(V_i),\tau(V_k))$, the path $V_i-V_j-V_k$ will not be oriented into $V_i\rightarrow V_j\leftarrow V_k$. However, this will happen in phase IIIb. Second, any node $V_j$ with $\tau(V_j)<\max (\tau(V_i),\tau(V_k))$ will not be considered a potential collider; this is consistent with the tiered background knowledge, which is assumed to be correct.

(b) Even though some candidate separating sets are skipped, at least one of the conditional independencies $X_{V_i}\indep X_{V_k}\mid\mathbf{X}_{\pa{\D}{V_i}}$ and $X_{V_i}\indep X_{V_k}\mid\mathbf{X}_{\pa{\D}{V_k}}$ will be tested. Since no conflicts arise when given oracle information, a single detected conditional independence is enough to correctly identify whether the triple $(V_i,V_j,V_k)$ forms a v-structure.

Phase III: Consider now the partially directed graph $\C'''$ from Algorithm 4. Any edge oriented in this phase is oriented according to the tiered background knowledge, which was assumed to be correct. Since we have established the adjacencies to be correct, these edges will be in $\D$. 

Phase IV: The edges in $\G$ are oriented according to Meek's rules, which were shown to be sound and complete \cite{meek1995}. This means that given a partially oriented graph $\C'''$ the same skeleton and v-structures as the true DAG $\D$, and possible additional oriented edges are also in $\D$, then applying Meek's rules to $\C'''$ will result in the MPDAG representing $\D$ and the background knowledge. 

\end{proof}

\subsection{Proof of Proposition \ref{prop:consistent}}

\begin{proof}
    The tiered background knowledge  does not concern adjacencies, only directions: It simply disallows all edges directed from a later to an earlier tier. Hence, any skeleton will be consistent with tiered background knowledge, and our main concern is whether any v-structures could be oriented against the tiered ordering. However, any candidate collider has to belong to the latest tier of the two potential parent nodes. Hence, no edge will be oriented from a later to an earlier tier in this procedure.
\end{proof}

\subsection{Proof of Proposition \ref{prop:stable}}

\begin{proof}
The tLMPC-stable algorithm is a modification of the LMPC-stable algorithm. Phases I, II, IIIa, and IV of the tLMPC-stable algorithm (Algorithms 2, 3 and 5) correspond to the phases I, II, III, and IV of the LMPC-stable algorithm, and these were shown to be stable by Colombo and Maathuis (2014).

Consider first phases I and II of the tLMPC-stable algorithm, which are identical to the same phases of the LMPC-stable algorithm expect for the fact that certain conditioning sets are skipped. Since the omitted sets do not depend on the sequence of the variables, the stability is preserved. 

Consider now phase IIIa of the tLMPC-stable algorithm, which is identical to phase III of the LMPC-stable algorithm expect for the fact that the tLMPC-stable algorithm excludes certain nodes from the conditioning sets. This exclusion is based on the tiered background knowledge, and not on the sequence in which the variables are visited, so the stability is preserved.

Next, consider phase IIIb of the tLMPC-stable algorithm. Since the edge orientation depends on background knowledge, it is independent of the order in which the variables are visited.

Finally, consider phase IV of the tLMPC-stable algorithm, which is identical to phase IV in the LMPC-stable algorithm in which Meek's rules are applied, except for the fact that we have added an additional rule. The order dependence does not occur through the rules, but rather through the way in which we apply them; since we have note modified this part of the phase, the stability is preserved.
\end{proof}

\end{document}